\documentclass[11pt,english]{article}
\usepackage{amssymb}
\usepackage{amsmath}
\usepackage{babel}
\usepackage{latexsym}
\usepackage{graphics}
\usepackage{amsfonts}
\usepackage{hyperref}

\allowdisplaybreaks[3]

\makeatletter
\@addtoreset{equation}{section}
\makeatother

\def\dalemb#1#2{{\vbox{\hrule height .#2pt
        \hbox{\vrule width.#2pt height#1pt \kern#1pt
                \vrule width.#2pt}
        \hrule height.#2pt}}}

\def\cA{{\cal A}}

\def\cM{{\cal M}}

\def\0{{\sst{(0)}}}
\def\1{{\sst{(1)}}}
\def\2{{\sst{(2)}}}
\def\3{{\sst{(3)}}}
\def\4{{\sst{(4)}}}
\def\5{{\sst{(5)}}}
\def\6{{\sst{(6)}}}
\def\7{{\sst{(7)}}}
\def\8{{\sst{(8)}}}
\def\n{{\sst{(n)}}}

\def\ep{\epsilon}
\def\td{\tilde}

\def\half{{\textstyle{\frac{1}{2}}}}
\def\hp{ \frac{1}{2}}

\let\a=\alpha \let\b=\beta \let\g=\gamma \let\d=\delta \let\e=\epsilon
    \let\k=\kappa
\let\l=\lambda \let\m=\mu \let\n=\nu  \let\r=\rho
\let\s=\sigma \let\t=\tau  \let\f=\phi  
\let\w=\omega  \let\D=\Delta  
   \let\F=\Phi 
 \let\W=\Omega   \let\G=\Gamma

\def\nn{\nonumber} \def\bd{\begin{document}} \def\ed{\end{document}}
\def\ds{\documentstyle} \let\fr=\frac \let\bl=\bigl \let\br=\bigr
\let\Br=\Bigr \let\Bl=\Bigl
\let\bm=\bibitem
\let\na=\nabla
\let\pa=\partial \let\ov=\overline
\newcommand{\be}{\begin{equation}}
\newcommand{\ee}{\end{equation}}
\def\ba{\begin{array}}
\def\ea{\end{array}}
\def\ft#1#2{{\textstyle{{\scriptstyle #1}\over {\scriptstyle #2}}}}
\def\fft#1#2{{#1 \over #2}}
\def\del{\partial}
\def\sst#1{{\scriptscriptstyle #1}}
 \def\oneone{\rlap 1\mkern4mu{\rm l}}
\def\ie{{\it i.e.\ }}
\def\via{{\it via}}
\def\semi{{\ltimes}}
\def\str{{\rm str}}
\def\Dm{{{D_{\sst{max}}}}}
\def\vac{ \left | 0 \right \rangle }
\def\kvac{ \left | k \right \rangle }

\def\sp{\; \; \;}

\def\bol{ \left | B (p^+) \right \rangle}
\def\bo1{ \left | B^0 (p^+) \right \rangle}

\def\bolt{ \left | B (p^+) \right \rangle_{\t}}

\def\boxl{ \left | B (x^-) \right \rangle}

\def\<{ \langle }
\def\>{ \rangle }

\def\vf{\varphi}

\def\ls{{(l,0)}}
\def\lv{{(l,\pm1)}}
\def\lt{{(l,\pm2)}}

\def\lse#1{{(l_{#1},0)}}
\def\lve#1{{(l_{#1},\pm1)}}
\def\lte#1{{(l_{#1},\pm2)}}

\def\lsg#1{{5(l_{#1},0)}}
\def\lvg#1{{5(l_{#1},\pm1)}}
\def\ltg#1{{5(l_{#1},\pm2)}}

\def\lsi#1{{5{(#1,0)}}}
\def\lvi#1{{5{(#1,\pm1)}}}
\def\lti#1{{5{(#1,\pm2)}}}

\def\lsr#1{{1{(#1,0)}}}
\def\lvr#1{{1{(#1,\pm1)}}}
\def\ltr#1{{1{(#1,\pm2)}}}

\def\cD{{\cal D}}
\def\cE{{\cal E}}
\def\cF{{\cal F}}
\def\cG{{\cal G}}
\def\cH{{\cal H}}
\def\cK{{\cal K}}
\def\cO{{\cal O}}
\def\cP{{\cal P}}
\def\cQ{{\cal Q}}
\def\cR{{\cal R}}
\def\cS{{\cal S}}
\def\cT{{\cal T}}
\def\cU{{\cal U}}
\def\cV{{\cal V}}
\def\cW{{\cal W}}

\newcommand{\nono}{\nonumber}
\newcommand{\dtilde}[1]{\tilde{\tilde{#1}}}
\newcommand{\hatb}[1]{\hat{\ov{#1}}}
\newcommand{\hatt}[1]{\hat{\tilde{#1}}}
\newcommand{\emnr}{{e_\m}^{\n\r}}
\newcommand{\sub}[1]{\phantom{}_{(#1)}\phantom{}}

\newcommand{\comment}[1]{}

\def\hna{\hat{\na}}

\newcommand{\hsp}{\hspace{0.5cm}}

\newcommand{\ho}[1]{$\, ^{#1}$}
\newcommand{\hoch}[1]{$\, ^{#1}$}
\newcommand{\bea}{\begin{eqnarray}}
\newcommand{\eea}{\end{eqnarray}}
\newcommand{\ra}{\rightarrow}
\newcommand{\lra}{\longrightarrow}
\newcommand{\Lra}{\Leftrightarrow}
\newcommand{\ap}{\alpha^\prime}
\newcommand{\bp}{\tilde \beta^\prime}
\newcommand{\tr}{{\rm tr} }
\newcommand{\Tr}{{\rm Tr} }
\newcommand{\NP}{Nucl. Phys. }

\newcommand{\ams}{{\it Institute for Theoretical Physics,
University of Amsterdam, \\
Valckenierstraat 65, 1018XE Amsterdam, The Netherlands} \\
{\tt I.R.G.Kanitscheider, K.Skenderis, M.Taylor@uva.nl}}
\newcommand{\auth}{{\large Ingmar Kanitscheider, Kostas Skenderis, Marika Taylor }}

\begin{document}
\begin{flushright}
\hfill{ITFA-2008-25}
\end{flushright}

\vspace{15pt}

\begin{center}

{\Large \bf Precision holography for non-conformal branes}

\vspace{20pt}

\auth

\vspace{15pt}

\vspace{8pt}

{\ams}

\vspace{15pt}

\underline{ABSTRACT}
\end{center}
We set up precision holography for the non-conformal branes preserving 16
supersymmetries.
The near-horizon limit of all such $p$-brane solutions
with $p \leq 4$, including the case of fundamental string solutions,
is conformal to $AdS_{p+2} \times S^{8-p}$ with a linear dilaton.
We develop holographic renormalization for all these cases.
In particular, we obtain the most general asymptotic solutions
with appropriate Dirichlet boundary conditions,
find the corresponding counterterms
and compute the holographic 1-point functions,
all in complete generality and at the full non-linear level.
The result for the stress energy tensor properly defines
the notion of mass for backgrounds with such asymptotics.
The analysis is done  both in the original formulation of
the method and also using a radial Hamiltonian analysis.
The latter formulation exhibits most clearly the existence
of an underlying generalized conformal structure. In the cases
of D$p$-branes,  the corresponding dual boundary theory,
the maximally supersymmetric  Yang-Mills theory SYM${}_{p+1}$,
indeed exhibits the generalized conformal structure found at strong coupling.
We compute the holographic 2-point functions of the
stress energy tensor and gluon operator and show they satisfy
the expected Ward identities and the constraints of
generalized conformal structure.
The holographic results are also manifestly compatible with the
M-theory uplift, with the asymptotic solutions, counterterms, one
and two point functions
etc. of the IIA F1 and D4 appropriately descending from those of M2 and M5
branes, respectively. We present a few applications including the
computation of condensates in
Witten's model of holographic YM{}$_4$ theory.

\pagebreak

{\small \setcounter{tocdepth}{2}
\tableofcontents }
\pagebreak

\section{Introduction}

The AdS/CFT correspondence \cite{Maldacena:1997re} is one of the most far
reaching and
important ideas to emerge in recent years. On the one hand it opens a
window into the strong coupling dynamics of gauge theories, whilst on the
other hand it provides a qualitatively new paradigm for gravitational
physics: spacetime is emergent, reconstructed from gauge theory data.
A key ingredient in using gravity/gauge theory duality in such a way
is the holographic dictionary. One needs to know the precise
relationship between bulk and boundary physics before one can use the
weakly coupled description on one side to compute quantities in the
other. In the case of asymptotically $AdS \times X$ backgrounds (with
$X$ compact) the underlying principles of the correspondence were laid
out in the foundational papers on the subject
\cite{Gubser:1998bc,Witten:1998qj}: for every bulk field
$\Phi$ there is a corresponding gauge invariant operator ${\cal
  O}_{\Phi}$ in the boundary theory, and the bulk partition function with given boundary
conditions for $\Phi$ acts as the generating functional for
correlation functions of this operator.

To promote the bulk/boundary correspondence from a formal relation to
a framework in which one can calculate, one needs to specify how
divergences on both sides are treated. In the boundary theory, these
are the UV divergences, which are dealt with by standard techniques of
renormalization. In the bulk, the divergences are due to the infinite
volume, and are thus IR divergences, which need to be dealt with by
holographic renormalization, the precise dual of standard QFT
renormalization \cite{Henningson:1998gx,Balasubramanian:1999re,deHaro:2000xn,
Skenderis:2000in,BFS1,BFS2,Papadimitriou:2004ap,Papadimitriou:2004rz};
for a review see \cite{Skenderis:2002wp}.
The procedure of holographic
renormalization in asymptotically AdS spacetimes allows one to
extract the renormalized one point functions for local gauge invariant
operators from the asymptotics of the spacetime; these can then be
functionally differentiated in the standard way to obtain higher
correlation functions.

By now there are many other conjectured examples of gravity/gauge
theory dualities
in string theory, which involve backgrounds with different
asymptotics. The case of interest for us is the dualities involving
non-conformal
branes \cite{Itzhaki:1998dd,Boonstra:1998mp}
which follow from decoupling limits, and are thus believed to hold, although
rather few quantitative checks of the dualities have been carried
out. It is important to develop our understanding of these dualities for a
number of reasons. First of all, a primary question in quantum gravity
is whether the theory is holographic. Examples such as AdS/CFT
indicate that the theory is indeed holographic for certain spacetime
asymptotics, but one wants to know whether this holds more
generally. Exploring cases where the asymptotics are different but one
has a proposal for the dual field theory is a first step to addressing
this question.

Secondly, the cases mentioned are interesting in their own right
and have many useful applications.
For example, one of the major aims of work in
gravity/gauge dualities is to find holographic models which capture
features of QCD. A simple model which includes confinement and chiral
symmetry breaking can be obtained from the decoupling limit of a
D4-brane background, with D8-branes added to include flavor, the
Witten-Sakai-Sugimoto model
\cite{Witten:1998zw,Sakai:2004cn,Sakai:2005yt}.
This model has been used extensively to
extract strong coupling behavior as a model for that in QCD.
More generally, non-conformal $p$-brane backgrounds with $p=0,1,2$ may have
interesting unexploited applications to condensed matter physics; the
conformal backgrounds have proved useful in modeling strong coupling
behavior of transport properties and the non-conformal examples may be
equally useful.

The non-conformal brane dualities have not been extensively tested,
although some checks of the duality can be found in
\cite{Hashimoto:1999xu,Antonuccio:1999iz,Hiller:2000nf,Morales:2002ys} whilst the papers
\cite{Jevicki:1998yr,Jevicki:1998qs,Jevicki:1998ub}
discuss the underlying symmetry
structure on both sides of the correspondence.
Recently, there has been progress in using lattice methods to
extract field theory quantities, particularly for the
D0-branes \cite{Catterall:2007fp}. Comparing these results to the holographic
predictions serves both to test the duality, and conversely to test lattice
techniques (if one assumes the duality holds).

\bigskip

Given the increasing interest in these gravity/gauge theory dualities,
one would like to develop precision holography for the non-conformal
branes,
following the same steps as in AdS: one wants to know exactly how
quantum field theory data is encoded in the asymptotics of the
spacetime. Precision holography has not previously been extensively
developed for non-conformal branes (see however
\cite{Cai:1999xg,Sekino:1999av,Gherghetta:2001iv,Asano:2003xp,Asano:2004vj}),
although as we will see the analysis is very
close to the analysis of the Asymptotically AdS case.
The reason is that the non-conformal branes admit a generalized
conformal symmetry \cite{Jevicki:1998yr,Jevicki:1998qs,Jevicki:1998ub}:
there is an underlying conformal symmetry
structure of the theory, provided that the string coupling (or in the
gauge theory, the Yang-Mills coupling) is transformed as a background
field of appropriate dimension under conformal transformations. Whilst
this is not a symmetry in the strict sense of the word, the underlying
structure can be used to derive Ward identities and perhaps even prove
non-renormalization theorems.

In this paper we develop in detail how quantum field
theory data can be extracted from the asymptotics of non-conformal
brane backgrounds. We begin in section \ref{two} by recalling the
correspondence between non-conformal brane backgrounds and quantum
field theories. We also introduce the dual frame, in which the near
horizon metric is $AdS_{p+2} \times S^{8-p}$. In section \ref{ldfe} we
give the field equations in the dual frame for both D-brane and
fundamental string solutions.

In the near horizon region of the
supergravity solutions conformal symmetry is broken only by the
dilaton profile. This means that the background admits a generalized
conformal structure: it is invariant under generalized conformal
transformations in which the string coupling is also transformed. This
generalized conformal structure and its implications are discussed in
section \ref{four}.

Next we proceed to set up precision holography. The basic idea is to
obtain the most general asymptotic solutions of the field equations
with appropriate Dirichlet boundary conditions. Given such solutions,
one can identify the divergences of the onshell action, find the
corresponding counterterms and compute the holographic 1-point
functions, in complete generality and at the non-linear level. This is
carried out in section \ref{five}. In particular, we give renormalized
one point functions for the stress energy tensor and the gluon
operator, in the presence of general sources, for all cases.

In section \ref{six} we proceed to develop a radial Hamiltonian
formulation for the holographic renormalization. As in the
asymptotically AdS case, the Hamiltonian formulation is more elegant
and exhibits clearly the underlying generalized conformal structure.
In the following sections, \ref{seven} and \ref{eight}, we give a
number of applications of the holographic formulae. In particular, in
section \ref{seven} we compute two point functions and in section
\ref{eight} we compute condensates in Witten's model of
holographic QCD and the renormalized action, mass etc. in a non-extremal
D1-brane background.

In section \ref{nine} we give conclusions and a summary of our results. The
appendices \ref{appA}, \ref{appb}, \ref{appB} and \ref{D4braneappendix} contain a
number of useful formulae and technical details. Appendix \ref{appA}
summarizes useful formulae for the expansion of the curvature whilst
appendix \ref{appb} discusses the holographic computation of the
stress energy tensor for asymptotically $AdS_{D+1}$, with $D
=4,6$; in the latter the derivation is streamlined, relative to
earlier discussions, and the previously unknown traceless, covariantly constant
contributions to the stress energy tensor in six dimensions are determined.
Appendix \ref{appB} contains the detailed relationship between
the M5-brane and D4-brane holographic analysis whilst appendix
\ref{D4braneappendix} gives explicit expressions for the asymptotic expansion of
momenta.

The results of this work have been reported at a number of recent conferences \cite{confe}.
As this paper was finalized we received \cite{Wiseman:2008qa} which contains related results.

\section{Non-conformal branes and the dual frame} \label{two}

Let us begin by recalling the brane solutions of supergravity, see
for example \cite{Skenderis:1999bs} for a review.
The relevant part of the supergravity action in the string frame is
\be \label{eff_act}
S = \frac{1}{(2 \pi)^7 \a'^4} \int d^{10}x \sqrt{-g} \left [ e^{-2
  \phi} (R + 4 (\pa \phi)^2 - \frac{1}{12} H_3^2 )  - \frac{1}{2
    (p+2)!} F_{p+2}^2 \right ].
\ee
The Dp-brane solutions can be written in the form:
\bea \label{D_sol}
ds^2 &=& (H^{-1/2} ds^2(E^{p,1}) + H^{1/2} ds^2(E^{9-p}) ); \\
e^{\phi} &=& g_s H^{(3-p)/4}; \nn \\
C_{0\cdots p} &=& g_{s}^{-1} (H^{-1} -1) \qquad {\rm or} \qquad
F_{8-p} = g_s^{-1} \ast_{9-p} dH, \nn
\eea
where the latter depends on whether the brane couples electrically or
magnetically to the field strength. Here $g_{s}$ is the string
coupling constant.
We are interested in the simplest supersymmetric
solutions, for which the defining function $H$ is harmonic on the flat space
$E^{9-p}$ transverse to the brane. Choosing a single-centered harmonic
function
\be
H = 1 + \frac{Q_p}{r^{7-p}},
\ee
then the parameter $Q_p$ for the brane solutions of interest is given by
$Q_p=d_p N g_s l_s^{7-p}$ with the constant $d_p$ equal  to
$d_p=(2 \sqrt{\pi})^{5-p} \G(\frac{7-p}{2})$, whilst $l_s^2 = \a'$ and $N$ denotes the integral
  quantized charge.

Soon after the AdS/CFT duality was proposed \cite{Maldacena:1997re},
it was suggested
that an analogous correspondence exists between the near-horizon limits
of non-conformal D-brane backgrounds and (non-conformal) quantum field
theories \cite{Itzhaki:1998dd}. More precisely, one
considers the field theory (or decoupling) limit to be:
\be \label{limit}
g_s \to 0, \qquad \a' \to 0, \qquad U \equiv \frac{r}{\alpha'} ={\rm fixed},
\qquad g_d^2 N ={\rm fixed},
\ee
where $g_d^2$ is the Yang-Mills coupling, related to the string coupling by
\be \label{gd_def}
g_{d}^2 = g_s (2 \pi)^{p-2} (\a')^{(p-3)/2}.
\ee
Note that $N$ can be arbitrary for $p<3$ but
(\ref{limit}) requires that $N \to \infty$ when $p>3$.
The decoupling limit implies that the constant part in the harmonic function is
negligible:
\be
H= 1 +\frac{D_p g_d^2 N}{\a'^2 U^{7-p}} \quad \Rightarrow \quad
\frac{1}{\a'^2} \frac{D_p g_d^2 N}{U^{7-p}},
\ee
where $D_p \equiv d_p (2 \pi)^{2-p}$.

The corresponding dual $(p+1)$-dimensional quantum field theory is
obtained by taking the low energy limit of the $(p+1)$-dimensional
worldvolume theory on $N$ branes. In the case of the $Dp$-branes this
theory is the dimensional reduction of ${\cal N} = 1$ SYM in ten
dimensions. Recall that the action of ten-dimensional SYM is given by
\be
S_{10} = \int d^{10} x \sqrt{-g} {\rm Tr} \left ( - \frac{1}{4
  g_{10}^2} F_{mn} F^{mn} + \frac{i}{2} \bar{\psi} \G^m [ D_{m}, \psi
] \right ),
\ee
 with $D_m = \pa_m - i A_m$. The dimensional reduction to $d$
 dimensions gives the bosonic terms
\be \label{YMd}
S_d =  \int d^{d} x \sqrt{-g} {\rm Tr} \left ( - \frac{1}{4
  g_{d}^2} F_{ij} F^{ij} - \frac{1}{2} D_{i} X D^{i} X
+ \frac{g_d^2}{4} [X,X]^2 \right )
\ee
where $i = 0, \cdots (d-1)$ and there are $(9-p)$ scalars $X$.
The fermionic part of the action will not play a role here. Note that the
Yang-Mills coupling in $d = (p+1)$ dimensions, $g_d^2$, has (length)
dimension $(p-3)$, and thus the theory is not renormalizable for
$p > 3$. Since the coupling constant is dimensionful, the effective
dimensionless coupling constant $g_{eff}^2(E)$ is
\be \label{eff_cou}
g^2_{eff}(E) = g_d^2 N E^{p-3}.
\ee
at a given energy scale $E$.

This discussion of the decoupling limit applies to
D-branes, but we will also be interested in fundamental strings.
The fundamental string solutions can be written in the form:
\bea \label{F1_sol}
ds^2 &=& (H^{-1} ds^2(E^{1,1}) + ds^2(E^{8}) ); \\
e^{\phi} &=& g_s H^{-1/2} ; \nn \\
B_{01} &=& (H^{-1} -1), \nn
\eea
where the harmonic function $H = 1 + Q_{F1}/r^6$ with $Q_{F1}=d_1 N
g_s^2 l_s^{6}$. For completeness, let us also mention that
the NS5-brane solutions can be written in the form:
\bea \label{NS5_sol}
ds^2 &=& (ds^2(E^{1,5}) + H ds^2(E^{4}) ); \\
e^{\phi} &=& g_s H^{1/2}; \nn \\
H_{3} &=& \ast_{4} dH, \nn
\eea
where the harmonic function $H = 1 + Q_{NS5}/r^2$ with
$Q_{NS5}=N l_s^{2}$.

Whilst the fundamental string solutions have a near
string region which is conformal to $AdS_3 \times S^7$ with a linear
dilaton, they do not appear to admit a decoupling limit like the
one in (\ref{limit}) which decouples the  asymptotically flat region
of the geometry and has a clear meaning from the worldsheet point of
view. Nonetheless one
can discuss holography for such conformally $AdS_3 \times S^7$ linear
dilaton backgrounds, using S duality and the relation to M2-branes:
IIB fundamental strings can be
included in the discussion by applying S duality to the D1 brane case,
and IIA fundamental strings by using the fact they are related to M2
branes wrapped on the M-theory circle.

In the cases of Dp-branes the decoupled region is conformal to
$AdS_{p+2} \times S^{8-p}$ and there is a non-vanishing dilaton. The
same holds for the near string region of the fundamental string
solutions. This implies that there is a
Weyl transformation such that the metric is exactly $AdS_{p+2} \times S^{8-p}$.
This Weyl transformation brings the string frame metric $g_{st}$ to the so-called
  {\it dual frame} metric $g_{dual}$ \cite{Boonstra:1998mp} and is given by
\be \label{dualf}
ds^2_{dual} = ( N e^{\phi})^{c} ds_{st}^2, \qquad
\ee
with
\be
c = - \frac{2}{(7-p)} \qquad {\rm Dp}. \qquad
\ee
In this frame the action is
\bea
S = \frac{N^2}{(2 \pi)^7 \a'^4} \int d^{10}x \sqrt{-g} (N e^{\phi})^{\gamma}
(R + 4 \frac{(p-1)(p-4)}{(7-p)^2} (\pa \phi)^2 - \frac{1}{2
    (8-p)! N^2} F_{8-p}^2 ).
\eea
with $\gamma = 2 (p-3)/(7-p)$. It is convenient to express the field strength magnetically; for
$p < 3$ this should be interpreted as $F_{p +2} = \ast F_{8-p}$, with
the Hodge dual being taken in the string frame metric.
The terminology dual frame has the following origin. Each $p$-brane
couples naturally to a $(p+1)$ potential. The corresponding (Hodge) dual
field strength is an $(8-p)$ form. In the dual frame this field strength
and the graviton couple to the dilaton in the same way. For example
the dual frame of the NS5 branes is the string frame:
the dual $(8-p)$ form is $H_3$ and the metric and $H_3$
couple the same way to the dilaton in the string frame, as can be
seen from (\ref{eff_act}).\footnote{The dual frame was originally introduced
in \cite{Duff:1994fg} and the rational
behind its introduction was the following. If one has a formulation
where the fundamental degrees of freedom are $p$-branes that couple
electrically to a $p$-form, then one expects there to exist
non-singular magnetic solitonic solutions. For example,
for perturbative strings, where the elementary objects
are strings, the corresponding magnetic objects, the NS5 branes,
indeed appear as solitonic objects. Moreover, the target space metric
and the $B$ field couple to the the dilaton in the same way, so 
the low energy effective action is in the string frame. 
In a formulation where the elementary degrees of freedom are 
$p$-branes one would
anticipate that there exist smooth solitonic $(6-p)$-brane solutions
of the effective action in the $p$-frame, which is
precisely the dual frame. Indeed, the spacetime
metric of $Dp$-branes when expressed in the dual frame 
is non-singular. We should note
though that there is currently no formulation of string theory where
$p$-branes appear to be the elementary degrees of freedom. Other special
properties of the dual frame solutions are discussed in \cite{Boonstra:1997dy,Boonstra:1998yu}.}

The D5-brane behaves qualitatively differently, as the solution in
  the dual frame is a linear dilaton background with metric $E^{5,1}
  \times R \times S^3$:
\bea
ds^2_{dual} &=&  ds^2 (E^{5,1}) + Q \left (\frac{dr^2}{r^2} +
  d\Omega_3^2 \right );
  \\
e^{\phi} &=& \frac{r}{\sqrt{Q}}; \qquad
F_3 = Q d \Omega_3. \nn
\eea
Holography for both D5 and NS5 branes involves such
linear dilaton background geometries, and will not be discussed further in this paper.

Here we will interested in precision holography for the cases where the geometry is conformal to
$AdS_{p+2} \times S^{8-p}$; this encompasses Dp-branes with $p=
0,1,2,3,4,6$.
In all such cases the dual frame solution takes the form
\bea
ds^2_{dual} &=& \a' d_p^{\frac{2}{(7-p)}} \left ( D_p^{-1} (g_d^2 N)^{-1} U^{5-p}
ds^2 (E^{p,1})+ \frac{dU^2}{U^2}   + d\Omega_{8-p}^2 \right ); \\
e^{\phi} &=&  \frac{1}{N} (2 \pi)^{2-p} D_p^{(3-p)/4}
\left ((g_d^2 N) U^{p-3} \right )^{(7-p)/4}, \nn
\eea
with the field strength being
\be \label{field-s}
F_{8-p} = (7- p) d_p N (\a')^{(7-p)/2}  d\Omega_{8-p}.
\ee
Note that the factors of $\a'$ cancel in the effective supergravity
action, with only dependence on the dimensionful 't Hooft coupling
and N remaining. 

Changing the variable,
\be
u^{2} =  \cR^{-2} (D_p g_d^2 N)^{-1} U^{5-p}, \qquad
\cR = \frac{2}{5-p},
\ee
brings the AdS metric into the standard form
\bea \label{dual_sol}
   ds^2_{dual} &=& \a' d_p^{\frac{2}{7-p}} \left[ \cR^2
\left( \frac{du^2}{u^2} + u^2 ds^2(E^{p,1})\right)
+ d\W_{8-p}^2\right], \\
   e^\phi &=& \frac{1}{N} (2 \pi)^{2-p} (g_d^2 N)^{\frac{(7-p)}{2(5-p)}}
   D_p^{\frac{(3-p)}{2(p-5)}} \left
(\cR^2 u^2 \right )^{\frac{(p-3)(p-7)}{4 (p-5)}}. \nn
\eea
with the field strength being (\ref{field-s}). Note that
by rescaling the metric, dilaton and field strength as
\bea
d{s}^2_{dual} = \a' d_p^{\frac{2}{7-p}} \td{ds}^2; \quad
N e^{\phi} = (2 \pi)^{2-p} (g_d^2 N)^{\frac{(7-p)}{2(5-p)}}
   D_p^{\frac{(3-p)}{2(p-5)}} e^{\td{\phi}}; \quad
F_{8-p} = d_p N (\a')^{(7-p)/2} \td{F}_{8-p}. \nn
\eea
the factors of $D_p$, $N$ and the 't Hooft coupling can be absorbed into the
overall normalization of the action.

It has been argued in \cite{Boonstra:1998mp} that the dual frame
is the holographic frame in the sense that the radial direction $u$
in this frame is identified with the
energy scale of the boundary theory,
\be \label{en-dist}
u \sim E.
\ee
More properly, as we will discuss later, the dilatations of the boundary
theory are identified with rescaling of the $u$ 
coordinate.
Using (\ref{en-dist}) and (\ref{eff_cou}) the dilaton in (\ref{dual_sol})
and for the case of D-branes becomes
\be \label{dil}
e^\f = \frac{1}{N} c_d \left(g^2_{eff}(u)\right)^{\frac{7-p}{2(5-p)}}, \qquad
c_d = (2 \pi)^{2-p} D_p^{\frac{(p-3)}{2 (5-p)}}
\cR^{\frac{(p-3)(7-p)}{2 (5-p)}}.
\ee
The validity of the various approximations
was discussed in \cite{Itzhaki:1998dd,Peet:1998wn,Boonstra:1998mp}. In particular,
we consider the large $N$ limit, keeping fixed the effective coupling constant
$g^2_{eff}$, so the dilaton is small in all cases (recall that the decoupling limit when
$p>3$ requires $N \to \infty$).
If $g^2_{eff} \ll 1$ then the perturbative SYM description is valid,
whereas in the opposite limit $g^2_{eff} \gg 1$ the supergravity approximation is valid.

As a consistency check, one can also derive (\ref{dil}) using the open
string description. The low energy description in the string frame
is given by
\be
S_{st}
= - \frac{1}{(2 \pi)^{p-2} (\a')^{(p-3)/2}} \int d^{p+1}x \sqrt{-g_{st}}
e^{-\phi} \frac{1}{4} \Tr (F_{ij} F_{kl}) g_{st}^{ik} g_{st}^{jl} + \cdots,
\ee
where we indicate explicitly that the metric involved is the string
frame metric. In the case of flat target spacetime, $g_{st}$ is the Minkowski
metric and $e^{\f} = g_s$ and we recover (\ref{gd_def}) by identifying the
 overall prefactor of $\Tr F^2$ with $1/(4 g_d^2)$.
In our case, transforming to the dual frame and using the form of
the metric in  (\ref{dual_sol}) we get
\be \label{sdual}
S_{dual} = - \frac{\cR^{p-3} d_p^{\frac{(p-3)}{(7-p)}}}{(2 \pi)^{p-2}}
\int d^{p+1}x  (N e^\f)^{\frac{2(p-5)}{(7-p)}} (N u^{p-3})
\frac{1}{4} (\Tr F^2)
 + \cdots
\ee
where now the Lorentz index contractions in $\Tr F^2$ are with the Minkowski
metric. Identifying now the overall prefactor of $\Tr F^2$ with $1/(4 g_d^2)$
is indeed equivalent to (\ref{dil}).

As mentioned above, we will also include fundamental strings
in our analysis, exploiting the relation to D1-branes and
M2-branes. In this case we focus on the near string geometry, dropping
the constant term in the harmonic function, and introduce a dual frame
metric $ds^2_{dual} = ( N e^{\phi})^{c} ds_{st}^2$ with
\be \label{f1-dual}
c = -\frac{2}{3} \qquad {\rm F1},
\ee
with the dual frame metric being $AdS_3 \times S^7$. The detailed form
of the effective action
in the dual frame will be given in the next section.

\bigskip

The aim of this paper will be to consider solutions which asymptote to the
decoupled non-conformal brane backgrounds and show how renormalized quantum
field theory information can be extracted from the geometry.
It may be useful to recall first how
the conformal case of $p=3$ works. Given the $AdS_5 \times S^5$
background, the spectrum of supergravity fluctuations about this background
corresponds to the spectrum of single trace gauge invariant chiral primary
operators in the dual ${\cal N} = 4$ SYM theory. The spectrum
includes stringy modes and D-branes, which correspond to other non
primary, high dimension and non-local operators in the dual ${\cal N}
= 4$ SYM theory.
Encoded in the asymptotics of any asymptotically
$AdS_5 \times S^5$ supergravity background are one point functions
of the chiral primary operators. These allow one
to extract the vacuum structure of the dual theory (its vevs and
deformation parameters), and if one switches on sources one can also
extract higher correlation functions.

The sphere in this background has
a radius which is of the same order as the $AdS$ radius, so
the higher KK modes are not suppressed relative to the zero modes
and one cannot ignore them. It is
nevertheless possible to only keep a subset of modes
when the equations of motion admit solutions with all modes
except the ones kept set equal to zero, i.e.
there exist consistent truncations.
The existence of such truncations signify the existence of a
subset of operators of the dual theory that are closed under OPEs.
The resulting theory is a $(d+1)$-dimensional gauged supergravity and such
gauged supergravity theories have been the starting point for
many investigations in AdS/CFT.
Gauged supergravity retains only the duals to low dimension chiral
primaries in SYM, those in the same multiplet as the stress energy
tensor. More recently, the method of Kaluza-Klein holography \cite{Skenderis:2006jq} has been
developed to extract systematically one point functions of all other
single trace chiral operators.

The goal here is to take the first step in holographic
renormalization for non-conformal branes. We will consistently truncate the bulk
theory to just the $(p+2)$-dimensional graviton and the dilaton, and
compute renormalized correlation functions in this sector. Unlike the
$p=3$ case one must retain the dilaton as it is running: the gauge
coupling of the dual theory is dimensionful and runs. Such a
truncation was considered already in \cite{Boonstra:1998mp} and we
will recall the resulting $(p+2)$-dimensional action in the next section.
Given an understanding of holographic renormalization in this
truncated sector, it is straightforward to generalize
this setup to include fields dual to other gauge theory operators.

\section{Lower dimensional field equations}
\label{ldfe}

The supergravity solutions for Dp-branes and fundamental strings in the decoupling limit can
be best analyzed by going to the {\it dual frame} reviewed in the previous section, (\ref{dualf})
and (\ref{f1-dual}). The dual frame is defined as $ds^2_{dual} = (N
e^{\phi})^c ds^2$, with $c = -2/(7-p)$ for Dp-branes and $c = - 2/3$
for fundamental strings. The
Weyl transformation to the dual frame in ten dimensions results in the following
action:
\bea
S = - \frac{N^2}{(2 \pi)^7 \a'^4} \int d^{10}x \sqrt{g} N^{\gamma} e^{\g \phi} [R +
  \beta (\pa \phi)^2 - \frac{1}{2 (8-p)! N^2} |F_{8-p}|^2 ]
\eea
where the constants $(\b,\g)$ are given below in (\ref{constdef}) for
Dp-branes and (\ref{const2}) for fundamental strings respectively.
Note that it is convenient to express the field strength magnetically; for
$p < 3$ this should be interpreted as $F_{p +2} = \ast F_{8-p}$. From
here onwards we will also work in Euclidean signature.

For $p \neq 5$, the field equations in this
frame admit $AdS_{p+2} \times S^{8-p}$ solutions with linear
dilaton. One can reduce the field equations over the sphere, truncating to the
$(p+2)$-dimensional graviton $\td{g}_{\m \n}$ and scalar $\td{\phi}$.
For the Dp-branes the reduction ansatz is
\bea
ds^2_{dual} &=& \a' d_{p}^{-c} ( \cR^2 \td{g}_{\m \n} (x^{\rho}) dx^{\m} dx^{\n} + d
\Omega_{8-p}^2);  \label{reduct} \\
F_{8-p} &=& (7-p) g_s^{-1} Q_p d\Omega_{8-p}; \nn \\
\qquad e^{\phi} &=& g_s (r_o^2\cR^2)^{(p-3)(7-p)/4(5-p)}e^{\td{\phi}}, \nn
\eea
with $r_o^{7-p} \equiv Q_p$ and $\cR = 2/(5-p)$.
The ten-dimensional metric is in the dual frame and prefactors
are chosen to absorb the radius and overall metric and dilaton
prefactors of the $AdS_{p+2}$ solution. For the fundamental string one
reduces the near horizon geometry as:
\bea
ds^2_{dual} &=& \a' (d_{1} N^{-1})^{1/3} ( \cR^2 \td{g}_{\m \n} (x^{\rho}) dx^{\m} dx^{\n} + d
\Omega_{7}^2); \\
H_{7} &=& 6 Q_{F1} d\Omega_{7}; \nn \\
\qquad e^{\phi} &=& g_s (r_o \cR )^{3/2} e^{\td{\phi}}, \nn
\eea
where $H_{7} = \ast H_3$, $r_o^6 \equiv Q_{F1}$ and $\cR = 2/(5-p)$.
It is then straightforward to show that the equations of motion for the
lower-dimensional fields
for both Dp-branes and fundamental strings follow from an action of the form:
\be
\label{Dpaction}
   S =  - L \int d^{d+1} x \sqrt{\tilde{g}}
e^{\g\tilde{\phi}} [\tilde{R} + \beta (\pa \tilde{\phi})^2 + C].
\ee
Here $d = p+1$ and the constants $(L,\beta,\g,C)$ depend on the case of
interest; since from here onwards we are interested only in
$(d+1)$-dimensional fields we suppress their tilde labeling.
For Dp-branes the constants are given by
\bea
\label{constdef}
   \g &=& \frac{2(p-3)}{7-p}, \qquad
   \b = \frac{4(p-1)(p-4)}{(7-p)^2}, \nono \\
   \cR &=& \frac{2}{5-p}, \qquad
   C =  \hp(9-p)(7-p)\cR^2, \\
L &=& \frac{\Omega_{8-p} r_o^{(7-p)^2/(5-p)}
\cR^{(9-p)/(5-p)}}{(2 \pi)^7 \a'^4}
=  \frac{(d_p N)^{(7-p)/(5-p)} g_d^{2(p-3)/(5-p)}
\cR^{(9-p)/(5-p)}}{64 \pi^{(5+p)/2}(2\pi)^{(p-3)(p-2)/(5-p)} \G(\frac{9-p}{2})}. \nono
\eea
For the fundamental string one gets instead:
\bea
\g &=& \frac{2}{3}, \qquad
\b =0, \qquad C = 6, \label{const2} \\
L &=& \frac{\Omega_7 r_o^9}{4 (2 \pi)^7 g_s^2 (\a')^4} = \frac{g_s N^{3/2}
  (\a')^{1/2}}{6 \sqrt{2}}, \nn
\eea
This expression is related to that for the D1-brane background by $g_s
\rightarrow 1/g_s$ with $\a' \rightarrow \a' g_s$, as one would expect from S duality.
The truncation is consistent, as one can show that any solution of the lower-dimensional
equations of motion also solves the ten-dimensional equations of
motion, using the reduction given in (\ref{reduct}). Note that more
   general reductions of type II theories on spheres to give gauged
   supergravity theories were discussed in
   \cite{Cvetic:2000ah}. These reductions would be relevant if one
   wants to include additional operators in the boundary theory,
   beyond the stress energy tensor and scalar operator.

In both cases the equations of motion admit an $AdS_{d+1}$ solution
\bea
\label{AdSsol}
   ds^2 &=& \frac{d\r^2}{4\r^2} + \frac{dx_i dx^i}{\r}; \\
   e^\phi &=& \r^\a, \nn
\eea
where $i=1, \ldots, d$. Note that $\r$ is related to the radial
coordinate $u$ used earlier by $\r=1/u^2$.
The constant $\a$ again depends on the case of
interest:
\bea
\a &=& -\frac{(p-7)(p-3)}{4(p-5)};  \qquad {\rm Dp} \\
\a &=& - \frac{3}{4}; \qquad {\rm F1}. \nn
\eea
Note that for computational convenience the metric and dilaton have
been rescaled relative to
\cite{Boonstra:1998mp} to set the AdS radius to one and to pull all
factors of $N$ and $g_{s}$ into an overall normalization factor. The radial
variable $\r$ then has length dimension 2
and $e^\phi$ has length dimension $2\a$.

For arbitrary $d,\b$ and $\g$, the field equations for
the metric and scalar field following from \eqref{Dpaction} are
\footnote{Our conventions for the Riemann and Ricci tensor are
${R^\s}_{\m\n\r} = - 2 \G^\s_{\m[\n,\r]} - 2 \G^\t_{\m[\n} \G^\s_{\r]\t},
R_{\m\n} = {R^\s}_{\m\s\n}$.}
\begin{gather}
    -R_{\m\n} + (\g^2-\b) \pa_\m \phi \pa_\n \phi + \g \na_\m \pa_\n
    \phi + \hp g_{\m\n}[R +
(\b-2\g^2) (\pa \phi)^2 - 2\g \na^2 \phi + C] = 0,  \nono \\
    \g R - \b \g (\pa \phi)^2 + C \g - 2 \b \na^2 \phi = 0. \label{Dpfelong}
\end{gather}
These equations admit an $AdS$ solution with linear dilaton provided
that $\a$ and $C$ satisfy
\be \label{adef}
  \a = - \frac{\g}{2(\g^2-\b)}, \qquad C = \frac{(d(\g^2-\b)+\g^2)(d(\g^2-\b)+\b)}{(\g^2-\b)^2}.
\ee
We can thus treat both Dp-brane and fundamental string cases
simultaneously, by processing the field equations for arbitrary
$(d,\b,\g)$ and writing $(\a,C)$ in terms of these parameters.
It might be interesting to consider whether other choices of
$(d,\b,\g)$ admit interesting physical interpretations.

By taking the trace of the first equation in
\eqref{Dpfelong} and combining it with the second one can obtain the more convenient three equations
\bea
    -R_{\m\n} + (\g^2-\b) \pa_\m \phi \pa_\n \phi +
\g \na_\m \pa_\n \phi  -\frac{\g^2 + d(\g^2-\b)}{\g^2-\b} g_{\m\n} &=&
0, \label{Dpfe} \\
    \na^2 \phi + \g (\pa \phi)^2  - \frac{\g(d(\g^2-\b)+\g^2)}{(\g^2-\b)^2} &=& 0, \nono \\
    R + \b (\pa \phi)^2 + \frac{(d(\g^2-\b) +\g^2)(d(\g^2-\b)-\b)}{(\g^2-\b)^2} &=& 0, \nn
\eea
where the last line follows from the first two.

\bigskip

The type IIA fundamental strings and D4-branes are related to the M
theory M2-branes and M5-branes respectively under dimensional
reduction along a worldvolume direction. The M brane theories fall
within the framework of AdS/CFT, with the correspondence being between $AdS_4 \times
S^7$ and $AdS_7 \times S^4$ geometries, respectively, and the still poorly
understood conformal worldvolume theories. Reducing on the spheres
gives four and seven dimensional gauged supergravity, respectively,
which can be truncated to Einstein gravity with negative cosmological
constant. That is, the effective actions are simply
\be \label{Ein}
S_M = - L_M \int d^{d+2}x \sqrt{G} \left (R(G) + d(d+1) \right ),
\ee
where $d=2$ for the M2-brane and $d=5$ for the M5-brane. The
normalization constant is
\be
L_{M2} = \frac{\sqrt{2} N^{3/2}}{24 \pi}; \qquad
L_{M5} = \frac{N^3}{ 3 \pi^3}.
\ee
and the action clearly admits an $AdS_{d+2}$-dimensional space with
unit radius as a solution:
\be
ds^2 = \frac{d\rho^2}{4 \rho^2} + \frac{1}{\rho} (dx_i dx^i + dy^2),
\ee
where $i = 1,\cdots,d$.

Now consider a diagonal dimensional reduction
of the $(d+2)$-dimensional solution over $y$, i.e. let the metric be
\be \label{reduc1}
ds^2 = g_{\m\n}(x) dx^{\m} dx^{\n} + e^{4\phi (x)/3} dy^2.
\ee
Substituting into the $(d+2)$-dimensional field equations gives
precisely the field equations following from the action
\eqref{Dpaction}; note that $\g = 2/3, \b = 0$ for both the fundamental string
and D4-branes. It may be useful to recall here that the standard dimensional
reduction of an M theory metric to a (string frame) type IIA metric
$g_{MN}$ is
\be \label{M_cir}
ds^2_{11} = e^{-2 \phi/3} g_{MN} dx^M dx^N  + e^{4\phi/3} dy_{11}^2.
\ee
The relation between dual frame and string frame metrics given in
(\ref{dualf}) leads to (\ref{reduc1}). Note that
\be
L = L_{M} (2 \pi R_y) = 2 \pi g_s l_s L_M,
\ee
where we use the standard relation for the radius of the M theory
circle.

The other Dp-branes of type IIA are of course also related to M theory
objects: the D0-brane background uplifts to a gravitational wave
background, the D6-brane background uplifts to a Kaluza-Klein monopole
background whilst the D2-branes are related to the reduction of
M2-branes transverse to the worldvolume. These connections will not
play a role in this paper. The uplifts reviewed above are useful here as
holographic renormalization for the conformal branes is well
understood, but holography for gravitational wave backgrounds and
Kaluza-Klein monopoles is less well understood than that for the
non-conformal branes.

One could use a different reduction and
truncation of the theory in the $AdS_4 \times S^7$ background to obtain the
action \eqref{Dpaction} for D2-branes. In this case one would embed
the M theory circle into the $S^7$, and then
truncate to only the four-dimensional graviton, along with the scalar
field associated with this M theory circle. This reduction will not
however be used here.

\section{Generalized conformal structure} \label{four}

In this section we will discuss the underlying generalized conformal structure
of the non-conformal brane dualities. Recall that the
corresponding worldvolume theory is SYM${}_{p+1}$.
We will be interesting in computing correlation
functions of gauge invariant operators in this theory.
Recall that gauge/gravity duality
maps bulk fields to boundary operators. In our discussion in the
previous section we truncated the bulk theory to gravity coupled to
a scalar field in $(d+1)$ dimensions. The bulk metric corresponds to the
stress energy tensor as usual,
while as we will see the scalar field corresponds to
a scalar operator of dimension four. As usual the fields that parametrize
their boundary conditions are identified with sources that couple
to gauge invariant operators.

Consider the following $(p+1)$-dimensional (Euclidean) action,
\bea \label{YM_s}
S_d[g_{(0)ij}(x), \F_{(0)}(x)] &=& - \int d^{d} x \sqrt{g_{(0)}}
\left ( -\F_{(0)} \frac{1}{4} {\rm Tr} F_{ij} F^{ij}
+ \frac{1}{2} {\rm Tr} \left(X (D^2 - \frac{(d-2)}{4 (d-1)} R) X\right)
\right. \nonumber \\
&& \left.\qquad \qquad \qquad \qquad  + \frac{1}{4 \F_{(0)}}  {\rm Tr} [X,X]^2 \right ).
\eea
where $g_{(0)ij}$ is a background metric $\F_{(0)}(x)$ is a scalar
background field. Setting
\be
g_{(0)ij}=\delta_{ij}, \qquad \F_{(0)}=\frac{1}{g_{d}^{2}},
\ee
the action (\ref{YM_s}) becomes equal to
the action of the SYM${}_{p+1}$ given in (\ref{YMd}) (here and it what follows
we suppress the fermionic terms).
The action (\ref{YM_s}) is invariant under the following Weyl transformations
\be
g_{(0)} \to e^{2 \s} g_{(0)}, \quad X \to e^{(1-\frac{d}{2}) \s} X, \quad
A_i \to A_i, \quad \F_{(0)} \to e^{-(d-4)\s} \F_{(0)}
\ee
Note that the combination
$P_1 = D^2 - \frac{d-2}{4 (d-1)} R$,
is the conformal Laplacian in $d$ dimensions, which transforms under Weyl
transformations as $P_1 \to e^{-(d/2+1)\s} P_1 e^{(d/2-1)\s}$.

Let us now define,
\be
T_{ij} = \frac{2}{\sqrt{g_{(0)}}} \frac{\d S_d}{\d g_{(0)}^{ij}}, \qquad
{\cal O}  = \frac{1}{\sqrt{g_{(0)}}} \frac{\d S_d}{\d \F_{(0)}}
\ee
They are  given by
\bea \label{tij}
T_{ij} &=& \Tr \left(\F_{(0)} F_{ik} F_j{}^k + D_i X D_j X
+ \frac{d-2}{4 (d-1)} (X^2 R_{ij} - D_i D_j X^2 + g_{(0)ij} D^2 X^2)
\right.\nonumber \\
&& \left.- g_{(0)ij} \left(\frac{1}{4} \F_{(0)} F^2
+ \frac{1}{2} (D X)^2 + \frac{(d-2)}{8 (d-1)} R X^2
- \frac{1}{4 \F_{(0)}} [X,X]^2\right)\right) \\
{\cal O} &=&  \Tr \left(\frac{1}{4} F^2
+ \frac{1}{4 \F_{(0)}^2} [X,X]^2\right).
\eea

Using standard manipulations, see for example \cite{BFS1,BFS2}, we obtain the
standard diffeomorphism and trace Ward identities,
\bea \label{WI}
&&\nabla^{j} \langle T_{ij} \rangle_J + \langle {\cal O}\rangle_J
\pa_i \Phi_{(0)} = 0, \\
&&\langle T^{i}_{i}\rangle_J + (d-4) \Phi_{(0)}
\langle {\cal O} \rangle_J = 0, \label{trace}
\eea
where $\langle B \rangle_J$ denotes an expectation value of $B$
in the presence of sources $J$. One can verify that these
relations are satisfied at the classical level,
i.e. by using (\ref{tij}) and the equations of motion that follow
from (\ref{YM_s}). Setting $g_{(0)ij}=\delta_{ij}, \F_{(0)} = g_d^{-2}$
one recovers the conservation of the energy momentum tensor of
the SYM${}_{d}$ theory and the
fact that conformal invariance is broken by the dimensionful
coupling constant. Note that the kinetic part of the scalar
field does not contribute to the breaking of conformal invariance
because this part of the action is conformally invariant
in any dimension (using the conformal Laplacian). This also dictates
the position of the coupling constant in (\ref{YMd}). In a flat background
one can change the position of the coupling constant by rescaling the
fields. For example, by rescaling $X \to X/g_d$ the coupling constant
becomes an overall constant. This is the normalization one gets from
worldvolume D-brane theory in the string frame. This action however
does not generalize naturally to a Weyl invariant action. Instead
it is (\ref{YMd}) (with the coupling constant promoted
to a background field)
that naturally couples to a metric in a Weyl invariant way.

The Ward identities (\ref{WI}) lead to an infinite
number of relations for correlation functions obtained
by differentiating with respect to the sources and
setting the sources to
$g_{(0)ij} = \eta_{ij}$, where $\eta_{ij}$ is the Minkowski metric and
$\Phi_{(0)}= 1/g_{d}^2$.
The first non-trivial relations are at the level of 2-point functions
($x \neq 0$).
\bea \label{WI-2pt}
&&\partial_x^{j} \langle T_{ij}(x) T_{kl}(0) \rangle =0, \qquad
\partial_x^{j} \langle T_{ij}(x) {\cal O}(0) \rangle =0 \\
&&\langle T^{i}_{i}(x) T_{kl}(0)\rangle + (p-3) \frac{1}{g_d^2}
\langle {\cal O}(x) T_{kl}(0) \rangle = 0 \nonumber \\
&&\langle T^{i}_{i}(x) {\cal O}(0)\rangle + (p-3) \frac{1}{g_d^2}
\langle {\cal O}(x) {\cal O}(0) \rangle = 0.
\nonumber
\eea

The Ward identities (\ref{WI}) were derived by formal path integral
manipulations and one should examine whether they really hold at the
quantum level. Firstly, for the case of the D4 brane the worldvolume
theory is non-renormalizable, so one might question whether the correlators
themselves are meaningful. At weak coupling, renormalizing the
correlators would require introducing new higher dimension operators
in the action, as well as counterterms that depend on the background
fields. This process should preserve diffeomorphism and supersymmetry,
but it may break the Weyl invariance. Introducing a new source
$\Phi_{(0)}^j$ for
every new higher dimension operator ${\cal O}_j$  added in the process of
renormalization would then modify the trace Ward identity as
\be
\< T^{i}_{i} \> - \sum_{j\geq0} (d-\D_j) \F_{(0)}^j \< {\cal O}_j \>
= {\cal A},
\ee
where $\D_j$ is the dimension of the operator ${\cal O}_i$
(with $\F_{(0)}^0 = \F_{(0)}, {\cal O}_0 ={\cal O}, \D_0=4$). Due to
supersymmetry one would anticipate that $\D_i$ are protected.
One would also anticipate that these operators are dual to the KK modes
of the reduction over the sphere $S^{8-p}$. As discussed in the
previous section, one can consistently truncate these modes at
strong coupling, so the gravitational computation should
lead to Ward identities of the
form (\ref{trace}), up to a possible quantum anomaly ${\cal A}$.
${\cal A}$ originates from the counterterms
that depend on the background fields only ($g_{(0)}, \Phi_{(0)}, \ldots$).
In general, ${\cal A}$ would be restricted by the Wess-Zumino consistency
and therefore should be built from generalized conformal
invariants. {\it We will show the extracted holographic Ward identities,
(\ref{ward-hol}), indeed agree with (\ref{WI})-(\ref{trace}))
with a quantum anomaly only for  $p=4$.}

\bigskip

In a $(p+1)$-dimensional conformal field theory, the entropy $S$ at finite
temperature $T_{H}$ necessarily scales as
\be
S = c(g^2_{YM} N,N, \cdots) V_{p} T_{H}^p
\ee
where $V_p$ is the spatial volume, $g_{YM}$ is the coupling, $N$ is
the rank of the gauge group, $g^2_{YM} N$ is the 't Hooft coupling constant
and the ellipses denote additional
dimensionless parameters.  $c(g^2_{YM} N,N, \cdots)$ denotes an arbitrary
function of these dimensionless parameters.
In the cases of interest here, scaling indicates that the entropy
behaves as
\be \label{SDp}
S = \td{c}( (g_{eff}^2(T_{H}),N, \cdots) V_{p} T_{H}^p,
\ee
where $g_{eff}^2(T_H) = g_{d}^2 N T_{H}^{p-3}$ is the effective
coupling constant and $\td{c}( (g_{d}^2 N T_{H}^{p-3}) ,N, \cdots)$
denotes a generic
function of the dimensionless parameters.

Next let us consider correlation functions, in particular of the
gluon operator ${\cal O} = -\frac{1}{4} \Tr (F^2 + \cdots)$. In a theory
which is conformally invariant the two point function of any operator of
dimension $\Delta$ behaves as
\be \label{conf-corr}
\< {\cal O} (x) {\cal O} (y) \> = f(g^2_{YM} N,N, \cdots)
\frac{1}{|x -y|^{2\Delta}},
\ee
where $f(g^2_{YM} N,N,\cdots)$ denotes an arbitrary function of the
dimensionless parameters. Now consider the constraints on a two point
function in a theory with generalized conformal invariance; these are
far less restrictive, with the correlator constrained to be of the form:
\be \label{g1-conf-corr}
\< {\cal O} (x) {\cal O} (0) \> = \td{f}(g_{eff}^2(x),N, \cdots)
\frac{1}{|x|^{2 \D}}.
\ee
where $g^2_{eff}(x)=g_{d}^2 N |x|^{3-p}$ and
$\td{f}(g^2_{eff}(x),N, \cdots)$ is an arbitrary function of
these (dimensionless) variables. Note that the
scaling dimension of the gluon operator as defined above is 4.
Both (\ref{conf-corr}) and (\ref{g1-conf-corr}) are
over-simplified as even in a conformal field theory the renormalized
correlators can depend on the renormalization group scale $\mu$. For
example, for $p=3$ the renormalized two point function of the
dimension four gluon operator  is
\be
\< {\cal O} (x) {\cal O} (0) \> = f(g^2_{YM} N,N) \Box^3 \left (
\frac{1}{|x|^2} \log (\mu^2 x^2) \right ),
\ee
where note that the renormalized version ${\cal R} \frac{1}{|x|^{8}}$
of $\frac{1}{|x|^{8}}$ is given by:
\be \label{ren-x}
{\cal R} \left ( \frac{1}{|x|^{8}} \right ) = - \frac{1}{3 \cdot 2^8}  \Box^3 \left (
\frac{1}{|x|^2} \log (\mu^2 x^2) \right ).
\ee
${\cal R} (\frac{1}{|x|^{8}})$
and $\frac{1}{|x|^{8}}$ are equal when $x \neq 0$ but they differ by
infinite renormalization at $x =0$. In particular, it is only
${\cal R} \frac{1}{|x|^{8}}$ that has a well defined Fourier transform,
given by $p^4 \log (p^2/\mu^2)$, which may be obtained using
the identity
\be
\int d^4 x e^{i p x} \frac{1}{|x|^{2}} \log (\mu^2 x^2) =
- \frac{4 \pi^2}{p^2} \log (p^2/\mu^2).
\ee
(see appendix A, \cite{Dan}). Thus the correlator in a theory with
generalized conformal invariance is
\be \label{g-conf-corr}
\< {\cal O} (x) {\cal O} (0) \> = {\cal R} \left (
\td{f}(g_{eff}^2(x),\mu|x|,N, \cdots) \frac{1}{|x|^{2 \D}} \right ).
\ee
Note that this is of the same form as a two point function of
an operator with definite scaling dimension in any quantum field
theory; the generalized conformal structure does not restrict it
further, although as discussed above the underlying structure does
relate two point functions via Ward identities.

The general form of the two point function (\ref{g-conf-corr})
is compatible with the holographic results discussed later. One
can also compute the two point function to leading (one loop) order in
perturbation theory, giving:
\be
\< {\cal O} (x) {\cal O} (0) \> =
\< : \Tr(F^2) (x): : \Tr(F^2) (0): \> \sim {\cal R} \left(
\frac{g_{eff}^4(x)}{|x|^{8}} \right ),
\ee
which is also compatible with the general form. (Note that although
the complete operator includes in addition other bosonic and
fermionic terms the latter do not contribute to the two point function
at one loop, whilst the former contribute only to the overall normalization.)
One shows this result as
follows. The gauge field propagator for $SU(N)$ in Feynman gauge in momentum space is
\be
\< A^{a}_{b\m} (k) A^{c}_{d\n} (-k)
\> = i g_d^2 (\d^{a}_{d} \d^{c}_{b} - \frac{1}{N} \d^{a}_{b} \d^{c}_d)
\frac{\eta_{\m\n}}{|k|^2},
\ee
where $(a,b)$ are color indices. Then the one loop contribution
to the correlation function in momentum space reduces (at large $N$) to
\be
\< {\cal O} (k) {\cal O} (-k) \> \sim N^2 (d-1) |k|^{4} \int {d^d
q} \frac{1}{|q|^2 |k-q|^2}.
\ee
Using the integral
\bea
I &=& \int d^{d}q \frac{1}{|q|^{2\a} |k-q|^{2 \beta}} \\
&=& \frac{\Gamma(\a + \b -d/2) \Gamma(d/2 -\beta) \Gamma(d/2 -\a)}{\Gamma(\a)
\Gamma(\b) \Gamma(d- \alpha - \beta)} |k|^{d - 2 \a - 2 \b}, \nn
\eea
one finds that
\be
\< {\cal O} (k) {\cal O} (-k) \> \sim N^2 (g_{d}^2)^2 (d-1) |k|^{d}
\frac{ \Gamma(2 -d/2) (\Gamma(d/2-1))^2}{\Gamma(d-2)}.
\ee
This is finite for $d$ odd, as expected given the general result that
odd loops are finite in odd dimensions; dimensional
regularization when $d$ is even
results in a two point function of the form
$N^2 g_{d}^4 |k|^d \log(|k^2|)$. Fourier transforming back
to position space results in
\be
\< {\cal O} (x) {\cal O} (0) \> \sim {\cal R} \left
(\frac{g_{eff}^4(x)}{|x|^{8}} \right ),
\ee
where again in even dimensions the renormalized expression is of the
type given in
(\ref{ren-x}). This is manifestly consistent with the form (\ref{g-conf-corr}).

The structure that we find at weak coupling is also visible at
strong coupling. The gravitational solution is the
linear
dilaton $AdS_{d+1}$ solutions in (\ref{AdSsol}) and conformal symmetry is
broken only by the dilaton profile. Therefore the background is
invariant under generalized conformal transformations in which one
also transforms the string coupling $g_s$ appropriately. This
{\it generalized conformal structure} was discussed in \cite{Jevicki:1998yr,
  Jevicki:1998qs,Jevicki:1998ub}, particularly in the context of
  D0-branes.

\section{Holographic renormalization} \label{five}

In this section we will determine how gauge theory data is extracted
from the asymptotics of the decoupled non-conformal brane
backgrounds, following the same steps as in the asymptotically AdS
case. In particular, one first fixes the non-normalizable part of the
asymptotics: we will consider solutions which asymptote to a linear
dilaton asymptotically locally AdS background. Next one needs to
analyze the field equations in the asymptotic region, to understand
the asymptotic structure of these backgrounds near the boundary.

Given this analysis, one is ready to proceed with
holographic renormalization. Recall that the aim of holographic
renormalization is to render well-defined the definition of
the correspondence: the onshell bulk action with given boundary values
$\Phi_{(0)}$ for the bulk fields acts as the generating functional for
the dual quantum field theory in the presence of sources $\Phi_{(0)}$
for operators ${\cal O}$. The
asymptotic analysis allows one to isolate the volume divergences of the
onshell action, which can then be removed with local covariant
counterterms, leading to a renormalized action. The latter allows one
to extract renormalized correlators for the quantum field theory.

\subsection{Asymptotic expansion}
\label{asymptfe}

In determining how gauge theory data is encoded in
the asymptotics of the non-conformal brane backgrounds the first step
is to understand the asymptotic structure of these backgrounds in the
asymptotic region near
$\rho = 0$ where the solution becomes a linear dilaton locally AdS background.
Let us expand the metric and dilaton as:
\bea
   ds^2 &=& \frac{d\r^2}{4\r^2} + \frac{g_{ij}(x,\r)dx^i dx^j}{\r},
\label{fg} \\
   \phi(x,\r) &=& \a \log \r + \frac{\k(x,\r)}{\g}, \nn
\eea
where we expand $g(x,\r)$ and $\k(x,\r)$ in powers of $\r$:
\bea
\label{fieldexp}
   g(x,\r) &=& g_{(0)}(x) + \r g_{(2)}(x) +\cdots  \\
   \k(x,\r) &=& \k_{(0)}(x) + \r \k_{(2)} (x) + \cdots \nono
\eea
For $p=3$ we should instead expand the scalar field as
\be
\phi(x,\r) = {\k}_{(0)} (x) + \r \k_{(2)} (x) + \cdots,
\ee
since $\a = \g=0$. Note that by allowing $(g_{(0)}, \kappa_{(0)})$ to
   be generic the spacetime is only asymptotically locally AdS.

Consider first the case of $p=3$, so that the action is Einstein
gravity in the presence of a negative cosmological constant, and a
massless scalar. The latter couples to the dimension four operator
${\rm Tr}(F^2)$. The metric is expanded in the
Fefferman-Graham form, with the scalar field expanded accordingly. By
the standard rules of AdS/CFT $g_{(0)}$ acts as the source for the
stress energy tensor and ${\k}_{(0)}$ acts as the source for the
dimension four operator, i.e. it corresponds to the Yang-Mills
coupling.  The vevs of these operators are captured by subleading
terms in the asymptotic expansion.

For general $p$ an analogous relationship should hold: $g_{(0)}$ sources
the stress energy tensor and the scalar field determines the
(dimensionful) gauge coupling. More precisely, the bulk field that
is dual to the operator ${\cal O}$ in (\ref{tij}) is
\bea \label{dualF}
\Phi(x,\r) &=& \exp \left(\chi \phi(x,\r)\right) =
\r^{-\half (p-3)}\left(\F_{(0)}(x) + \r \F_{(2)}(x) + \cdots \right) \\
\F_{(0)}(x) &=& \exp \left(-\frac{(p-5)}{(p-3)} \k_{(0)}(x)\right)
\eea
The $\F_{(0)}$ appearing here is identified with $\F_{(0)}$  in (\ref{YM_s}).
It will be convenient however to work on the gravitational side
with $\phi(x,\r)$ instead of $\Phi(x,\r)$.

In the asymptotic expansion we fix the
non-normalizable part of the asymptotics, and the vevs should be captured by
subleading terms. One now needs to show
that such an expansion is consistent with the equations of motion,
and what terms occur in the expansion for given $(\a,\b,\g)$.

Substituting the scalar and the metric given in (\ref{fg}) into the
field equations \eqref{Dpfe} gives
\bea
&&\hspace{-0.5cm}    -\frac{1}{4} \Tr(g^{-1}g')^2 + \hp \Tr g^{-1} g''
+ \k'' + (1-\frac{\b}{\g^2})(\k')^2 = 0, \label{tra-eq} \\
&& \hspace{-0.5cm}    -\hp\na^i {g'}_{ij} + \hp \na_j
(\Tr g^{-1} g') + (1-\frac{\b}{\g^2})\pa_j \k \k' +
\pa_j \k' - \hp {g'_j}^k \pa_k \k = 0, \label{div-equ} \\
&& \hspace{-0.5cm}    \left[-Ric(g) -(d-2-2\a\g)g' - \Tr(g^{-1}g')g + \r(2g''
- 2g'g^{-1}g'+ \Tr(g^{-1}g')g')\right]_{ij}    \nono \\
&&    + \na_i \pa_j \k +
    (1-\frac{\b}{\g^2})\pa_i \k
\pa_j \k -2(g_{ij} - \r g'_{ij}) \k' = 0, \label{Ricasympt} \\
&&\hspace{-0.5cm}
4\r(\k'' + (\k')^2) +(8\a\g +2(2-d))\k' + \na^2 \k + (\pa \k)^2
    + 2\Tr(g^{-1} g')(\a\g + \r\k') = 0 \label{scalasympt},
\eea
where differentiation with respect to $\r$ is denoted with a prime,
$\na_i$ is the covariant derivative constructed from the metric $g$
and $d=p+1$ is the dimension of the space orthogonal to $\r$.
Note that coefficients
in these equations are polynomials in $\rho$  implying that this
system of equations admits solutions with $g(x,\rho)$ and $\k(x,\r)$
being regular
functions of $\rho$ and this justifies (\ref{fieldexp}). To solve
these equations one may successively differentiate the equations w.r.t.
$\rho$ and then set $\rho=0$.

\bigskip

Let us first recall how these equations are solved in the
pure gravity, asymptotically locally $AdS_{d+1}$ case, i.e. when the scalar is trivial. Then the
equations become
\bea
&&     -\frac{1}{4} \Tr(g^{-1}g')^2 + \hp \Tr g^{-1} g'' = 0; \qquad
    -\hp\na^i {g'}_{ij} + \hp \na_j
(\Tr g^{-1} g') = 0 \label{fg2} \\
&&    \left[-Ric(g) -(d-2 )g' - \Tr(g^{-1}g')g + \r(2g'' - 2g'g^{-1}g' +
      \Tr(g^{-1}g')g')\right]_{ij} = 0, \nn
\eea
The structure of the expansions depends on whether $d$ is even or
odd. For $d$ odd, the expansion is of the form
\be
   g(x,\r) = g_{(0)}(x) + \r g_{(2)}(x) + \cdots + \r^{d/2}
   g_{(d)} (x) + \cdots.
\ee
Terms with integral powers of $\r$ in the expansion are determined
locally in terms of $g_{(0)}$ but $g_{(d)}(x)$ is not determined by
$g_{(0)}$, except for its trace
and divergence, i.e. $g_{(0)}^{ij} g_{(d)ij}$ and
$\nabla^i g_{(d)ij}$, which are forced by
the field equations to vanish. In this case $g_{(d)}(x)$ determines
the vev of the dual stress energy tensor, whose trace must vanish as
the theory is conformal and there is no conformal anomaly in odd dimensions.
The fact that $g_{(d)}$ is divergenceless leads to the conservation of the
stress energy tensor.

For $d$ even, the structure is rather different:
\be
   g(x,\r) = g_{(0)}(x) + \r g_{(2)}(x) + \cdots + \r^{d/2}(
   g_{(d)} (x) +  h_{(d)}(x) \log \rho) + \cdots.
\ee
In this case one needs to include a logarithmic term to satisfy the field
equations; the coefficient of this term is determined by $g_{(0)}$
whilst only the trace and divergence of $g_{(d)}(x)$
are determined by $g_{(0)}$.
This structure reflects the fact that the trace of the stress energy
tensor of an even-dimensional conformal field theory on a curved
background is non-zero and
picks up an anomaly determined in terms of $g_{(0)}$; the explicit
expression for the stress energy tensor in terms of
$(g_{(0)},g_{(d)})$ is rather more complicated than in the other case
but it is such that the divergence of $g_{(d)}$ leads again to conservation
of the stress energy tensor.

\bigskip

Let us return now to the cases of interest. As mentioned above, the
field equations are solved by successively differentiating the equations
w.r.t. $\rho$ and then setting $\rho$ to zero.
This procedure leads to equations of the form
\be
c(n,d) g_{(2n)ij} = f(g_{(2k)ij}, \k_{(2k)}), \qquad k<n
\ee
where the right hand side depends on the lower order coefficients and
$c(n,d)$ is a numerical coefficient that depends on $n$ and $d$. If
this coefficient is non-zero, one can solve this equation to determine
$g_{(n)ij}$. However, in some cases this coefficient is zero
and one has to include a logarithmic term at this order
 for the equations to have a solution.
An example of this is the case of pure gravity with $d$ even,
where $c(d/2,d)=0$.
Furthermore, note that since in \eqref{Ricasympt} -\eqref{scalasympt} only
integral powers of $\r$ enter, likewise only integral powers in
\eqref{fieldexp} will depend on $g_{(0)}$ and $\k_{(0)}$. In general
however non-integral powers can also appear at some order and one must
determine these terms separately. An example of this is the case of pure gravity with $d$ odd
reviewed above, where a half integral power of $\r$ appears at order
$\r^{d/2}$.

Let us first consider when one needs to include non-integral powers in
the expansion. Let us assume that $\rho^\s$ is the lowest non-integral power
that appears in the asymptotic expansion
\bea
  \k(x,\r) &=& \k_{(0)} + \r \k_{(2)} + \cdots + \r^\s \k_{(2\s)}
+ \cdots \\
g_{ij}(x,\r) &=& = g_{(0)ij} + \r g_{(2)ij} + \cdots +  \r^\s g_{(2\s)ij}
+ \cdots \nonumber
\eea
Differentiating  the scalar equation \eqref{scalasympt}
$[\s]$ times, where $[\s]$ is the integer part of $\s$,
and taking $\r \rightarrow 0$ after multiplying with
$\r^{1+[\s]-\s}$ one obtains
\be
\label{kappanu2}
   (2 \s +4 \a\g- d) \k_{(2\s)} + \a\g \Tr g_{(2\s)} = 0,
\ee
Similarly, equation \eqref{Ricasympt} yields,
\be
\label{gsigma}
(2 \s - d + 2 \a \g) g_{(2\s)ij} - (\Tr g_{(2\s)} + 2\k_{(2\s)})g_{(0)ij} =0.
\ee
which upon taking the trace becomes
\be \label{trgsigma}
-d \k_{(2 \s)} + (\s -d + \a \g) \Tr g_{(2\s)} = 0,
\ee
If the determinant of the coefficients of the system of equation
(\ref{kappanu2})-(\ref{trgsigma}) is non-zero,
\be
D = (2 \s +4 \a\g- d)  (\s -d + \a \g)  + \a \g d \neq 0
\ee
the only solution of these equations is
\be
\Tr g_{(2\s)} = \k_{(2\s)} = 0
\ee
which then using (\ref{gsigma}) implies
\be
g_{(2 \s)ij}=0
\ee
i.e. in these cases no non-integral power appears in the expansion.

On the other hand, when $D=0$  equations (\ref{trgsigma})-(\ref{kappanu2})
admit a non-trivial solution.
The two solution of $D=0$  are $\s_1 = d/2 - \a \g$ and
$\s_2 =2 (d/2 - \a \g)$. Clearly, $\s_2>\s_1$ and when $\s_2$
in non-integer so is $\s_1$, so a non-integer power first appears
at:
\be \label{nints}
\s= \frac{d}{2} - \a \g
\ee
When this holds equations (\ref{kappanu2})-(\ref{trgsigma})
reduce to
\be \label{oldtraceconstr}
\Tr g_{(2\s)} + 2\k_{(2\s)} = 0.
\ee
and the coefficient of $g_{(2\s)ij}$ in (\ref{gsigma}) vanishes,
so apart from its trace, these equations leave
$g_{(2 \s)ij}$ undetermined.
The remaining Einstein equation (\ref{div-equ}) also imposes a
constraint on the divergence of the terms occurring at this order,
as will be discussed later. To summarize, the expansion contains
a non-integer power of $\rho^{\s}$ in the following cases
\be
\s = \frac{p-7}{p-5} \Rightarrow \qquad
D0: \s = 7/5; \qquad D1,F1: \s  = 3/2; \qquad D2: \s = 5/3,
\ee
and the coefficient multiplying this power in only partly constrained.
As we will see, this category is the analogue of even dimensional
asymptotically AdS backgrounds, which are dual to odd dimensional boundary
theories.

The second case to discuss is the case of only integral powers.
In this case the undetermined term occurs at an integral power
$\rho^{\s}$ with
\be
\s = \frac{p-7}{p-5} \Rightarrow \qquad D3: \s = 2; \qquad D4: \s=3,
\ee
and logarithmic terms need to be included in the expansions. In these
cases the combination $(\Tr g_{(2\s)} + 2\k_{(2\s)})$ is determined by
$g_{(0)}$ and $\k_{(0)}$. This category is analogous to odd-dimensional
asymptotically $AdS$ backgrounds, which are dual to
even-dimensional boundary theories. The remaining Einstein
equation (\ref{div-equ}) also imposes a
constraint on the divergence of the terms occurring at this order.

Actually one can see on rather general grounds why the undetermined
terms occur at these powers: the undetermined terms will relate to the
vev of the stress energy tensor, which is of dimension $(p+1)$ for a
$(p+1)$-dimensional field theory. However, the overall normalization
of the action behaves as $l_s^{(p-3)^2/(5-p)}$, and therefore on
dimensional grounds the vev
should sit in the $g_{(2 \s)} \rho^{\s}$ term where
\be
\s = (p+1) + \frac{(p-3)^2}{(5-p)} = \frac{ (p-7)}{(p-5)},
\ee
which agrees with the discussion above.
Put differently  we can compare the power of the first undetermined term to
pure AdS and notice that it is shifted by $-\a\g =
-\frac{(p-3)^2}{2(p-5)}$ (for both Dp-branes and the fundamental
string). This is just what is needed to offset the
background value of the $e^{\g\phi}$ term multiplying the
Einstein-Hilbert action in \eqref{Dpaction}, in order to ensure that
all divergent terms in the action are still determined by the asymptotic field equations.

One should note here that the case of $p=6$ is outside the
computational framework discussed above. In this case the prefactor in
the action is of positive mass dimension nine, whilst the stress
energy tensor in the dual seven-dimensional theory must be of dimension
seven. Therefore one finds a (meaningless) negative value for $\s$,
indicating that one is not making the correct asymptotic expansion.
In other words, one finds that the ``subleading terms'' are more singular
than the leading term.

\subsection{Explicit expressions for expansion coefficients}

In all cases of interest $2 \s > 2$ and thus there are $g_{(2)}$ and
$\k_{(2)}$ terms. Evaluating \eqref{scalasympt}
and \eqref{Ricasympt} at $\r=0$ gives in the case of
$\beta =0$ and $2 \a  \g = -1$
(relevant for D1-branes, fundamental strings and D4-branes):
\bea
\label{firstcoeffs}
   \k_{(2)} &=& \frac{1}{2d}(\na^2 \k_{(0)} + g_{(0)}^{ij}
\pa_i \k_{(0)} \pa_j \k_{(0)} +\frac{1}{2(d-1)} R\sub{0}),  \\
   g_{(2)ij} &=& \frac{1}{d-1}(-R_{(0)ij}
+ \frac{1}{2d}R_{(0)} g_{(0)ij} + (\na_{\{i} \pa_{j\}} \k)_{(0)} +
\pa_{\{i} \k_{(0)} \pa_{j\}} \k_{(0)}) \nn
\eea
Here the parentheses in a quantity $A_{\{ a b \} }$ denote the traceless symmetric
tensor and $\nabla_i$ is the covariant derivative in the metric
$g_{(0) ij}$.

If $\b \neq 0$, as for $p=0,2$, the expressions are slightly
more involved:
\bea
\label{gen_firstcoeffs}
\k\sub{2} &=& -\frac{1}{M}\left (2 \a\g R\sub{0} -2(d-1)\na^2 \k\sub{0} +
(\frac{2\a\b}{\g} -2d+2)(g_{(0)}^{ij} \pa_i \k \sub{0} \pa_j \k
\sub{0}) \right ), \nono \\
g\sub{2}_{ij} &=& \frac{1}{d-2\a\g -2}\left(-R\sub{0}_{ij}
+ \na_i \pa_j \k\sub{0} +(1-\frac{\b}{\g^2})\pa_i\k\sub{0} \pa_j \k\sub{0}
\right.\\
  && \left.+\frac{\g^2 -\b}{2(\g^2 d - \b d +\b)}
g\sub{0}_{ij}\left(R_{(0)} - 2 \nabla^2 \k_{(0)} - 2 (1-\frac{\b}{2 \g^2})
(g_{(0)}^{ij} \pa_i \k \sub{0} \pa_j \k\sub{0})\right) \right), \nono \\
  M &\equiv& 16 \a^2 \b -2(d-1)(8\a\g + 4-2d)= \frac{16(9-p)}{(5-p)^2}. \nn
\eea
The final equality, expressing the coefficient $M$ in terms of $p$,
 holds for the Dp-branes of interest here.

\subsubsection{Category 1: undetermined terms at non-integral order}

Let us first consider the case where the undetermined terms occur at
non-integral order.

In the cases of $p=0,1,2$ the terms given above in (\ref{gen_firstcoeffs})
are the only determined terms. The underdetermined terms appear
at order $\rho^{(p-7)/(p-5)}$ and satisfy the constraints
\bea
&&2 \k_{(2 \s)} + \Tr g_{(2 \s)} = 0, \qquad \s = \frac{p-7}{p-5}
\label{trace-id} \\
&&\nabla^{i} g_{(2 \s) ij} - 2 (1 - \frac{\beta}{\g^2}) \pa_j \k_{(0)}
\k_{(2 \s)} + g_{(2 \s) ij} \pa^i \k_{(0)} = 0. \label{div-id}
\eea
We will see that the trace and divergent constraints translate into
conformal and diffeomorphism Ward identities respectively.

\subsubsection{Category 2: undetermined terms at integral order}

Let us next consider the case where the undetermined terms occur at
integral order: this includes the D3 and D4 branes.
Explicit expressions for the conformal cases, including the case of
D3-branes, are given in \cite{deHaro:2000xn}.
For the D4-branes, the equations at next order can be solved to
determine $\k_{(4)}$ and $g_{(4)ij}$:
\bea
\label{secondcoeffs}
   \k_{(4)} &=& \frac{1}{8}((\na^2 \k)_{(2)} + 6 \k_{(2)}^2 +
   (\pa \k)^2_{(2)}  + \half \Tr g_{(2)}^2 +
2 \k_{(2)} \Tr g_{(2)}), \\
   g_{(4)ij} &=& \frac{1}{4}[ (2 \kappa_{(2)}^2
+ \half \Tr g_{(2)}^2) g_{(0)ij}
- R_{(2)ij} - 2 (g_{(2)}^2)_{ij} + (\na_i \pa_j
     \k)_{(2)} + 2 \pa_i \k_{(2)} \pa_j \k_{(0)} ]. \nn
\eea
where we introduce the notation
\be
A[g(x,\r),\k(x,\r)] = A_{(0)}(x) + \rho A_{(2)}(x) + \rho^2 A_{(4)}(x) + \cdots
\ee
for composite quantities $A[g,\k]$ of $g(x,\r)$ and $\k(x,\r)$.
For (\ref{secondcoeffs}) we need the coefficients of
$A=\{\na^2 \k, (\pa \k)^2, R_{ij}\}$.
The explicit expression for these coefficients
can be worked out straightforwardly using the asymptotic expansion
of $g(x,\r)$ and $\k(x,\r)$ and we give these expressions for the
Christoffel connections and curvature
coefficients 
in appendix \ref{appA}. Note also that we use the compact notation
\be
(g_{(2)}^2)_{ij} \equiv (g_{(2)} g_{(0)}^{-1} g_{(2)} )_{ij}, \qquad
\Tr (g_{(2n)}) \equiv \Tr (g_{(0)}^{-1} g_{(2n)}).
\ee

Proceeding to the next order, one finds that
the expansion coefficients $\k_{(6)}$ and $g_{(6)ij}$
cannot be determined independently in terms of lower order
coefficients because after further differentiating the highest
derivative terms in \eqref{Ricasympt} \emph{and} \eqref{scalasympt}
both vanish. Only the combination $(2\k_{(6)} + \Tr g_{(6)})$ is fixed,
along with a constraint on the divergence.
Furthermore one has to introduce logarithmic terms
in \eqref{fieldexp} for the equations to be satisfied, namely
\bea
   g(x,\r) &=& g_{(0)}(x) + \r g_{(2)}(x) + \r^2 g_{(4)}(x) + \r^3
   g_{(6)}(x) + \r^3 \log (\r) h_{(6)} (x) + \cdots \\
   \k(x,\r) &=& \k_{(0)}(x) + \r \k_{(2)} (x) + \r^2 \k_{(4)}(x) + \r^3
   \k_{(6)}(x) + \r^3 \log (\r) \td{\k}_{(6)} (x) + \cdots \nn
\eea

For the logarithmic terms one finds
\bea
\label{logcoeffs}
\tilde{\k}_{(6)} &=& -\frac{1}{12}[(\na^2 \k)_{(4)} +
     (\pa \k)^2_{(4)} + 20 \k_{(2)} \k_{(4)} - \half \Tr g_{(2)}^3
+ \Tr g_{(2)} g_{(4)} \\
   && + 2 \k_{(2)} (-\Tr g_{(2)}^2 + 2 \Tr g_{(4)}) + 4 \k_{(4)} \Tr g_{(2)} ], \nono \\
   h_{(6)ij} &=& -\frac{1}{12} [ -2R_{(4)ij} +
(- \Tr g_{(2)}^3+ 2 \Tr g_{(2)} g_{(4)} +8 \k_{(2)} \k_{(4)}) g_{(0)ij}
+ 2 \Tr g_{(2)} g_{(4)ij}  \nono \\
   && - 8 (g_{(4)} g_{(2)} )_{ij} - 8 (g_{(2)} g_{(4)} )_{ij} + 4 g^3_{(2)ij} + 2 (\na_i \pa_j
\k)_{(4)} + 2 (\pa_i \k \pa_j \k)_{(4)} + 4 \k_{(2)} g_{(4)ij}], \nn
\eea
Note that these coefficients satisfy the following identities
\bea \label{log-id}
&&\Tr h_{(6)} + 2 \tilde{\k}_{(6)} = 0, \\
&&g^{ki}_{(0)} (\nabla_{k} h_{(6) ij} + h_{(6) ij} \pa_{k} \kappa_{(0)})
- 2 \pa_{j} \kappa_{(0)} \td{\kappa}_{(6)} = 0. \nonumber
\eea
Furthermore, $\k_{(6)}, \Tr g_{(6)}$ and $\nabla^{i} g_{(6) ij}$ are
constrained by the following equations,
\bea \label{tr-div}
&&2 \k_{(6)} + \Tr g_{(6)}
= -\frac{1}{6} (-4 \Tr g_{(2)} g_{(4)}
+ \Tr g_{(2)} ^3 + 8 \k_{(2)}\k_{(4)}), \\
&&\nabla^{i} g_{(6) ij} - 2 \pa_j \k_{(0)}
\k_{(6)} + g_{(6) ij} \pa^i \k_{(0)} = T_{j}, \nn
\eea
where $T_{j}$ is locally determined in terms of $(g_{(2n)},\k_{(2n)})$ with $n
\le 2$,
\bea
T_j &=& \nabla^{i} A_{ij} - 2 \pa_j \k_{(0)}
(A -\frac{2}{3}\k^3_{(2)} - 2 \k\sub{2}\k\sub{4})  + A_{ij} \pa^i \k_{(0)} \\
  && + \frac{1}{6} \Tr (g_{(4)} \nabla_j
g_{(2)}) + \frac{2}{3} (\kappa_{(4)} + \kappa_{(2)}^2) \pa_j
\kappa_{(2)}, \nono
\eea
with
\bea
A_{ij} &=&  \frac{1}{3} \left ((2  g_{(2)} g_{(4)} + g_{(4)}
g_{(2)})_{ij} - (g_{(2)}^3)_{ij} \right . \\
&& \qquad + \frac{1}{8} (\Tr (g_{(2)}^2) - \Tr
g_{(2)} ( \Tr g_{(2)}+ 4 \kappa_{(2)}) ) g_{(2) ij} \nono \\
&& \qquad - ( \Tr g_{(2)}+ 2 \kappa_{(2)}) ( g_{(4) ij} - \half
(g_{(2)}^2)_{ij}) \nono \\
&&  \qquad - (\frac{1}{8} \Tr g_{(2)} \Tr g_{(2)}^2 -
\frac{1}{24} (\Tr g_{(2)})^3 - \frac{1}{6} \Tr g_{(2)}^3 + \frac{1}{2}
\Tr g_{(2)} g_{(4)} ) g_{(0) ij} \nn \\
&& \qquad \left . + \left ( \frac{1}{4} \kappa_{(2)} ( (\Tr g_{(2)})^2 - \Tr
g_{(2)}^2)- \frac{4}{3} \kappa_{(2)}^3  - 2 \kappa_{(2)} \kappa_{(4)}
\right ) g_{(0) ij} \right ) \nn \\
A &=& \frac{1}{6} \left (  - \left (\frac{1}{8} \Tr g_{(2)} \Tr g_{(2)}^2 -
\frac{1}{24} (\Tr g_{(2)})^3 - \frac{1}{6} \Tr g_{(2)}^3 + \frac{1}{2}
\Tr g_{(2)} g_{(4)} \right ) \right . \nn \\
&& \qquad \left . - \frac{32}{3} \kappa_{(2)}^3 - 6  \kappa_{(2)} \kappa_{(4)}
-  \kappa_{(2)}^2 \Tr g_{(2)} -2 \kappa_{(4)} \Tr g_{(2)}  \right ). \nn
\eea
We would now like to integrate the equations (\ref{tr-div}).
Following the steps in \cite{deHaro:2000xn}, it is convenient
to express $g_{(6)ij}$ and $\k_{(6)}$ as
\bea
g_{(6) ij} &=& A_{ij} - \frac{1}{24} S_{ij} + t_{ij}; \label{t-def} \\
\kappa_{(6)} &=& A - \frac{1}{24} S - 2 \kappa_{(2)} \kappa_{(4)}
- \frac{2}{3} \kappa_{(2)}^3 +\varphi , \nn
\eea
where  $(S_{ij},S)$ are local functions of $g_{(0)}, \k_{(0)}$,
\bea
S_{ij} &=& (\na^2 + \pa^m \k\sub{0} \na_m)I_{ij} -2 \pa^m \k\sub{0}
\pa_{(i} \k\sub{0} I_{j)m} + 4 \pa_i \k\sub{0} \pa_j \k\sub{0} I \\
&& + 2 R_{kilj} I^{kl} -4I(\na_i \pa_j \k\sub{0} + \pa_i \k\sub{0}
\pa_j \k\sub{0})
+ 4 (g\sub{2}g\sub{4}-g\sub{4}g\sub{2})_{ij} \nono \\
&& + \frac{1}{10}(\na_i \pa_j B - g\sub{0}_{ij} (\na^2 + \pa^m \k\sub{0} \pa_m)B) \nono \\
&& + \frac{2}{5}B + g\sub{0}_{ij}(-\frac{2}{3} \Tr g_{(2)}^3 -
\frac{4}{15} (\Tr g\sub{2})^3
+ \frac{3}{5} \Tr g\sub{2} \Tr g_{(2)}^2 \nono \\
&& -\frac{8}{3} \k^3_{(2)} -\frac{8}{5} \k\sub{2} (\Tr g\sub{2})^2
-\frac{4}{5} \k^2_{(2)} \Tr g\sub{2} + \frac{6}{5} \k\sub{2} \Tr g_{(2)}^2), \nono \\
S &=& (\na^2 + \pa^m \k\sub{0} \pa_m) I+ \pa_i \k\sub{0} \pa_j \k\sub{0} I^{ij} - 2 (\pa\k)^2_{(0)} I \\
&& - (\na_k \pa_l \k\sub{0} + \pa_k \k\sub{0} \pa_l \k\sub{0} ) I^{kl}
- \frac{1}{20} (\na^2 + \pa^m \k\sub{0} \pa_m)B  \nono \\
&&  + \frac{2}{5} B \k\sub{2} - \frac{4}{3} \k^3_{(2)} - \frac{4}{5}
\k\sub{2} (\Tr g\sub{2})^2 - \frac{2}{5} \k^2_{(2)} \Tr g\sub{2} + \frac{3}{5} \k_{(2)} 
\Tr g\sub{2}^2, \nono \\
I_{ij} &=& (g\sub{4} - \hp g^2_{(2)} + \frac{1}{4} g\sub{2}(\Tr
g\sub{2} + 2\k\sub{2}))_{ij} + \frac{1}{8} g\sub{0}_{ij} B, \nono \\
I &=& \k\sub{4}+ \hp \k^2_{(2)} + \frac{1}{4} \k\sub{2} \Tr g\sub{2} + \frac{B}{16}, \nono \\
B &=& \Tr g^2_{(2)} - \Tr g\sub{2}(\Tr g\sub{2} + 4 \k\sub{2}). \nono
\eea
Note that these definitions imply the following identities
\bea
\nabla^{i} S_{ij} - 2 \pa_j \k_{(0)}
S  + S_{ij} \pa^i \k_{(0)} &=& - 4 \left (\Tr (g_{(4)} \nabla_j
g_{(2)}) + 4 (\kappa_{(4)} + \kappa_{(2)}^2) \pa_j
\kappa_{(2)} \right ); \\
\Tr (S_{ij}) + 2 S &=& -8 \Tr (g_{(2)} g_{(4)}  - 32 \kappa_{(2)}
(\kappa_{(2)}^2 + \kappa_{(4)}). \nn
\eea
Now, these definitions imply that $t_{ij}$ defined in (\ref{t-def})
is a symmetric tensor: $A_{ij}$ contains an antisymmetric part
but this is canceled by a corresponding antisymmetric part in $S_{ij}$.
Inserting (\ref{t-def}) in (\ref{tr-div}) one finds that
the quantities $(t_{ij},\varphi)$ satisfy the following divergence and trace
constraints:
\bea
\nabla^{i} t_{ij} &=&  2 \pa_j \k_{(0)} \varphi  - t_{ij} \pa^i \k_{(0)}; \\
\Tr t  + 2 \varphi &=& - \frac{1}{3} \left ( \frac{1}{8}
(\Tr g_{(2)})^3 - \frac{3}{8} \Tr g_{(2)} \Tr g_{(2)}^2 + \frac{1}{2}
\Tr g_{(2)}^3 - \Tr g_{(2)} g_{(4)} \right . \nn \\
&& \left . - \frac{3}{4} \k_{(2)} (\Tr g_{(2)}^2 - (\Tr g\sub{2})^2) -
4 \kappa_{(2)} \kappa_{(4)} + 2\kappa_{(2)}^3 \right ). \nn
\eea
We will find that the one point functions are expressed in terms of
$(t_{ij},\varphi)$ and these constraints translate into the conformal
and diffeomorphism Ward identities.

\subsection{Reduction of M-branes} \label{redM}

The D4-brane and type IIA fundamental string solutions
are obtained from the reduction along a
worldvolume direction of the M5 and M2
brane solutions respectively. The boundary conditions for the
supergravity solutions also descend directly from dimensional
reduction: diagonal reduction on a circle of an asymptotically (locally)
$AdS_{d+2}$ spacetime results in an asymptotically (locally)
$AdS_{d+1}$ spacetime with linear dilaton. Therefore the rather complicated
results for the asymptotic expansions in the D4 and fundamental string cases
should follow directly from the previously derived results for
$AdS_7$ and $AdS_4$ given in \cite{deHaro:2000xn}, and we show
that this is indeed the case in this subsection.

As discussed in section \ref{ldfe}, solutions of the field equations
of (\ref{Ein}) are related to solutions of the field equations of the action
\eqref{Dpaction} via the reduction formula (\ref{reduc1}).
In the cases of F1 and D4 branes this means in particular
\be
e^{4 \phi/3} = \frac{1}{\rho} e^{2 \k},
\ee
where in comparing with (\ref{fg}) one should note that $\a=-3/4, \g = 2/3$
for both F1 and D4. This implies that the $(d+2)$ solution is automatically
in the Fefferman-Graham gauge:
\be \label{oxi}
ds_{d+2}^2 =  \frac{d\rho^2}{4 \rho^2} + \frac{1}{\rho} (g_{ij} dx^i dx^j +
e^{2 \k} dy^2).
\ee

Recall that for an asymptotically $AdS_{d+2}$ Einstein manifold, the
asymptotic expansion in the Fefferman-Graham gauge is
\be \label{d2}
ds^2_{d+2} = \frac{d\rho^2}{4 \rho^2} + \frac{1}{\rho} G_{ab} dx^a dx^b
\ee
where $a = 1,\ldots,(d+1)$ and
\be
G = G_{(0)}(x) + \r G_{(2)}(x) + \cdots + \r^{(d+1)/2}
   G_{(d+1)/2}(x) + \r^{(d+1)/2} \log (\r) H_{(d+1)/2} (x) + \cdots,
\ee
with the logarithmic term present only when $(d+1)$ is even. The
explicit expression for $G_{(2)}(x)$ in terms of $G_{(0)}(x)$
is\footnote{Note that the conventions for the curvature used here
  differ by an overall sign from those in \cite{deHaro:2000xn}.}
\be \label{hssg2}
G_{(2)}{}_{ab}  =  \frac{1}{d - 1} \left( - R_{ab} + \frac{1}{2d}
R G_{(0)}{}_{ab} \right).
\ee
where the $R_{ab}$ is the Ricci tensor of $G_{(0)}$, etc.

Comparing (\ref{oxi}) with (\ref{d2}) one obtains
\be \label{mred1}
G_{ij} = g_{ij}; \qquad G_{yy} = e^{2 \k}.
\ee
In particular $G_{(0) ij} = g_{(0) ij}$ and $G_{(0) yy} =  e^{2 \k_{(0)}}$,
so
\bea
R[G_{(0)}]_{ij} &=&
R_{(0) ij} - \nabla_{i} \pa_j \k_{(0)} - \pa_i \k_{(0)} \pa_j
\k_{(0)}; \\
R[G_{(0)}]_{yy} &=& e^{2 \k_{(0)}} ( - \nabla^{i} \pa_i \k_{(0)} - \pa_i \k_{(0)} \pa^i
\k_{(0)}),  \nn
\eea
with $R[G_{(0)}]_{yi} =0$. Substituting into (\ref{hssg2}) gives
\bea
G_{(2)}{}_{ij} &=& \frac{1}{d - 1} \left ( - R_{(0) ij}  +
\frac{1}{2d}  R_{(0)} g_{(0) ij} + (\na_{\{i} \pa_{j\}} \k)_{(0)} +
\pa_{\{i} \k_{(0)} \pa_{j\}} \k_{(0)} \right ); \\
G_{(2)}{}_{yy} &=&  e^{2 \k_{(0)}} \left (
\frac{1}{2d (d - 1)} R_{(0)} + \frac{1}{d} (\na^2 \k_{(0)} +
(\pa \k_{(0)})^2 ) \right ), \nn
\eea
with $G_{(2)}{}_{yi} = 0$. We thus find exact agreement between
$G_{(2)ij}$ and $g_{(2)ij}$ in (\ref{firstcoeffs}).
Now using
\be
G_{yy} = e^{2\k} = e^{(2 \k_{(0)} + 2 \rho \k_{(2)} + \cdots)} = e^{2
  \k_{(0)}} (1 + 2 \rho \k_{(2)} + \cdots)
\ee
one determines $\k_{(2)}$ to be exactly the expression given in
(\ref{firstcoeffs}).

Now restrict to the asymptotically $AdS_4$ case; the next coefficient in the asymptotic
expansion occurs at order $\rho^{3/2}$, in $G_{(3) ab}$, and is
undetermined except for the vanishing of its trace and divergence:
\be
G_{(0)}^{ab} G_{(3) ab} = 0; \qquad
D^{a}   G_{(3) ab} = 0.
\ee
Reducing these constraints leads immediately to
\bea
&& g_{(0)}^{ij} g_{(3) ij} + 2 \k_{(3)} = 0; \\
&& \nabla^{i} g_{(3) ij} - 2 \pa_j \k_{(0)}
\k_{(3)} + g_{(3) ij} \pa^i \k_{(0)} = 0, \nn
\eea
in agreement with (\ref{trace-id}) and (\ref{div-id}).

Similarly if one considers the asymptotically $AdS_7$ case, the
determined coefficients $G_{(4)}$ and $H_{(6)}$ reduce to give $(g_{(4)},
\k_{(4)})$ and $(h_{(6)}, \td{\k}_{(6)})$ respectively. Furthermore,
the trace of $G_{(6)}$ fixes the combination
$(2 \k_{(6)} + \Tr g_{(6)})$. One
can show that all explicit formulae agree precisely with the
dimensional reduction of the formulae in \cite{deHaro:2000xn}; the
details are discussed in appendix \ref{appB}.

\subsection{Renormalization of the action}

Having derived the general form of the asymptotic expansion one can
now proceed to holographic renormalization, following the discussion
in \cite{deHaro:2000xn}. In this method one substitutes the
asymptotic expansions back into the regulated action and then introduces local
covariant counterterms to cancel the divergences and renormalise the
action. Whilst this method is conceptually very simple, in practice it
is rather cumbersome for explicit computations. A more efficient
method based on a
radial Hamiltonian formalism \cite{Papadimitriou:2004ap,Papadimitriou:2004rz}
will be discussed in the next section.

\bigskip

Let us choose an illustrative yet simple example to demonstrate this
method of holographic renormalization: we will work out the renormalised on-shell action and compute the
one-point function of the energy-momentum tensor and the operator
$\cO$ for the case $p=1$, both fundamental strings and D1-branes.

Since in this case $\b =0$, $\hat{\F}\equiv e^{\g \phi}$ behaves
like a Lagrange
multiplier and the bulk part of the action vanishes on-shell. The only
non-trivial contribution comes then from the Gibbons-Hawking boundary term:
\be
\label{bdryactp1}
   S_{boundary} = - L \int_{\r=\e} d^{2} x \sqrt{h} 2 \hat{\F}   K,
\ee
where $h_{ij}$ is the induced metric on the boundary and $K$ is the
trace of the extrinsic curvature. Since \eqref{bdryactp1} is divergent
we regularise the action by evaluating it at $\r=\e$.

We would like now to find counterterms to remove the divergences in
\eqref{bdryactp1}. From the discussion in section \ref{asymptfe} we
know the asymptotic
expansion for $\Phi$ and $h_{ij}(x,\r) = g_{ij}(x,\r)/\r$:
\bea
\label{fieldexpp1}
\hat{\F}  &=& \frac{e^{\k_{(0)}}}{\sqrt{\r}}(1 + \r \k_{(2)} + \r^{3/2} \k_{(3)} + \cdots), \\
   h   &=& \frac{1}{\r} (g_{(0)} + \r g_{(2)} + \r^{3/2} g_{(3)} +
   \cdots), \nono
\eea
where $\k_{(3)}$ and $g_{(3)}$ are the lowest undetermined
coefficients. Note that the expansions are the same for both
fundamental strings and D1-branes, since in both cases $\a \g = -1/2$. Inserting the
expansion \eqref{fieldexpp1} in \eqref{bdryactp1} we find for the divergent part
\be
\label{Sdivp1}
   S_{div}= - 4 L \int_{\r=\e} d^2 x e^{\k_{(0)}}
\sqrt{ g_{(0)}} (\e^{-3/2} + \e^{-1/2}\k_{(2)}),
\ee
using the formula
\be
   K = d - \r \Tr (g^{-1} g')
\ee
for the trace of the extrinsic curvature in the asymptotically
$AdS_{d+1}$ background. The trace term here
cancels against the one in the expansion of the determinant.

From \eqref{fieldexpp1} and \eqref{firstcoeffs} we find
\be
   \sqrt{g_{(0)}} = \r \sqrt{h} ( 1 + \frac{1}{4(d-1)} R[h]),
\ee
which allows us to write the counterterms in a gauge-invariant form:
\be
   S_{ct} = -S_{div}
= 4 L \int_{\r=\e} d^2 x \sqrt{h} \hat{\F} (1 + \frac{1}{4} R[h]).
\ee
The renormalised action is then
\be
S_{ren}[g_{(0)},\k_{(0)}] = \lim_{\e \to 0}
S_{sub}[h(x,\e),\hat{\F}(x,\e));\e]
\ee
where
\bea
   S_{sub} &=& S_{bulk} + S_{boundary} + S_{ct} \nono \\
   &=& - L [ \int_{\r \geq \e} d^3 x \sqrt{g} \hat{\Phi} (R+C) + \int_{\r=\e}
d^2 x \sqrt{h} \hat{\Phi} ( 2K -4 - R[h])].
\eea
This allows us to compute the renormalised vevs of the operator dual
to $\hat\Phi$ and the stress-energy tensor. For the former, only the
boundary part contributes, since $R+C =0$ from the equation of motion
for $\hat{\Phi}$. It can be easily checked that the divergent parts cancel and
we obtain the finite result
\be \label{scalar-d1}
    \< \cO\> =  \frac{1}{\sqrt{g_{(0)}}} \frac{\d S_{sub}}{\d \Phi_{(0)}}=
-\frac{1}{2} e^{3 \k_{(0)}}
\lim_{\e \rightarrow 0}
    (\frac{1}{\e^{3/2}\sqrt{h}} \frac{\d S_{sub}}{\d \hat{\Phi}})
= \frac{3}{2} e^{3 \k_{(0)}} L \Tr g_{(3)} = - 3 e^{3 \k_0}
L \k_{(3)}.
\ee
where we used (\ref{dualF}) and the definition of $\hat{\F}$.
The vev of the stress-energy tensor $\< T_{ij} \> = \lim_{\e
   \rightarrow 0} T_{ij}[h]$ gets a contribution from the bulk term
as well. We can split it into the contribution of the regularised action
and the counterterms
\be
    T_{ij}[h] = T_{ij}^{reg} + T_{ij}^{ct},
\ee
where
\bea
   T_{ij}^{reg}[h] &=& 2 L [ \hat{\Phi} (K h_{ij} - K_{ij}) - 2 \r \pa_\r
\hat{\Phi} h_{ij}], \\
   T_{ij}^{ct}[h] &=& 2 L [ \hat{\Phi} (R_{ij} - \hp R h_{ij} - 2 h_{ij}) +
\na^2 \hat{\Phi} h_{ij} - \na_i \pa_j \hat{\Phi}]. \nono
\eea
One can again check that the divergent terms cancel and obtain the
finite contribution
\be \label{set-d1}
    \< T_{ij} \> = \lim_{\e \rightarrow 0} ( \frac{2}{\sqrt{h}}
\frac{\d S_{ren}}{\d h^{ij}}) = 3 L e^{\k_{(0)}}  g_{(3)ij}.
\ee
Note that the expressions for the vevs take the same form for both
D1-brane and fundamental string cases. The one point functions satisfy
the following Ward
identities:
\bea \label{d1-ward}
 \< T^i_{i} \> - 2 \F_{(0)} \< \cO\> &=& 0. \\
 \nabla^{i} \< T_{ij} \> +  \pa_{j} \F_{(0)} \< \cO \> &=&
 0. \nn
\eea
To derive these one needs the trace and divergence identities given in
(\ref{trace-id}) and (\ref{div-id}) and the relation
$\F_{(0)} = e^{-2 \k_{(0)}}$
(see (\ref{dualF})). These Ward identities indeed
agree exactly with what we derived on the QFT side, (\ref{WI})-(\ref{trace}).

The first variation of the renormalized action yields the relation
between the 1-point functions and non-linear combinations of the asymptotic coefficients.
The one point functions are obtained in the presence of sources, so higher
point functions can be obtained by further functional differentiation
with respect to sources.

One should note here that the local boundary counterterms are required, irrespectively of
the issue of finiteness, by the more fundamental requirement of the
well-posedness of the appropriate variational problem
\cite{Papadimitriou:2005ii}. The conformal boundary of
asymptotically $AdS$ spacetimes
has a well-defined conformal class of metric rather than an induced metric.
This means that the appropriate variational problem involves keeping fixed
a conformal class and not an induced metric as in the usual Dirichlet problem
for gravity in a spacetime with a boundary.  The new variational problem requires the addition
of further boundary terms, on top of the Gibbons-Hawking term. In the
context of asymptotically $AdS$ spacetimes (with no linear dilaton)
these turn out to be precisely the boundary counterterms, see
\cite{Papadimitriou:2005ii} for the details and a discussion of
the subtleties related to conformal anomalies.

\subsection{Relation to M2 theory}

In the case of fundamental strings these formulae again follow directly from
dimensional reduction of the $AdS_4$ case, since for the latter the
renormalized stress energy tensor is \cite{deHaro:2000xn}
\be
 \< T_{ab} \> = 3 L_{M} G_{(3) ab}.
\ee
Recalling the dimensional reduction formula (\ref{mred1}), and noting
that
\be
L_{M} = L e^{\k_{0}},
\ee
one finds immediately that
\be
\< T_{ij} \> = 3 L e^{\k_0} g_{(3) ij},
\ee
in agreement with (\ref{set-d1}). Noting that $G_{yy} = e^{4 \phi/3}
\rho = \hat{\Phi}^2 \rho$ one finds
\be
\< T_{yy} \> = 6 L e^{3 \k_0} \k_{(3)} = -2 \< {\cal O}\>,
\ee
in agreement with (\ref{scalar-d1}). The first Ward identity in (\ref{d1-ward})
is thus an immediate consequence of the conformal Ward identity of the M2
brane theory, i.e. the tracelessness of the stress energy tensor. The
second Ward identity in (\ref{d1-ward}) similarly follows from the vanishing
divergence of the stress energy tensor in the M2-brane theory.

\subsection{Formulae for other Dp-branes}

It is straightforward to derive analogous formulae for the other
Dp-branes. Note that in general there is also a bulk contribution to the on-shell action
\be
\label{genonshell}
   S_{on-shell} = L\frac{4\a\b(d-2\a\g)}{h} \int_{\r \geq \e}
   d^{d+1}x  \sqrt{g} e^{\g \phi} + L \int_{\r =\e} d^dx \sqrt{h} e^{\g \phi} 2K
\ee
where $h_{ij}$ is the induced metric on the boundary, $K$ is the
trace of the extrinsic curvature and the action is regularised at
$\r=\e$. Focusing first on the cases $p < 3$ the divergent terms are:
\bea
   S_{div}&=&  - L \int_{\r=\e} d^d x \sqrt{ g_{(0)}} e^{\k_{(0)}}
   \e^{-d/2 + \a \g} \left (2d-\frac{4\a\b}\g +
   (-\frac{4\a\b(d-2\a\g)}{\g(d-2\a\g-2)}  +2d)\r \k_{(2)}\right. \nn \\
   && \left.+ (-\frac{2\a\b(d-2\a\g)}{\g(d-2\a\g-2)} + d-2) \r
   \Tr g_{(2)} \right ),
\eea
which can be removed with the counterterm action
\bea
S_{ct} &=& L \int_{\r=\e} d^d x \sqrt{h} e^{\g \phi} (2d
- \frac{4\a\b}{\g}+ C_R(\hat{R}[h] + \b (\pa_i \phi)^2))
   \label{ct-general} \\
       &=& L \int_{\r=\e} d^d x \sqrt{h} e^{\g \phi}
       \left (\frac{2(9-p)}{5-p}+ \frac{5-p}{4}(\hat{R}[h] + \b (\pa_i
       \phi)^2) \right ) \nono \\
C_R &\equiv& \frac{\g^2-\b}{d\g^2-d\b-\g^2+2\b} = \frac{5-p}{4}. \nn
\eea
Again for convenience we give the formulae both in terms of
$(\alpha,\beta,\gamma)$ and for the specific cases of interest here, the Dp-branes.
The renormalised vevs of the operator\footnote{Note that
$\<\cO_\phi\> = \chi \Phi_{(0)} \<\cO\>$. This is obtained using
       (\ref{dualF}) and the chain rule. \label{chi-O}} ${\cal O}_\f$ dual
to $\phi$ and the stress-energy tensor can now be computed giving:
\bea \label{other-vev}
\<\cO_\phi\> &=&
2 \s L e^{\k\sub{0}} \frac{1}{\a} \k\sub{2 \s},
\\
   \<T_{ij}\> &=& 2 \s  Le^{\k\sub{0}}g\sub{2\s}_{ij}.
\nn
\eea

Using (\ref{trace-id}) and (\ref{div-id}) one obtains
\bea
&&0=\<T^i_i\> + 2 \a \<\cO_\phi\>
= \<T^i_i\> +(p-3) \F_{(0)} \<\cO \> \label{ward-hol} \\
&&0=\nabla^i \<T_{ij}\> - \frac{1}{\g} \pa_j \k_{(0)} \<\cO_\phi\>
=\nabla^i \<T_{ij}\> + \pa_j \F_{(0)} \<\cO\>,
\eea
where in the second equality we use the relation between $\k_{(0)}$ and
$\F_{(0)}$ in (\ref{dualF}) which implies in particular that
$\<\cO_\phi\> = \chi \Phi_{(0)} \<\cO\>$.
These are the anticipated dilatation and diffeomorphism Ward identities.

Next let us consider the case of D4-branes, for which one needs more counterterms:
\bea
S_{ct} &=& L \int_{\r=\e} d^5x \sqrt{h} e^{\g \phi} (10 + \frac{1}{4}
\hat{R}[h]
+ \frac{1}{32} (\hat{R}[h]_{ij} - \gamma (\hat{\nabla}_i
\partial_j \phi + \pa_i \phi \pa_j \phi))^2
\label{d4-ct1} \\
&& \qquad + \frac{1}{32} \gamma^2 (\hat{\nabla}^2 \phi + (\pa_i \phi)^2)^2
- \frac{3}{320} (\hat{R}[h] - 2 \gamma (\hat{\nabla}^2
\phi + (\pa_i \phi)^2) )^2 + a_{(6)} \log \ep) , \nn
\eea
where the coefficient of the logarithmic term $a_{(6)}$ is given by
\bea
a_{(6)} &=& 6 \Tr h_{(6)}; \nn \\
&=& \frac{1}{8} (\Tr g_{(2)})^3 - \frac{3}{8} \Tr g_{(2)} \Tr
g_{(2)}^2 + \frac{1}{2} \Tr g_{(2)}^3 -  \Tr g_{(2)} g_{(4)} \\
&& - \frac{3}{4} \k_{(2)} \Tr g_{(2)}^2 + \frac{3}{4}  \k_{(2)} (\Tr
g_{(2)})^2 - 4 \kappa_{(2)} \kappa_{(4)} - 2 \kappa_{(2)}^3. \nn
\eea
Note that in cases such as the D4-brane, where one needs to compute many
counterterms, it is rather more convenient
to use the Hamiltonian formalism, which will be discussed in the next section.
We will also discuss the structure of this anomaly further in the
following section.

The renormalised vevs of the operator dual
to $\phi$ and the stress-energy tensor can now be computed giving:
\bea \label{dppp4}
   \<\cO_\phi\> &=& - L e^{\k\sub{0}} (8 \varphi + \frac{44}{3} \td{\kappa}_{(6)}), \\
   \<T_{ij}\> &=& L e^{\k\sub{0}} (6 t_{ij} + 11 h_{(6) ij}), \nn
\eea
where $(t_{ij},\varphi)$ are defined in (\ref{t-def}). Note that the contributions proportional to
$\td{\kappa}_{(6)}, h_{(6) ij}$ are scheme dependent; one can remove these contributions by
adding finite local boundary terms.

The dilatation Ward identity is
\be \label{d4-dilatt}
\<T^i_{i}\> + \F_{(0)} \<\cO\> = - 2 L e^{\k\sub{0}} a_{(6)},
\ee
whilst the diffeomorphism Ward identity is
\be
\nabla^{i} \<T_{ij}\> + \pa_{j} \F_{(0)} \<\cO\> = 0.
\ee
The terms involving $( h_{(6) ij},\td{\kappa}_{(6)})$ drop out of the
Ward identities because of the trace and divergence identities given
in (\ref{log-id}).

These formulae are as expected consistent with the reduction of
the M5 brane formulae given in \cite{deHaro:2000xn}. This computation
of the renormalized stress energy tensor for the M5-brane case is
reviewed in appendix \ref{appb}. In fact in \cite{deHaro:2000xn} the
renormalized stress energy tensor for the $AdS_7$ case was given
only up to scheme dependent traceless, covariantly constant terms, proportional to the
coefficient $H_{(6) ab}$ of the logarithmic term in the asymptotic
expansion. In appendix \ref{appb} we determine these contributions to
the stress energy tensor, with the resulting stress energy tensor
being \eqref{em-six}:
\be
\< T_{ab} \> =  \frac{N^3}{3 \pi^3} (6 t_{ab} + 11  H\sub{6}_{ab}).
\ee
The streamlined method of derivation of the renormalized stress energy tensor
given in appendix \ref{appb} is also useful in the explicit derivation
of the D4-brane formulae given in \eqref{dppp4}.
Dimensional reduction of the $t_{ab}$ term in the stress energy tensor
results in the
$(t_{ij},\varphi)$ terms in the D4-brane operator vevs, whilst reduction of
the $H\sub{6}_{ab}$ term gives the terms involving $( h_{(6)
  ij},\td{\kappa}_{(6)})$. The details of this dimensional reduction
are discussed in appendix \ref{appB}.

\section{Hamiltonian formulation} \label{six}

In the previous section we showed how
correlation functions can be computed using the basic holographic dictionary
that relates the on-shell gravitational action to the generating
functional of correlators, and we renormalized the action with
counterterms to obtain finite expressions. This method of
holographic renormalization is conceptually very simple but does not
exploit all the structure of the theory.

The underlying structure of the correlators is best exhibited in the
radial Hamiltonian formalism, which is a Hamiltonian formulation
with the radius playing the role of time. The Hamilton-Jacobi theory,
introduced in this context in \cite{de Boer:1999xf},
relates the variation of the on-shell action w.r.t. boundary conditions,
thus the holographic 1-point functions, to radial canonical momenta.
It follows that one can bypass the on-shell
action and directly compute renormalized correlators using radial
canonical momenta $\pi$, as was developed for asymptotically AdS
spacetimes in \cite{Papadimitriou:2004ap,Papadimitriou:2004rz}.

A fundamental property of asymptotically (locally) AdS spacetimes is that
dilatations are part of their asymptotic symmetries. This implies that
all covariant quantities can be decomposed into a sum of terms
each of which has definite scaling. These coefficients
are in 1-1 correspondence with the asymptotic coefficients in
(\ref{fg}) with the exact relation being in general non-linear.
The advantage of working with dilatation eigenvalues rather than
with asymptotic coefficients is that the former are manifestly covariant
while the latter in general are not: the asymptotic expansion
(\ref{fg}) singles out one coordinate so it is not covariant.
Holographic 1-point functions can be expressed most compactly in terms
of eigenfunctions of the dilatation operator, and this explains the
non-linearities found in explicit computations of 1-point functions.

\subsection{Hamiltonian method for non-conformal branes}
\label{Hammeth}

We now develop a Hamiltonian version of the holographic
renormalization of these backgrounds following closely the steps of
\cite{Papadimitriou:2004ap,Papadimitriou:2004rz}.
We consider the action \eqref{Dpaction} with the Gibbons-Hawking
boundary term added to ensure that the action depends only on
first radial derivatives (as we will see shortly),
so a radial Hamiltonian formalism can be set up:
\be
\label{Dpactionbdry}
   S = - L \int_{AdS_{d+1}} d^{d+1} x \sqrt{g}  e^{\g\phi}
[R + \beta (\pa \phi)^2 + C] - 2L \int_{\pa AdS_{d+1}} d^dx \sqrt{h}e^{\g\phi} K.
\ee
Note that we are again working in Euclidean signature.
Next we introduce a radial Hamiltonian formulation.
In the usual Hamiltonian formulation of gravity in the ADM formalism
one foliates spacetime by hypersurfaces of constant time. Here
analogously we introduce a family of hypersurfaces $\Sigma_r$ of
constant radius $r$ near the boundary and denote by $n^\mu$ their unit normal.
For asymptotically locally AdS manifolds there always exists a radial
function normal to the boundary which can be used to foliate the space
in such radial slices, at least in a neighborhood of the boundary.

In order to give a Hamiltonian description of the dynamics, one needs
to express the action  \eqref{Dpaction} in terms of quantities on
$\Sigma_r$. In particular, this means that the Ricci
scalar in the action \eqref{Dpaction} should be expressed in terms of
expressions which only contain first derivatives in the radial
variable. The induced metric on the hypersurface $\Sigma_r$ can be
expressed as $h_{\s\m} = g_{\s\m} - n_\s n_{\m}$, with
$h^{\r}_{\m} \equiv g^{\r \s} h_{\s \m}$. Now let us define the radial
flow vector field $r^{\m}$ by the relation $r^{\m} \pa_{\m} r = 1$,
such that the components of $r^{\m}$ tangent and normal to $\Sigma_r$
define shift and lapse functions respectively:
\be
r^{\mu}_{\parallel} = h^{\mu}_{\r} r^{\r} \equiv N^{\mu}; \qquad
r^{\mu}_{\perp} = N n^{\mu}.
\ee
Thus the metric is decomposed as
\be
ds^2 = (N^2 + N_{\m} N^{\m}) dr^2 + 2 N_{\m} dx^{\m} dr + h_{\m\n}
dx^{\m} dx^{\n},
\ee
analogously to the usual ADM decomposition.

A useful tool in our analysis is
the extrinsic curvature $K_{\m\n}$ of the hypersurface given by the covariant derivative of the unit normal
\be
   K_{\m\n} = h_{\s(\m} \na^\s n_{\n)}.
\ee
The geometric {\it Gauss-Codazzi} equations
(in the contracted form of
\cite{Papadimitriou:2004ap,Papadimitriou:2004rz}) can be used to express
the curvature of the embedding space in terms of extrinsic and
intrinsic curvatures on the hypersurface\footnote{The Lie derivative in our conventions is defined as
$\pounds_n K_{\m\n} = n^\s K_{\m\n,\s} - 2{n^\s}_{,(\m}K_{\n)\s}$.} :
\bea
\label{GaussCod}
   && K^2 - K_{\m\n} K^{\m\n} = \hat{R} + 2 G_{\m\n}n^\m n^\n, \\
   && \hat{\na}_\m K_\n^\m - \pa_\n K = G_{\r\s} h_\n^\r n^\s, \nono \\
   && \pounds_n K_{\m\n} + K K_{\m\n} - 2 K_\m^\r K_{\r\n} =
   \hat{R}_{\m\n} - h_\m^\r h_\n^\s R_{\r\s}, \nono
\eea
where $G_{\m\n}$ is the Einstein tensor in the embedding spacetime,
$K$ is the trace of the extrinsic curvature, $\hat{R}_{\m\n}$ is the
intrinsic curvature and $\hat{\na}$ is the covariant derivative on the hypersurface.

Combining the first equation in \eqref{GaussCod} with the Ricci
identity $R_{\m\n}n^\m n^\n = n^\n (\na_\s \na_\n - \na_\n \na_\s)
n^\s$ the Ricci scalar can be expressed as
\be
   R = K^2-K_{\m\n} K^{\m\n} + \hat{R} - 2 \na_\m(n^\m \na_\n n^\n) + 2 \na_\n (n^\m \na_\m n^\n),
\ee
Inserting this expression into the action \eqref{Dpaction}, the last two
terms cancel the Gibbons-Hawking boundary term in \eqref{Dpaction}
after partial integration and the remaining term is
\bea
\label{Dpsingledaction}
   S &=& - L \int d^{d+1} x \sqrt{g} e^{\g \phi} [ \hat{R} + K^2 -
     K_{\m\n} K^{\m\n} + \b (\pa \phi)^2 + C  \\
     && \qquad + 2 \g \pa_\m \phi n^\m  \na_\n n^\n - 2 \g \pa_\n \phi n^\m
     \na_\m n^\n]. \nn
\eea
Note that the extrinsic curvature can be expressed as
\be
K_{\m\n} = \frac{1}{2N} (\partial_{r} h_{\m\n} - \hat{\nabla}_{\m}
     N_{\n} - \hat{\nabla}_{\n} N_{\m} ),
\ee
and thus the action can be expressed entirely in terms of the fields $(h_{\m\n},
     N^{\m},N)$ and the scalar field $\phi$, and their
     derivatives. The canonical momenta conjugate to these fields are
     given by
\be
\pi^{\m\n} \equiv \frac{\delta L}{\delta \dot{h}_{\m\n}}, \qquad
\pi_{\phi} \equiv \frac{\delta L}{\delta \dot{\phi}},
\ee
where $\dot{f} \equiv \pa_r f$ and the momenta conjugate to the lapse
     and shift functions vanish identically. The corresponding
     equations of motion in the canonical formalism become
     constraints, which are precisely those obtained from the first
     two equations in (\ref{GaussCod}) and are the Hamiltonian and
     momentum constraints respectively.

The diffeomorphism gauge is most naturally fixed by choosing Gaussian normal
     coordinates ($N^{\m} = 0$ and $N=1$), such that
\bea \label{G_nor}
   ds^2 &=& dr^2 + h_{ij}(r,x) dx^i dx^j, \qquad K_{ij} = \hp \dot{h}_{ij} \\
   n^\m &=& \d^\m_r, \qquad \na_\m n^\m = K, \qquad n^\m \na_\m n^\n =
     0, \nn
\eea
where the dot denotes differentiation with respect to $r$.
The action becomes
\be
\label{Dpsingledactiongf}
   S = - L \int d^{d+1} x \sqrt{h} e^{\g \phi} [ \hat{R} +
K^2 - K_{ij} K^{ij} + \b (\dot{\phi}^2 + (\pa_i \phi)^2) + C
+ 2 \g \dot{\phi} K].
\ee
and the canonical momenta are given by
\bea
\label{canmom}
   \pi_\phi &=& 2 B \,(\b \dot{\phi} +  \g K), \qquad
B \equiv - L e^{\g \phi} \sqrt{h}. \\
   \pi^{ij} &=& B \,(K h^{ij} -K^{ij}  + \g \dot{\phi} h^{ij}), \nono
\eea
The Gauss-Codazzi identities in this gauge become:
\bea
\label{GaussCodgf}
   K^2 - K_{ij} K^{ij} &=& \hat{R} + 2G_{rr}, \\
   D_i K^i_j - D_j K &=& G_{jr}, \nono \\
   \dot{K}^i_j + K K^i_j &=& \hat{R}^i_j - R^i_j, \nn
\eea
Now consider the regulated manifold $\cM_{r_0}$ defined as the
submanifold of $\cM$ bounded by the
hypersurface $\Sigma_{r_0}$. The values of the induced fields on
   $\Sigma_{r_0}$ become boundary conditions for the action, and
   therefore the momenta on the regulating surface can be obtained
   from variations of the on-shell action with respect to the boundary
   values of the induced fields.
The Hamilton-Jacobi identities thus allow the momenta \eqref{canmom}
on the regulating surface to be expressed in terms of the on-shell action by
\be
\label{HamJac1}
\pi^{ij}(r_0,x) = \frac{\d S_{on-shell}}{\d h_{ij}(r_0,x)},
\qquad \pi_\phi(r_0,x) = \frac{\d S_{on-shell}}{\d \phi(r_0,x)}.
\ee
Since the choice of the regulator $r_0$ is arbitrary, the equations
\eqref{sonshell} and \eqref{HamJac1} can be used
not just to compute the on-shell action and momentum on the regulating
surface $\Sigma_{r_0}$ but on any radial surface ${\Sigma_r}$.

Now to calculate the regulated on-shell action one uses the first of
the Gauss-Codazzi identities, together with the field equations \eqref{Dpfe}:
\be \label{sonshell_prel}
   S_{on-shell} = - 2 L \int_{\cM_{r_0}} d^{d+1}x \sqrt{h}
e^{\g \phi}[ \hat{R} + \b (\pa_i \phi)^2  + C].
\ee
However, since the field equations follow from the variation of the
bulk part of the action, the radial derivative of the on-shell action
can be expressed as a purely boundary term,
\be
   \dot{S}_{on-shell} = - 2  L \int_{\Sigma_{r_0}} d^{d}x \sqrt{h}
e^{\g \phi} [ \hat{R} + \b (\pa_i \phi)^2  + C].
\ee
From this expression follows that
the regulated on-shell action can itself also be written as a
   $d$-dimensional integral by introducing a covariant variable $\l$,
\be
\label{sonshell}
   S_{on-shell} = - 2  L \int_{\Sigma_{r_0}} d^{d}x \sqrt{h} e^{\g \phi} [ K - \l],
\ee
where $\l$ satisfies the equation
\begin{gather}
\label{lambdadot}
   \dot{\l} + \l(K+\g \dot{\phi}) + E=0, \\
   E = \frac{(\g^2+d(\g^2-\b))\b}{(\g^2-\b)^2} =
   -\frac{2(p-1)(p-4)(p-7)}{(p-5)^2}, \nn
\end{gather}
and the trace of the third equation in \eqref{GaussCodgf} is used,
along with the field equations \eqref{Dpfe}. Note that since
$\Sigma_{r_0}$ is compact $\l$ is defined only up to a total
divergence.

The Hamilton-Jacobi identities then imply that:
\be
\label{HamJac2}
   \pi^{ij} \d h_{ij} + \pi_\phi \d \phi = - 2
L \d [ \sqrt{h} e^{\g \phi} (K- \l)],
\ee
up to a total derivative. One can always use the total divergence
ambiguity in $\l$ to ensure that this expression holds without
integrating it over $\Sigma_{r}$. First one chooses any $\l$
satisfying (\ref{lambdadot}), and then one calculates the variation
$\d [ \sqrt{h} e^{\g \phi} (K- \l)]$. This variation necessarily gives the left
hand side of (\ref{HamJac2}), up to total derivative terms, which can
be absorbed into the definition of $\l$. (Strictly speaking, this
argument applies only to the local terms in $\l$; the finite part of
$\l$ as $r \rightarrow \infty$ is actually non-local in the sources,
and only the integrated identity holds for this part.)

\subsection{Holographic renormalization}

We next turn to the formulation of a Hamiltonian method of holographic
renormalization. In the earlier sections, we discussed holographic
renormalization by solving asymptotically the field equations, as a
function of sources. Here we will instead use the equations of motion
to determine the asymptotic form of the momenta as functionals of
induced fields. Such a procedure is manifestly covariant at all
stages, with the Ward identities being manifest and the one point
functions of dual operators being naturally expressed in terms of the
momenta.

An important tool in the Hamiltonian method is the
dilatation operator, whose eigenfunctions are covariant expressions on the
hypersurface $\Sigma_r$, and which asymptotically behaves like the
radial derivative. The radial derivative acting on the
on-shell action and on the momenta can be represented
as a functional derivative, since by means of the field equations the
on-shell action and the momenta are given as functionals of
$h_{ij}$ and $\phi$:
\be
\label{radfunct}
   \pa_r = \int d^d x (2 K_{ij}[h,\phi] \frac{\d}{\d h_{ij}}
+ \dot{\phi}[h,\phi] \frac{\d}{\d \phi})
\ee
where we used (\ref{G_nor}).
Now, recall that the dilatation operator for a $d$-dimensional theory on a
curved background containing sources for operators of dimension $\D$
is given by
\be
\d_D \equiv \int d^d x ( 2 h_{ij} \frac{\d}{\d h_{ij}} + (\D-d) \Phi
   \frac{\d}{\d \Phi})
\ee
In our case, the field $\Phi=\exp \frac{2(p-5)}{(7-p)} \phi$ couples to ${\cal O}$ which has dimension $4$.
Using the chain rule we obtain
\be
\label{defdilop}
   \d_D \equiv \int d^d x ( 2 h_{ij} \frac{\d}{\d h_{ij}} - 2 \a
   \frac{\d}{\d \phi}) = \pa_r + \cO(e^{-2r}),
\ee
so indeed the radial derivative can be asymptotically identified
with the dilatation operator since asymptotically
$\dot{\phi} \rightarrow - 2 \alpha$ and $\dot{h}_{ij}
   \rightarrow 2 h_{ij}$.

The next key observation is that the momenta and on-shell action can be
 expanded asymptotically in terms of eigenfunctions of the dilatation
 operator $\d_{D}$. The structure one expects in these expansions of $K^i_j$, $\l$ and
$\dot{\phi}$ in terms of weights of the dilatation operator is
similar to the radial expansions \eqref{fieldexp}, except that
every term in the expansion also contains terms subleading in $e^{-2r}$:
\bea
\label{dilexp}
   K^i_j[h,\phi] &=& {K_{(0)}}^i_j + {K_{(2)}}^i_j + \cdots +
   {K_{(d-2\a\g)}}^i_j + {\tilde{K}\sub{d-2\a\g}}^i_j \log e^{-2r}, \\
   \l[h, \phi] &=& \l_{(0)} + \l_{(2)} + \cdots + \l_{(d-2\a\g)}
+ \tilde{\l}_{(d-2\a\g)} \log e^{-2r}, \nono \\
   \dot{\phi}[h, \phi] &=& p^{\phi}_{(0)} + p^{\phi}_{(2)} +
   \cdots + p^{\phi}_{(d-2\a\g)} + \tilde{p}^{\phi}_{(d-2\a\g)}
   \log e^{-2r}. \nn
\eea
(We will see that the logarithmic terms appear only if $(d- 2 \a \g)$ is an even
integer, i.e. for $p=3, 4$.)
The transformation properties of these terms under the dilatation operator are:
\bea
\label{diltrafo}
   \d_D K\sub{n}^i_j &=& -n K\sub{n}^i_j, \qquad \d_D
   \tilde{K}\sub{d-2\a\g}^i_j = -(d-2\a\g)\tilde{K}\sub{d-2\a\g}^i_j, \\
   \d_D K\sub{d-2\a\g}^i_j &=& -(d-2\a\g)K \sub{d-2\a\g}^i_j -
   2\tilde{K}\sub{d-2\a\g}^i_j, \nn
\eea
and similarly for $\l_k$ and $p^{\phi}_k$. Thus terms with weight $n < (d- 2
   \a\g)$ transform homogeneously, whilst terms with weight $n = (d-2
   \a\g)$ transform inhomogeneously, indicating that these terms
   depend non-locally on the induced fields.
As we will see below, the terms with weight $n < d-2\a\g$
are algebraically (locally) determined in terms of the asymptotics, while the
weight $(d-2\a\g)$ terms are undetermined up to some constraints. 
The latter will be identified with the renormalized one point
   functions and the on-shell action, which are non-local functionals
   of the sources.
Given a solution from which one wishes to extract the 1-point function
dual to a given field, one simply subtracts  the
lower weight terms in the dilatation expansion of the corresponding
momentum. We will show below how these lower weight terms can be
determined recursively in terms of the asymptotic data.

Although it is as mentioned above not necessary to compute the
renormalised action to obtain renormalised 1-point functions, the
Hamiltonian method is more efficient at determining the
counterterms. The divergences in the on-shell action can be expressed
as terms in the expansions which are divergent as $r_{0}
\rightarrow \infty$. These divergences can be removed by a counterterm
action which consists of these divergent terms in the expansions,
namely:
\be
   I_{ct} = 2 L \int_{\Sigma_{r_0}} \sqrt{h} e^{\g \phi} ( \sum_{0 \leq n
     < d-2\a\g}
(K\sub{n} - \l\sub{n}) + (\tilde{K}\sub{n} - \tilde{\l}\sub{n}) \log
   e^{-2r_{0}}). \label{Ctaction}
\ee
This counterterm action also leads through the Hamilton-Jacobi
relations to the covariant counterterms of the momenta.
The renormalised action is then given by the terms of
appropriate weight in the on-shell action \eqref{sonshell}:
\be
\label{renaction}
   I_{ren} = - 2 L \int_{\Sigma_{r_0}} d^{d} x \sqrt{h} e^{\g \phi}[K\sub{d-2\a\g} - \l\sub{d-2\a\g}].
\ee
The gravity/gauge theory prescription identifies this with the
generating functional in the dual field theory, and so, in particular,
the first derivatives of this action with respect to the sources
correspond to the one point functions of the dual operators. Since
the Hamilton-Jacobi relations identify these first derivatives with
the non-local terms in the expansions of the momenta one obtains
immediately the relations:
\be
\< T_{ij} \> = \pi_{(d- 2 \a\g) ij}; \qquad
\< {\cal O}_{\phi} \> = (\pi_{\phi})_{(d -2 \a\g)}.
\ee
From \eqref{canmom} one sees that the one-point functions are given by:
\bea
\label{onepoint}
   \< \cO_\phi \> &=& - 2 L e^{\g \phi} (\b p^{\phi}\sub{d-2\a\g} + \g K\sub{d-2\a\g}), \\
   \< T_{ij} \> &=& 2  L e^{\g \phi} ((K\sub{d-2\a\g} + \g
   p^{\phi}\sub{d-2 \a\g})h_{ij} - K\sub{d-2\a\g}_{ij}). \nn
\eea
Thus to obtain both the counterterms and the one-point functions one
needs to solve for the terms in the dilatation expansions.

\subsection{Ward identities}

The diffeomorphism Ward identity can be derived from the momentum constraint,
the second Gauss-Codazzi equation in \eqref{GaussCodgf}:
\be
   \hat{\na}_i K^i_j - \hat{\na}_j K = G_{jr} = (\g^2-\b) \pa_j \phi
   \dot{\phi} + \g \pa_j \dot{\phi} - \g K_j^i \pa_i \phi.
\ee
Using \eqref{canmom} this can easily be expressed in terms of momenta:
\be
   \hat{\na}_i (\frac{\pi^{ij}}{\sqrt{h}}) = \frac{1}{2 \sqrt{h}} \pa^j \phi \pi_\phi.
\ee
Expressing this identity at weight $(d-2\a\g)$ in terms of one-point functions yields the Ward identity
\be
   \hat{\na}_i \< T^{ij} \> - \g^{-1} \< \cO_\phi \> \pa^j \k\sub{0} =0.
\ee
which becomes of the standard QFT form (\ref{WI}) upon expressing it in terms
of $\< O\>$ and $\F_{(0)}$.
To determine the dilatation Ward identity one computes the infinitesimal Weyl
transformation of the renormalised action \eqref{renaction}
\be
\label{infweyl1}
   \d_\s I_{ren} = 4  L \int_{\Sigma_r} d^{d} x \sqrt{h} (Ne^\phi)^\g
[\tilde{K}\sub{d-2\a\g} - \tilde{\l}\sub{d-2\a\g}]\d\s,
\ee
where one uses the non-diagonal behaviour of $K\sub{d-2\a\g}$ and
$\l\sub{d-2\a\g}$ under the dilatation operator exhibited in
\eqref{diltrafo}. However,
this infinitesimal Weyl transformation is also given by the
renormalised version of the Hamilton-Jacobi relations \eqref{HamJac1}
given by\footnote{We define e.g. $\pi_\phi\sub{d-2\a\g}$ to be the
  weight $(d-2\a\g)$ part of $\pi_\phi/\sqrt{h}$.}
\be
\label{infweyl2}
   \d_\s I_{ren} = - \int_{\Sigma_r} d^{d} x \sqrt{h}
[2 \pi\sub{d-2\a\g}^i_i - 2\a \pi_\phi\sub{d-2\a\g}] \d\s.
\ee
Since these identities hold for arbitrary $\d\s$ we can infer the conformal Ward identity
\be
\label{confwi}
   \<T^i_i \> + 2\a \< \cO_\phi \> = \cA,
\ee
where the anomaly is given by
\be \label{anomdef}
   \cA = -4 L [\tilde{K}\sub{d-2\a\g} - \tilde{\l}\sub{d-2\a\g}].
\ee
The anomaly for the D4-brane will be computed below.
Again this becomes the standard Ward identity (\ref{trace}) (with an anomaly) upon replacing
$\< O_\f \>$ by $\chi \Phi_{(0)} \< O \>$ (see footnote \ref{chi-O}).

\subsection{Evaluation of terms in the dilatation expansion}

Let us now discuss how to evaluate the local terms in the dilatation
expansion. In the previous section we have derived a number of
identities which can be solved recursively to determine terms in the
expansions. In particular,
applying the Hamilton-Jacobi identity \eqref{HamJac2} to dilatations gives
\be
\label{lambdadil}
   (1+\d_D) K - (d-2\a\g + \d_D) \l - (d\g - 2\a\b) \dot{\phi} = 0.
\ee
The Hamilton-Jacobi relations \eqref{HamJac1} and \eqref{canmom} also
imply expressions for the extrinsic curvature and scalar field
momenta:
\bea \label{Kijvar}
(K h^{ij} - K^{ij} + \gamma \dot{\phi} h^{ij} ) &=& \frac{2}{e^{\gamma
    \phi} \sqrt{h}} \frac{\delta}{\delta h_{ij}} \int_{\Sigma_{r_0}}
d^d x \sqrt{h} e^{\gamma \phi} (K - \lambda); \\
(\beta \dot{\phi} + \gamma K) &=& \frac{1}{e^{\gamma
    \phi} \sqrt{h}} \frac{\delta}{\delta \phi} \int_{\Sigma_{r_0}}
d^d x \sqrt{h} e^{\gamma \phi} (K - \lambda). \nn
\eea
Next one has the Einstein equations, rewritten as the Gauss-Codazzi
equations \eqref{GaussCodgf}. Note that the Hamiltonian constraint in
\eqref{GaussCodgf} can be written as
\be
\label{Hamilconst}
   K^2 - K_{ij}K^{ij} = \hat{R} - \beta \dot{\phi}^2 + (\beta-2\g^2)
   (\pa_i \phi)^2 -
2\g \hat{\na}^2 \phi - 2\g K \dot{\phi} + C,
\ee
where the field equations \eqref{Dpfe} are used on the
right hand side, and the double radial derivative terms  $\ddot{\phi}$
are eliminated using the scalar equation of motion.
One can also use the scalar equation of motion
(the second equation in \eqref{Dpfe}), which in Gaussian normal coordinates reads
\be
\label{scaleom}
   \ddot{\phi} + \hat{\na}^2 \phi + K \dot{\phi} + \g \dot{\phi}^2 +
   \g (\pa_i \phi)^2  - \frac{\g(d(\g^2-\b)+\g^2)}{(\g^2-\b)^2} = 0.
\ee
as well as the differential equation for $\l$ \eqref{lambdadot}.
Not all of these identities are necessary in order to recursively
determine the lower terms in the dilatation expansion.

In practice it is convenient to first use the Hamilton-Jacobi identity
\eqref{lambdadil} to express the local coefficients of $\l$ in terms
of those in $K$ and $\dot{\phi}$:
\be
\label{lambdan}
   \l\sub{2n} = \frac{(1-2n)K\sub{2n} -
   (d\g-2\a\b)p^{\phi}\sub{2n}}{d-2\a\g -2n}.
\ee
Thus this identity ensures that all counterterms are expressed in
terms of the momenta.

Next one needs to solve for the momenta, using the Hamilton-Jacobi
relations, Gauss-Codazzi relations and the scalar equation of motion.
Consider first the Hamilton constraint \eqref{Hamilconst};
this equation can be expanded into terms of given dilatation weight,
   and solving at each weight yields a recursion relation for terms in
   the expansion of $K_{ij}$ and $\dot{\phi}$.
At dilatation weight zero this constraint
yields merely a check of the background solution. Noting that $K_{(0)}
= K_{(0) ij} K^{ij}_{(0)} = d $ the zero weight constraint is
\be
   d (d-1) = -\beta (p^{\phi}_{(0)})^2 - 2\g d p^{\phi}\sub{0} + C,
\ee
which is satisfied given that $p^{\phi}\sub{0} = -2 \a$ and the
definition of $\a$ in terms of $(\b,d,C)$.

At higher dilatation weight one
obtains a recursion relation for a linear combination for $K\sub{2n}$
and $p^{\phi}\sub{2n}$ at a given weight $2n$:
\bea
\label{Krec}
    K\sub{2} + \g p^{\phi}\sub{2} &=& \frac{1}{2(d-2\a\g
    -1)}[\hat{R} + (\beta -2 \g^2) (\pa_i \phi)^2 - 2\g \hat{\na}^2 \phi], \\
   K\sub{2n} + \g p^{\phi}\sub{2n} &=& \frac{1}{2(d-2\a\g-1)}[\sum_{m=1}^{n-1} (
     K\sub{2m}_j^i K
\sub{2n-2m}_i^j - K\sub{2m}K\sub{2n-2m} ) \nono \\
      &-& \sum_{m=1}^{n-1} (\b p^{\phi}\sub{2m} p^{\phi}\sub{2n-2m} + 2\g K\sub{2m}
     p^{\phi}\sub{2n-2m})].  \nn
\eea
Note that if $(d-2\a\g)$ is not an even integer one immediately finds the relation
\be
\label{traceconstr}
   K\sub{d-2\a\g} + \g p^{\phi}\sub{d-2\a\g} = 0,
\ee
since no terms on the right hand side can contribute at this
weight. This relation precisely corresponds to \eqref{oldtraceconstr}
in the old formalism, in the case where the undetermined term appears at a non-integral power of $\r$.

Consider next the scalar equation of motion;
to express this in terms of terms of given dilatation weight,
   it is necessary to expand $\ddot{\phi}$ in terms of eigenfunctions
   of the dilatation operator. (Note that eliminating $\ddot{\phi}$ using the
other field equations does not give an identity which is
independent of \eqref{Hamilconst}.) The additional radial
derivative in $\ddot{\phi}$ can be expressed
in terms of the dilatation operator by keeping higher terms in the expansion of the radial derivative:
\bea
   \pa_r &=& \int d^d x (2 K_{ij} \frac{\d}{\d h_{ij}} + \dot{\phi} \frac{\d}{\d \phi}) \\
   &=& \d_D + \sum_{n \ge 1} \int d^d x (2K\sub{2n}_{ij} \frac{\d}{\d h_{ij}}
+ p^{\phi}\sub{2n} \frac{\d}{\d \phi} ) \equiv \d_D + \sum_{n \ge 1}
   \d\sub{2n}. \nn
\eea
Given the transformation properties
(\ref{diltrafo}) of the expansion coefficients of the momenta,
the subleading terms in the expansion of $\pa_r$
must satisfy the commutation relation $[\d_D, \d\sub{2n}]=-2n\d\sub{2n}$.

Solving the scalar field equation at weight zero,
\eqref{scaleom} is automatically satisfied given the leading
asymptotic behavior. At higher weights $2n$ with $n > 1$ a
recursion relation for a distinct linear combination of $K\sub{2n}$
and $p^{\phi}\sub{2n}$ is obtained:
\bea
\label{scalrec}
   (d-2-4\a\g) p^{\phi}\sub{2} - 2\a K\sub{2}  &=& - \hat{\na}^2 \phi - \g (\pa_i \phi)^2, \\
   (d-2n-4\a\g) p^{\phi}\sub{2n} - 2\a K\sub{2n} &=& - \sum_{m=1}^{n-1}
( \d\sub{2m} p^{\phi}\sub{2n-2m} + K\sub{2m} p^{\phi}\sub{2n-2m}) \nn.
\eea
In the case that $(d - 2 \a \g)$ is not an even integer, the relevant
   term in the recursion relation becomes
\be
- 2 \a ( K\sub{d-2\a\g} + \g p^{\phi}\sub{d-2\a\g}) = 0,
\ee
since no terms on the right hand side can contribute at this
weight, and thus reproduces the trace constraint (\ref{traceconstr}).

The Hamiltonian constraint \eqref{Krec}
together with the scalar equation \eqref{scalrec} thus constitutes a linear
system of equations which allows one to express $K\sub{2n}$ and
$p^{\phi}\sub{2n}$ in terms of lower order coefficients. One can then
determine $\l\sub{2n}$ from \eqref{lambdan}, and use the
Hamilton-Jacobi relations \eqref{Kijvar} to determine the extrinsic
curvature $K\sub{2n}^i_j$. This is all information needed to proceed
in the recursion.

It is useful to recall here the equation
\eqref{lambdadot} for the variable $\l$, which determines the on-shell
action. Here again the radial derivative can be expressed in
terms of the dilatation operator, giving:
\be
   (\d_D + \sum_{n=1}^{d/2-\a\g} \d\sub{2n}) \l + \l (K + \g \dot{\phi}) + E = 0.
\ee
Note that in the case of $E=0$,
i.e. for F1,D1 and D4 branes $\l=0$ solves the differential equation,
and thus the coefficients $\l_{(2n)}$ consist only of total derivative terms which are
determined by (\ref{lambdan}).

\subsubsection{Category 1: Undetermined terms at non-integral order}

Let us consider first the case where the undetermined terms occur at
non-integral order, namely $p < 3$, and obtain the counterterms and
one point functions.

The Hamiltonian constraint \eqref{Krec} together with the scalar
equation \eqref{scalrec} can be solved at first order to give:
\bea \label{mom-2}
K\sub{2} &=& \frac{1}{2 (d - 2 \a \g -1) (d - 2 \a \g -2)}
\left (  (d -2 - 4 \a\g) (\hat{R} + \beta (\partial \phi)^2) + 2 (1 +
2 \a \g) e^{-\gamma \phi} \hat{\nabla}^2 (e^{\gamma \phi}) \right ); \nn \\
p^{\phi}\sub{2} &=&   \frac{1}{\gamma (d - 2 \a \g -1) (d - 2 \a \g -2)}
\left (  \gamma \alpha (\hat{R} + \beta (\partial \phi)^2) - (d-1)
e^{- \gamma \phi} \hat{\nabla}^2 (e^{\gamma \phi}) \right );
\eea
Next note that the counterterms $\l\sub{2n}$ follow from
\eqref{lambdan}, and are given by
\bea
\l\sub{0} &=& - \frac{2 \a \b}{\g}; \\
\l\sub{2} &=& - \frac{K\sub{2} + ( d \gamma  - 2 \alpha \beta)
  p^{\phi}\sub{2}} { (d - 2 \a \g -2)}. \nn
\eea
For the cases $p < 3$ one only needs to solve up to this order to
obtain all counterterms, with the counterterm action being:
\bea
   I_{ct} &=&  L \int_{\Sigma_{r_0}} \sqrt{h} e^{\g \phi} \left ( 2d
   - \frac{4 \a \b}{\g} + \frac{\g^2 - \b}{ (d-1)\g^2 + \b (2-d)}
   (\hat{R} + \beta (\pa \phi)^2) \right ) \\
&& - L \int_{\Sigma_{r_0}} \sqrt{h} \frac{d}{(d -2 \a \g -2)}
   \hat{\nabla}^2 (e^{\gamma \phi}). \nn
\eea
This coincides with the counterterm action found earlier in
\eqref{ct-general}, up to the (irrelevant) total derivative term in
the second line.

Next consider the one point functions. To apply the general formula
\eqref{onepoint}, one needs to relate the momentum coefficients with
terms in the asymptotic expansion of the metric and the scalar field.
In the case that $(d-2\a\g)$ is not an even integer, this
identification turns out to be very simple. Recall that in the
original method of holographic renormalization one expanded the
induced metric asymptotically in the radial coordinate $\rho =
e^{-2r}$ as
\be
h_{ij} = \frac{1}{\rho} (g_{(0) ij} + \r g_{(2) ij} + \cdots +
\rho^{\half (d - 2 \a\g)} g_{(d -2 \a \g) ij} + \rho^{\half (d - 2 \a
  \g)} \ln \rho h_{(d -2 \a \g) ij} + \cdots ),
\ee
where the logarithmic term is included when $(d - 2 \a \g)$ is an even
integer. Differentiating with respect to $r$ gives
\bea
K_{ij} &=& \half \dot{h}_{ij} = \frac{1}{\rho} g_{(0) ij} -  \r g_{(4) ij}
+ \cdots + \rho^{\half (d - 2 \a\g  - 2)} \left ( (1 - \half (d - 2
\a\g))  g_{(d -2 \a \g) ij} - h_{(d -2 \a \g) ij} \right ) \nn \\
&& + \rho^{\half (d - 2 \a\g  - 2)}  \ln \rho (1 - \half (d - 2
\a\g)) h_{(d -2 \a \g) ij} + \cdots  \label{cov-exp2}
\eea
However, each term in the covariant expansion of the extrinsic
curvature is a functional of $h_{ij}$ and can be expanded as:
\bea
K_{(0) ij} [h] &=& h_{ij} = \frac{1}{\rho} \left (g_{(0) ij} + \r g_{(2) ij} + \cdots +
\rho^{\half (d - 2 \a\g)} g_{(d -2 \a \g) ij} \right . \nn \\
&& \qquad \left . + \rho^{\half (d - 2 \a
  \g)} \ln \rho h_{(d -2 \a \g) ij} + \cdots \right ); \nn \\
K_{(2) ij} [h] &=& K_{(2) ij} [g_{(0)}] + \rho \int d^d x g_{(2) kl}
\frac{\delta K_{(2) ij} } {\delta g_{(0) kl}} + \cdots;  \\
K_{(d - 2 \a \g) ij} [h]  &=& \rho^{\half (d - 2 \a\g  - 2)}
K_{(d - 2 \a \g) ij} [g_{(0)}] + \cdots; \nn \\
\td{K}_{(d - 2 \a \g) ij} [h]  &=& \rho^{\half (d - 2 \a\g  - 2)}
\td{K}_{(d - 2 \a \g) ij} [g_{(0)}]
+ \cdots. \nn
\eea
Inserting these expressions into the expansion and comparing with
\eqref{cov-exp2} implies:
\bea
K_{(0) ij} [g_{(0)}] &=& g_{(0) ij}; \\
K_{(2) ij} [g_{(0)}] &=& - g_{(2) ij}; \nn \\
K_{(d - 2 \a \g) ij} [g_{(0)}] &=& - \half (d - 2
\a\g) g_{(d -2 \a \g) ij} - h_{(d -2 \a \g) ij} + \cdots; \nn \\
\td{K}_{(d - 2 \a \g) ij} [g_{(0)}] &=& - \half (d - 2
\a\g) h_{(d -2 \a \g) ij}. \nn
\eea
Here the ellipses denote terms involving functional derivatives with
respect to $g_{(0) ij}$ of lower order coefficients $g_{(2n)
  ij}[g_{(0)}]$.

The formulae are thus simplified in the case where $(d -2 \a \g)$ is
not an even integer, since no lower weight terms
can contribute and we obtain $K\sub{d-2\a\g}_{ij} =
 -(\frac{d}{2} - \a\g) g\sub{d-2\a\g}_{ij}$. Similarly treating the
 scalar field expansion, one finds that
\be
\g p^{\phi} \sub{d-2\a\g} = -(d-2\a\g) \k\sub{d-2\a\g},
\ee
which yields for the one point functions:
\bea
\label{onepointsimple}
   \< \cO_\phi \> &=& (d-2\a\g)(\g-\frac{\b}{\g}) L e^{\k\sub{0}} \Tr
   g\sub{d-2\a\g}, \\
   \< T_{ij} \> &=& (d-2\a\g) L e^{\k\sub{0}} g\sub{d-2\a\g}_{ij}, \nn
\eea
where we used the constraint \eqref{traceconstr} in the last equation. Note that the mixing of
$K$ and $\dot{\phi}$ in the momenta conspires to ensure that the
expectation value of the energy-momentum tensor is proportional
to just $g\sub{d-2\a\g}_{ij}$, without involving $\Tr g\sub{d-2\a\g}$.
These formulas exactly agree with the ones in (\ref{other-vev}) we derived earlier
(upon use of (\ref{nints}) and (\ref{adef})).

The D4-branes are the only case under consideration where
$(d-2\a\g)$ is an even integer; here the lower weight terms
do contribute and the expressions for the vevs are considerably more complicated.
We thus turn next to the evaluation of the momentum coefficients in this
case.

\subsubsection{Category 2: The D4-brane}
\label{D4anomaly}

In this section we will consider the case of the D4-branes,
where $(d-2\a\g)$ is an even integer, and derive the counterterms;
the anomaly term $\cA$ in the dilatation Ward identity \eqref{confwi}
and the one point functions.
Note that the anomaly appears only if $(d-2\a\g)$ is an even integer,
since only then do we
need nonzero coefficients $\tilde{K}\sub{d-2\a\g}$ and
$\tilde{p}^{\phi}\sub{d-2\a\g}$ of the logarithmic terms in \eqref{dilexp}
to fulfill the field equations. For the branes of interest, only the
cases of $p=3$ and $p=4$ have anomalies, and the
coefficients can be calculated from the counterterms. The
case $p=3$ was discussed already in
\cite{Henningson:1998gx,deHaro:2000xn} and will not be discussed
further here.

The counterterms and the anomaly are found
by recursively computing the momentum coefficients. The
Hamiltonian constraint \eqref{Krec} along with the scalar equation
\eqref{scalrec} provides a system of equations to determine
$K\sub{2n}$ and $p^{\phi}\sub{2n}$, whilst
the uncontracted Hamilton-Jacobi identity \eqref{Kijvar} can
be used to obtain $K\sub{2n}^i_j$. Recall that in this case $E=0$, and
thus $\l$ is zero, up to total
derivatives. This means in particular that the dilatation equation
\eqref{lambdan} can always be written as
\be
\label{betzrec1}
   (1-2n)K\sub{2n} - d \g \dot{\phi}\sub{2n} = (d - 2 \a \g - 2 n )
\l_{(2n)} \equiv \hat{\Phi}^{-1} \hna_l Y^l_{(2n)},
\ee
where $\hat{\Phi} \equiv e^{\g\phi}$. As $\l$ is zero, up to these total
derivatives, the only counterterms needed are the $K\sub{2n}$, along
with the logarithmic counterterm $\tilde{K}\sub{6}$.
Explicit expressions for the momenta found by solving the recursion
relations are given in appendix
\ref{D4braneappendix}, with the terms $K_{(2)}$ and $K_{(4)}$ agreeing
with the (non-logarithmic) counterterms found previously, see \eqref{d4-ct1}.

At weight $(d-2\a\g)= 6$ the dilatation equation \eqref{lambdan}
breaks down and only a linear combination of $K\sub{6}$ and
$p^{\phi}\sub{6}$ can thus be determined. This however is
sufficient to determine the anomaly
\be
\<T^i_i \> - \frac{3}{2} \< \cO_\phi \> = \cA,
\ee
which is given by
\be
   \cA = -4 L  \tilde{K}\sub{6}  = 2 L d (K\sub{6} + \g p^{\phi}\sub{6}),
\ee
where the right hand side is the combination of $K\sub{6}$ and
$p^{\phi}\sub{6}$ which is determined by \eqref{Krec} in terms of
lower counterterms. The anomaly in terms of the momentum coefficients
is therefore:
\bea \label{an-d4}
   \cA &=& 10 L (K\sub{6} + \g p^{\phi}\sub{6}) \\
       &=& 2L ( K\sub{2}^i_j K\sub{4}_i^j - K\sub{2}K\sub{4} -
K\sub{2} \g p^{\phi}\sub{4} - K\sub{4} \g p^{\phi}\sub{2}). \nono
\eea
Explicit expressions for each of these terms are given in appendix
\ref{D4braneappendix}; the total anomaly can then be written as
\bea
    \cA   &=& - 
\frac{N^2}{192 \pi^4} (g_5^2 N) 
\left[ - R^{ljki} \cR_{lk} \cR _{ij} -
2 \hat{\Phi}^{-2} \na^2 \hat{\Phi} \na_i \pa_j \hat{\Phi} \cR^{ij}
\right. \label{anomad4} \\
       &&  + \hp\cR(\cR_{ij}\cR^{ij} + \hat{\Phi}^{-2} (\na^2 \hat{\Phi})^2) -
\frac{3}{50} \cR^3 + \frac{1}{5}\cR_{ij}\na^i \pa^j \cR  \nono \\
       && + \frac{1}{20} \cR (\na^2 + \hat{\Phi}^{-1} \pa^i \hat{\Phi} \pa_i)\cR \nono \\
       && -\hp \cR_{ij} [ (\na^2 + \hat{\Phi}^{-1} \pa^l \hat{\Phi} \na_l)
\cR^{ij} - 2 \hat{\Phi}^{-2} \pa_l \hat{\Phi} \pa^{(i} \hat{\Phi} \cR^{j) l}
- 2 \hat{\Phi}^{-3} \pa^i \hat{\Phi} \pa^j \hat{\Phi} \na^2 \hat{\Phi}] \nono \\
       && \left. + \hp \hat{\Phi}^{-1} \na^2 \hat{\Phi} [ -(\na^2 +
\hat{\Phi}^{-1} \pa^i \hat{\Phi} \pa_i)(\hat{\Phi}^{-1} \na^2 \hat{\Phi}) + 2 \hat{\Phi}^{-2}
\pa_i \hat{\Phi} \pa_j \hat{\Phi} \cR^{ij} + 2 \hat{\Phi}^{-3} \pa_i
\hat{\Phi} \pa^i \hat{\Phi} \na^2 \hat{\Phi}] \nono \right],
\eea
where $\na$ is the covariant derivative in the five-dimensional metric and
\bea
   \cR &\equiv& R - 2 \hat{\Phi}^{-1} \na^2 \hat{\Phi}, \\
   \cR_{ij} &\equiv& R_{ij} - \hat{\Phi}^{-1} \na_i \pa_j \hat{\Phi}. \nono
\eea
Note that for notational simplicity we dropped the hats for the covariant
derivative and curvature of the boundary metric.
Here the anomaly has been expressed in such a way to demonstrate that
it agrees with the dimensional reduction of the anomaly of the M5-brane theory
found in \cite{Henningson:1998gx,deHaro:2000xn}. The latter is
given in terms of the six-dimensional curvature $R_{abcd}(G)$ of the
six-dimensional metric $G_{ab}$ by
\bea
\< T^{a}_a \> &=&  \frac{N^3}{96 \pi^3} \left ( R^{ab} R^{cd} R_{abcd} -
\half R R^{ab} R_{ab}   + \frac{3}{50} R^3 \right . \\
&& \left . \qquad + \frac{1}{5} R^{ab} D_a D_b R - \half R^{ab} \Box
R_{ab} + \frac{1}{20} R \Box R \right ). \nn
\eea
In particular, the anomaly vanishes for a Ricci flat manifold (more generally it vanishes
for conformally Einstein manifolds). Now
recall that on diagonal reduction  the six-dimensional Ricci tensor
$R(G)_{ab}$ can be written as:
\be
R(G)_{ij} = R_{ij} -  \hat{\Phi}^{-1} \na_i \pa_j \hat{\Phi}; \qquad
R(G)_{yy} = -  \hat{\Phi}^{-1} \na^2 \hat{\Phi}.
\ee
Clearly,
\be
R_{ij} = \na_i \pa_j \hat{\Phi}  = 0,
\ee
in the reduced theory implies that the six dimensional manifold is Ricci flat.
Comparing with (\ref{anomad4}) one sees that
indeed the anomaly vanishes under these conditions.

The anomaly of the six-dimensional theory can be expressed in terms of
conformal invariants, such that it is of the form
\be
{\cal A} = a N^3 (E_{(6)} + I_{(6)} + D_a J^{a}_{(5)}),
\ee
where $a$ is an appropriate constant, $E_{(6)}$ is proportional to
the six-dimensional Euler density
(type A anomaly), $I_{(6)}$ is a conformal invariant (type B anomaly)
and the $D_a J^{a}_{(5)}$ terms are scheme dependent, as they can
always be canceled by adding finite counterterms.

The D4 anomaly can necessarily be expressed in terms of invariants of the generalized
conformal structure: dimensional reduction of each of the
six-dimensional conformal invariants gives a generalized conformal
invariant. Note however that the reduction of the six-dimensional
Euler density will give an invariant which is not topological with
respect to the five-dimensional background. It is also not clear that
the basis of generalized conformal invariants obtained by dimensional
reduction would be irreducible; it would be interesting to explore
this issue further.

\bigskip

The general one point functions in this case are given by evaluating
the expressions:
\bea
   \< \cO_\phi \> &=& - 2 L e^{\g \phi} (\g K\sub{d-2\a\g}), \\
   \< T_{ij} \> &=& 2  L e^{\g \phi} \left ((K\sub{d-2\a\g} + \g
   p^{\phi}_{(d - 2 \a \g)} )h_{ij} - K\sub{d-2\a\g}_{ij} \right ). \nn
\eea
The resulting expressions are as found before, see \eqref{dppp4}:
\be
\< \cO_\phi \> = - L e^{\kappa_{(0)}} ( 8 \varphi + \frac{44}{3}
\td{\k}_{(6)} ); \qquad
\< T_{ij} \> = L e^{\kappa_{(0)}} ( 6 t_{ij} + 11 h_{(6) ij}),
\ee
where $(\varphi,t_{ij})$ are given in (\ref{t-def}).

\section{Two-point functions} \label{seven}

In this section we will discuss the computation of 2-point functions
for backgrounds with the asymptotics of the non-conformal branes.
Transforming to the dual frame, these become Asymptotically locally
AdS backgrounds with a linear dilaton and this implies that their
analysis is essentially the same as the analysis of the more familiar
holographic RG flows with conformal asymptotics \cite{BFS1, BFS2,Papadimitriou:2004rz}.
In the next subsection we briefly review the basic principles of the computation
of 2-point functions, mostly following the discussion in \cite{BFS1}.
Then we compute the 2-point functions for the D-branes
in subsection \ref{2ptsub} and finally we will discuss the computation for the general
case in subsection \ref{2ptgen}.

\subsection{Generalities} \label{generalities}

Let us start by recalling the basic formula relating bulk and boundary quantities:
\be
\<\exp (-S_{QFT}[g_{(0)},\F_{(0)}]) \> = \exp (-S_{SG}[g_{(0)},\F_{(0)}]).
\ee
The left hand side denotes the functional integration involving the field theory
action $S_{QFT}$ coupled to background metric $g_{(0)}$ and sources $\F_{(0)}$ that
couple to composite operators. For the case of Dp-branes the action $S_{QFT}$ is
given in (\ref{YM_s}). On the right hand side $S_{SG}[g_{(0)},\F_{(0)}]$ is the
bulk supergravity action evaluated on classical solutions with boundary data
$g_{(0)}, \F_{(0)}$. For the cases at hand this action is given in (\ref{Dpaction}).
As discussed extensively in previous sections, this relation needs to be renormalized
and we have determined the renormalized action $S_{ren}$ for all cases. By definition
the variation of the renormalized action is given by
\be
\d S_{ren}[g_{(0)},\F_{(0)}]= \int d^{d+1} x \sqrt{g_{(0)}}
\left(\frac{1}{2} \< T_{ij} \> \d g^{ij}_{(0)} + \< {\cal O} \> \delta \F_{(0)} \right).
\ee
Higher point functions are determined by further differentiation of the 1-point functions,
e.g. for the case of Dp-branes
\be
\< {\cal O}(x) {\cal O}(y) \> = - \left.\frac{1}{\sqrt{g_{(0)}}} \frac{\d \< {\cal O}(x) \>}{\d \F_{(0)}(y)}
\right|_{g_{(0)ij}=\d_{ij}, \F_{(0)}=g_{d}^{-2}}.
\ee

As we have shown in earlier sections, the 1-point functions in the presence of sources
are expressed in terms of the asymptotic coefficients in the near-boundary expansion of the
bulk solution. In particular, they depend on the coefficients that the asymptotic analysis
does not determine. To obtain those we need exact regular solutions with prescribed
boundary conditions. On general grounds, regularity in the interior should fix the
relation between the asymptotically undetermined coefficients and the boundary data.
Having obtained such relations one can then proceed to compute the holographic $n$-point
functions. To date, this program has only been possible to carry out perturbatively
around given solutions. In particular, linearized solutions determine 2-point functions,
second order perturbations determine 3-point functions etc. Here we will discuss the 2-point functions
involving the stress energy tensor $T_{ij}$ and the scalar operator ${\cal O}$.

Let us decompose the metric perturbation as,
\be
\d g_{(0)ij}(x) = \d h^T_{(0)ij} + \nabla_{(i} \d h^L_{(0)j)} + g_{(0)ij} \frac{1}{d-1} \d f_{(0)}
- \nabla_i \nabla_j \d H_{(0)}
\ee
where
\be
\nabla^i h_{(0)ij}^T=0, \qquad h_{(0)i}^{T\ i}=0, \qquad \nabla^i h_{(0)i}^L =0.
\ee
All covariant derivatives are that of $g_{(0)}$. Then the different components
source different irreducible components of the stress energy tensor,
\bea  \label{sren_s}
\d S_{ren}[g_{(0)},\F_{(0)}]&=& \int d^{d+1} x \sqrt{g_{(0)}}
\left(\< {\cal O} \> \delta \F_{(0)}-\frac{1}{2} \< T_{ij} \> \d h^{T\ ij}_{(0)}
- \frac{1}{2 (d-1)} \< T_{i}^i \>  \d f_{(0)} \right. \nonumber \\
&&\left.+ \nabla ^i \< T_{ij} \> \d h_{(0)}^{L\ j} + \nabla^i \nabla^j \<T_{ij}\> \delta H_{(0)}
 \right)
\eea
Now, recall that in the cases we discuss here we have already established that the holographic
Ward identities,
\bea
&&\nabla^{j} \langle T_{ij} \rangle_J + \langle {\cal O}\rangle_J
\pa_i \Phi_{(0)} = 0, \\
&&\langle T^{i}_{i}\rangle_J + (d-4) \Phi_{(0)}
\langle {\cal O} \rangle_J = {\cal A},
\eea
where there is an anomaly only for $p=4$. These and the fact that $\F_{(0)}$ in the background
solution is a constant imply that the second line in (\ref{sren_s}) does not contribute to 2-point functions.
Note also that the source for the trace of stress energy tensor is
$-f_{(0)}/(2(d-1))$.

We will be interested in cases with $g_{(0)ij} = \delta_{ij}$ (or somewhat more generally the cases with
$g_{(0)}$ being conformally flat). The two-point functions of $T_{ij}$ and ${\cal O}$
have the following standard representation in momentum space,
\bea
\<T_{ij}(q) T_{kl}(-q)\> &=& \Pi^{TT}_{ijkl} A(q^2) + \pi_{ij} \pi_{kl} B(q^2) \nonumber \\
\<T_{ij}(q) {\cal O}(-q) \> &=& \pi_{ij} C(q^2) \nonumber \\
\<{\cal O}(q) {\cal O}(-q) \> &=& D(q^2) \label{stfo}
\eea
where $A, B, C, D$ are functions of $q^2$ and
\bea
\pi_{ij} &=& \d_{ij} - \frac{q_i q_j}{q^2} \\
\Pi_{ijkl}^{TT} &=& - \frac{\d h_{(0)ij}^{TT}}{\d h_{(0)}^{TT \ kl}} =
\frac{1}{2} (\pi_{ik} \pi_{jl} + \pi_{il} \pi_{jk}) - \frac{1}{d-1} \pi_{ij} \pi_{kl} \nonumber
\eea
are transverse and transverse traceless projectors, respectively. The trace Ward identity implies
\bea
\<T_{ij}(q) T^k_{k}(-q)\> &=& -\frac{1}{g_{d}^2} (d-4)\<T_{ij}(q) {\cal O}(-q)\> \nonumber \\
\<T^i_i(q) {\cal O}(-q)\> &=& -\frac{1}{g_{d}^2} (d-4)\<{\cal O}(q) {\cal O}(-q)\>
\eea
which then leads to the relations,
\be \label{WI_coe}
B(q^2)= - \frac{1}{g_d^2} \frac{(d-4)}{(d-1)} C(q^2) =
\left(\frac{1}{g_d^2} \frac{(d-4)}{(d-1)}\right)^2 D(q^2)
\ee
Furthermore, the coefficient $D(q^2)$ is also constrained by the generalized conformal invariance
as discussed in section \ref{four}.

\subsection{Holographic 2-point functions for the brane backgrounds} \label{2ptsub}

We next discuss the computation of the 2-point functions in the backgrounds
of the non-conformal branes. Earlier discussions of the 2-point
functions in the D0-brane background can be found in
\cite{Sekino:1999av} and for Dp-brane backgrounds they were discussed in
\cite{Hashimoto:1999xu,Antonuccio:1999iz,Gherghetta:2001iv}. 

We need to solve for small fluctuations
around the background solution given in (\ref{AdSsol}). We thus consider
a solution of the form
\bea
   ds^2 &=& \frac{d\r^2}{4\r^2} + \frac{g_{ij}(x,\r)dx^i dx^j}{\r},
\label{fg1} \\
   \phi(x,\r) &=& \a \log \r + \varphi(x,\r), \qquad \varphi(x,\r) \equiv \frac{\k(x,\r)}{\g}, \nn
\eea
with
\be \label{glin}
g_{ij}(x,\r) = \d_{ij} + \g_{ij}(x,\r).
\ee
and $\varphi, \g_{ij}$ considered infinitesimal.
The background metric is translationally invariant, so it is convenient to
Fourier transform. The fluctuation $\g_{ij}(q,\r)$ can be decomposed
into irreducible pieces as
\be \label{hdec}
\g_{ij}(q,\r) = e_{ij}(q,\r) + \frac{d}{d-1}\left(\frac{1}{d} \d_{ij} -
\frac{q_i q_j}{q^2} \right)f(q,\r) + \frac{q_i q_j}{q^2} S(q,\r),
\ee
Let us also express the transverse traceless part as
$e_{ij}(q,\r) \equiv h_{(0)ij}^T(q) h(q,\r)$, where $h(q,\r)$ is
normalized to go to 1 as $\r \to 0$.
The field theory sources $h_{(0)ij}^T(q), f_{(0)}(q), S_{(0)}(q)$
are the leading $\r$ independent parts of
$e_{ij}(q,\r), f(q,\r), S(q,r)$.
Relative to the discussion in the previous subsection, we have gauged away
the longitudinal vector perturbation $h_i^L$ and traded $H$ for
$S = \frac{d}{d-1} f + p^2 H$.

The linearized equations are now obtained by inserting
(\ref{glin})-(\ref{hdec}) into (\ref{tra-eq})-(\ref{scalasympt})
and treating $\k,  h, f, S$ as infinitesimal variables.
This leads to the following equations:
\bea
&& \frac{1}{2} S'' + \k''=0; \label{tra-eq_l} \\
&& \frac{1}{2} f' + \k' = 0; \label{div-equ_l} \\
&& 2 \r h'' - (d-2 - 2 \a \g) h'
- \frac{1}{2} q^2 h  =0; \label{Ricasympt_l-tt} \\
&& 2 \r S'' + (2 \a \g + 2 - 2 d) S' - 2 d \k' - q^2 (\k + f) =0;
\label{Ricasympt_l-tr} \\
&&  4 \r \k'' + (8 \a \g + 4 - 2 d) \k' + 2 \a \g  S' - q^2 \k  =0,
\label{scalasympt_l}
\eea
where the equations are listed in the same order as in
(\ref{tra-eq})-(\ref{scalasympt}) with (\ref{Ricasympt_l-tt}) and
(\ref{Ricasympt_l-tr}) being the transverse traceless and trace
part of (\ref{Ricasympt}). Equation (\ref{Ricasympt_l-tt}) is already
diagonal. The remaining equations can be diagonalized by elementary
manipulations leading to the following expressions,
\bea
\k(q,\r) &=& 2 \a \g v_0(q) + v_1(q) \chi(q,\r) \\
f(q,\r) &=& -2 (d-1) v_0(q) - 2 v_1(q)  \chi(q,\r), \nonumber \\
S(q,\r) &=& v_2(q) + \r q^2 v_0(q) - 2 v_1(q) \chi(q,\r) \nonumber
\eea
where $v_0, v_1, v_2$ are integration constants, which can be expressed
in terms of the sources as
\be
v_0 = \frac{ 2 \g \f_{(0)} + f_{(0)}}{2(1-2 \s)}, \quad 
v_1 = \frac{(d-1) \g \f_{(0)}+ \a \g f_{(0)}}{2 \s -1}, \quad 
v_2 = S_{(0)} + 2 v_1,
\ee
where $\s = d/2 - \a \g = (p-7)/(p-5)$ and
$\f_{(0)}=\k_{(0)}/\g$ with $\k_{(0)}$ the $\rho$ independent part
of $\k(q,\r)$. $\chi(q,r)$ is normalized to go to 1 as $\r \to 0$ and satisfies
the same differential equation as the transverse traceless mode, namely
\be
2 \r \chi'' - 2 (\s -1) \chi' - \frac{1}{2} q^2 \chi  =0
\ee
The solution of this equation that is regular in the interior
is given in terms of the modified Bessel function of the second kind,
\be \label{chi-func}
\chi_\s(q,\r) = c(\s) x^\s K_\s(x), \qquad x = \sqrt{q^2 \r}, \qquad
\s = \frac{p-7}{p-5},
\ee
where the normalization coefficient $c(\s)$ is chosen such that
$\chi(q,\r)$ approaches 1 as $\r \to 0$. In our case,
$\s=\{7/5, 3/2, 5/3, 3\}$ for $p=\{0,1,2,4\}$.

\subsubsection{Non-integral cases}

The asymptotic expansion for non-integer values of $\s$ is
\be
\chi_\s(q,\r) =
1 + \frac{1}{4 (1-\s)} q^2 \r + \cdots
+ \tilde{\chi}_{(2 \s)}(q) \r^\s + \cdots \qquad
(\nu \  {\rm non{-}integer})
\ee
where
\be \label{tilde-chi}
\tilde{\chi}_{(2 \s)}(q) = - \frac{\G(1-\s)}{2^{2 \s} \G(1+\s)} (q^{2})^\s.
\ee
One can verify that the leading order terms in the exact linearized
solution indeed agree with the
linearization of the asymptotic coefficients derived earlier and
furthermore one can obtain the coefficient that the asymptotic analysis
left undetermined. Combining the previous formulas we obtain,
\bea
\k_{(2 \s)} &=& v_1(q) \tilde{\chi}_{(2 \s)}(q^2)  \\
g_{(2 \s)ij} &=&\left(h_{(0)ij}^T(q) -\frac{2}{(d-1)} v_1(q) \pi_{ij} \right)
 \tilde{\chi}_{(2 \s)}(q^2) \nonumber
\eea
which indeed satisfy the linearization of (\ref{trace-id})-(\ref{div-id}).
Thus the 1-point functions (\ref{other-vev}) to linear order in the
sources are then given by
\bea
\<\cO_\phi\> &=& \frac{2 \s  L \g (d-1)}{\a (2 \s-1)}
\left(\f_{(0)} -2 \a \left(-\frac{f_{(0)}}{2 (d-1)}\right) \right)
\tilde{\chi}_{(2 \s)}(q^2), \label{1pt_O}\\
   \<T_{ij}\> &=&  2 \s L
   \left( h_{(0)ij}^T - \frac{2 \g}{(2 \s-1)}
\left(\f_{(0)} -2 \a \left(-\frac{f_{(0)}}{2 (d-1)} \right)\right)
\pi_{ij}\right)
 \tilde{\chi}_{(2 \s)}(q^2). \label{1pt_t}
\eea

It follows that the 2-point functions are given by
\bea
\<T_{ij}(q) T_{kl}(-q)\> &=& \Pi^{TT}_{ijkl}
\left( 4 \s L \tilde{\chi}_{(2 \s)}(q^2) \right) +
\pi_{ij} \pi_{kl}
\left(-\frac{2 \a}{(d-1)}\right)^2 \<{\cal O}_\f(q) {\cal O}_\f(-q) \>
 \nonumber \\
\<T_{ij}(q) {\cal O}_\f(-q) \> &=& \pi_{ij}
\left(-\frac{2 \a}{(d-1)}\right) \<{\cal O}_\f(q) {\cal O}_\f(-q) \>
\label{2pt-0} \\
\<{\cal O}_\f(q) {\cal O}_\f(-q) \> &=& -
\frac{2 \s L \g (d-1)}{\a (2 \s-1)} \tilde{\chi}_{(2 \s)}(q^2) \nonumber
\eea
These relations are of the form (\ref{stfo}) with the coefficients
$B, C$ related to the $D$ coefficient as dictated by the
trace Ward identity (with the relation becoming
(\ref{WI_coe}) once we pass from $\cO_\f$ to $\cO$).
Thus we only need discuss the transverse traceless part of the
2-point function of $T_{ij}$ and the scalar 2-point function.

We now Fourier transform to position space using
\be
\int d^{d} q e^{-i q x} (q^2)^\s = \pi^{d/2} 2^{d+2 \s}
\frac{\G(d/2-\s)}{\G(-\s)} \frac{1}{|x|^{d+2 \s}},
\ee
which is valid when $\s \neq -(d/2 + k)$, where $k$ is an integer.
Let us first discuss the case of D$p$-branes.
The scalar two function becomes
\bea \label{2pgen}
 \< \cO_{\phi}(x) \cO_{\phi}(0)\> &=& C_{\f} N^{(7-p)/(5-p)}
 (g_d^2)^{(p-3)/(5-p)} |x|^{\frac{p^2 -19 -2p}{5-p}}, \\
&=& C_{\f}
N^2 \frac{(g_{eff}^2(x))^{\frac{(p-3)}{(5-p)}}}{|x|^{2d}} \nonumber
\eea
where $ C_{\f}$ is a positive numerical constant (obtained
by collecting all numerical constants in previous formulas).
Note that the characteristic scale in this case is $x$ and therefore
the effective coupling constant is $g_{eff}^2(x) = g_d^2 N |x|^{3-p}$.
The $g_d$ and $x$ dependence is consistent with the constraints
of generalized conformal invariance discussed in section \ref{four}.
Recall also that the operator $\cO_\f$ at weak coupling has
dimension $d$ (and $\cO$ has dimension 4). So going from weak to
strong coupling we find that the dimension is protected but the
2-point function itself gets corrections.
The overall factor of $N^2$ reflects the
fact that this is a tree level computation.
Similarly, the transverse traceless part of the
2-point function of the stress energy tensor is given by
\be
\<T_{ij}(x) T_{kl}(0)\>_{TT}
=  C_{T} \Pi^{TT}_{ijkl}
\frac{N^2 (g_{eff}^2(x))^{ \frac{(p-3)}{(5-p)}}}{|x|^{2 d}}
\ee
with  $C_{T}$ a positive constant. In this case the dimension is
protected because $T_{ij}$ is conserved.
We can trust these results provided
\be
g_{eff}^2(x) \gg 1 \quad \Rightarrow \quad |x| \gg (g_d^2 N)^{-1/(3-p)}
\ee

For the fundamental string background we obtain
\bea \label{2ptf1}
\< \cO_{\phi}(x) \cO_{\phi}(0)\> &\sim &
N^{3/2} g_s (\a')^{1/2} \frac{1}{|x|^5}, \\
\<T_{ij}(x) T_{kl}(0)\>_{TT} &\sim&
N^{3/2} g_s (\a')^{1/2} \Pi^{TT}_{ijkl} \frac{1}{|x|^5}
\eea
In the IIB case S-duality relates the fundamental string solution to the
D1 brane solution. Indeed, the 2-point function (\ref{2ptf1}) becomes
equal the $p=1$ case in (\ref{2pgen})
under S-duality, $g_s \to 1/g_s, \a' \to \a' g_s$.

In the IIA case the fundamental string lifts to the M2 brane.
As discussed in section \ref{redM}, the source for the stress energy tensor
of the M2 theory is simply related to the sources for the stress energy tensor
of the string and the operator $\cO_\f$, see (\ref{mred1}). Taking into account
that the worldvolume theories are related  by reduction over the
M-theory circle and so their actions are related by the factor of $R_{11}$, the radius
of the M-theory circle, we find (up to numerical constants)
\be
T^{M2}_{ij} \sim R^{-1}_{11} T_{ij}, \qquad T^{M2}_{yy} \sim R^{-1}_{11} \cO_\f
\ee
Using $R_{11} = g_s l_s$ we get
\be \label{M2T}
\<T^{M2}_{yy}(x) T^{M2}_{yy}(0)\>_
= \frac{1}{R_{11}^2} \< \cO_\f(x) \cO_\f(0)\>
\sim \frac{N^{3/2}}{R_{11} |x|^5}
\ee
with similar results for the other correlators. The stress energy tensor of the M2 theory has dimension 3,
so one expects the correlator to scale as $|x|^{-6}$. However, one of the worldvolume directions is
compactified with radius $R_{11}$. Smearing out over the compactified direction indeed results in the
fall off in (\ref{M2T}). Finally the $N$ scaling is the well-known $N^{3/2}$ scaling of the
M2 theory.

\subsubsection{The D4 case}

For the $\s =3$ case corresponding to D4 branes we have
\be
\chi_3(q,\r)=
1 - \frac{1}{8} q^2 \r + \cdots + \rho^3 (\tilde{\chi}_{(6)}(q)
+  \frac{1}{768} q^6 \log \r ) + \cdots
\ee
where
\be \label{tilde-chi2}
\tilde{\chi}_{(6)}(q) = \frac{1}{384} q^6 (\frac{1}{2} \log q^2 - \log 2
+\g - \frac{11}{12})
\ee
and $\g$ is the Euler constant (not to be confused with the $\gamma$
used in other parts of this paper).
The terms without $\log q^2$ are scheme dependent and will be omitted
in what follows. The one point functions and two point functions are
then given by \eqref{1pt_O}, \eqref{1pt_t} and \eqref{2pt-0}
respectively. In particular,
\be
\<{\cal O}_\f(q) {\cal O}_\f(-q) \> = \frac{L}{180} q^6 \ln q^2.
\ee
Fourier transforming back to position space, the scalar two function becomes
\be \label{2pd4}
 \< \cO_{\phi}(x) \cO_{\phi}(0)\>
= C_{\f} N^2 {\cal R} \left ( \frac{g_{eff}^2(x)}{|x|^{10}}  \right ),
\ee
where $ C_{\f}$ is a positive numerical constant (obtained
by collecting all numerical constants) and,
as in section \ref{four}, ${\cal R} (1/|x|^a)$ denotes the renormalised
version of $(1/|x|^a)$. The effective coupling constant is
$g_{eff}^2(x) = g_d^2 N/ |x|$,
and the $g_d$ and $x$ dependence is consistent with the constraints
of generalized conformal invariance discussed in section \ref{four}.

This result is also consistent with the uplift to the M5-brane
results. The source for the stress energy tensor
of the M5 theory is simply related to the sources for the stress energy tensor
of the D4-brane and the operator $\cO_\f$. Taking into account
that the worldvolume theories are related  by reduction over
M-theory circle and so their actions are related by the factor of $R_{11}$, the radius
of the M-theory circle, we find (up to numerical constants)
\be
T^{M5}_{ij} \sim R^{-1}_{11} T_{ij}, \qquad T^{M5}_{yy} \sim R^{-1}_{11} \cO_\f
\ee
Using $R_{11} = g_s l_s$ we then get
\be \label{M5T}
\<T^{M5}_{yy}(x) T^{M5}_{yy}(0)\>_
= \frac{1}{R_{11}^2} \< \cO_\f(x) \cO_\f(0)\>
\sim \frac{N^3}{R_{11}} {\cal R} \left ( \frac{1}{|x|^{11}}  \right )
\ee
with similar results for the other correlators.
The stress energy tensor of the M5 theory has dimension six,
and the correlator of the six-dimensional theory
behaves as ${\cal R} |x|^{-12}$. Here one of the worldvolume directions is
compactified with radius $R_{11}$ and smearing out over the compactified direction indeed results in the
fall off in (\ref{M5T}). Note that the $N$ scaling is the well-known $N^{3}$ scaling of the
M5-brane theory.

\subsection{General case} \label{2ptgen}

In the simple case discussed above, it was straightforward to
solve the equations for linear perturbations, but in more general
backgrounds the diagonalisation of the fluctuation equations is more
involved. To treat the general case, it is convenient to use the analysis
\cite{DeWolfe:2000xi,BFS1,Papadimitriou:2004rz} of linear fluctuations around
background solutions of a single scalar field coupled to
gravity; in these paper the fluctuation equations were diagonalised for
a general domain wall scalar system.

In this section we will explain a general method for computing the two point
functions which exploits this analysis. As discussed in section \ref{generalities} we need to
determine the one point functions to linear order in the sources and
in the Hamiltonian method this corresponds to determining the momenta
to linear order in the sources. So, as in the previous section, let us
begin by considering linear fluctuations around the background of interest
in the dual frame:
\bea
   h_{ij} &=& h_{ij}^B(r) + \g_{ij}(r,x) =
e^{2A(r)} \d_{ij} + \g_{ij}({r},x), \label{dual-frame-fl} \\
   {\phi} &=& \phi_B ({r}) + \varphi ({r},x). \nn
\eea
Note that the metric fluctuation has already been put into axial
gauge. Next we will express the canonical momenta in terms of these
fluctuations. To do this, first note that the extrinsic
curvature of constant ${r}$ hypersurfaces can be expressed as:
\be
   K^i_j = \dot{A}\d^i_j + \hp \dot{S}^i_j,
\ee
where $S^i_j \equiv h_B^{ik} \g_{kj}$. $S^i_j$ can be decomposed into irreducible components as
\be
\label{decSij}
   S^i_j =  e^i_j + \frac{d}{d-1}(\frac{1}{d} \d^i_j -
\frac{\pa^i \pa_j}{\na^2_B})f + \frac{\pa^i \pa_j}{\na^2_B}S,
\ee
where $\pa_i e^i_j = e^i_i =0$, $S=S^i_i$, indices are raised with the
inverse background metric $e^{-2A}\d^{ij}$ and $\na^2_B = e^{-2A}
\na^2 = e^{-2A} \d^{ij} \pa_i \pa_j$. Here
the diffeomorphism invariance of the
transverse space was used to set the vector component to zero.

The momenta \eqref{canmom} up to linear order in the fluctuations are then given by
\bea
\label{scalmomfluct}
   \pi_{{\phi}} &=& 2B( \beta \pa_r {\phi} + \g K) =
   \pi_{{\phi}}^B + B( 2 \beta \pa_r \varphi + \g \pa_r S), \\
   \pi^i_i &=& \pi^{i,B}_i - \half B (d-1) \pa_r S + B d \gamma \pa_r \varphi,
\qquad \pi^i_{j,TT} = \pi^{i,B}_{j,TT} - \half B \pa_r e^i_j, \nono
\eea
where $\pi_{\tilde{\phi}}^B$, $\pi^{i,B}_i$ and $\pi^{i,B}_{j,TT}$ are the background values, in the absence
of fluctuations, and TT stands for transverse and traceless. The one
   point functions are obtained by extracting the components of
   appropriate dilatation weight from these momenta. So we need to determine
   $\pa_r \varphi, \pa_r S, \pa_r e^i_j$. 

To obtain these momenta, however, we would need to diagonalise the
equations of motion for the linear fluctuations, and solve for $\pa_r
\varphi$ etc. Diagonalising such fluctuation equations is in general
rather difficult, and thus it is convenient to exploit the analysis of
\cite{BFS1,Papadimitriou:2004rz}, where the
fluctuation equations were diagonalised for a generic domain wall
dilaton background. In the latter work, however, an Einstein frame
bulk action was used, so we will first need to
transform our backgrounds
to the Einstein frame, and then map our fluctuation equations to the set
of equations which were diagonalised in full generality in
\cite{BFS1,Papadimitriou:2004rz}.

The analysis of \cite{BFS1,Papadimitriou:2004rz} begins with an Einstein frame bulk action:
\be
\label{dwaction}
   S =  - \int d^{d+1}x \sqrt{G_E}(\frac{1}{2\k^2}  R_E -\hp (\pa \td{\phi})^2 - V(\td{\phi})).
\ee
and then one considers domain wall solutions of the form
\be
\label{Poincdw}
   ds^2_B = d\tilde{r}^2 + e^{2A(\tilde{r})} dx_i dx^i, \qquad \td{\phi} = \td{\phi}_B(\tilde{r}),
\ee
which preserve Poincar\'e symmetry in the transverse directions. Here
the subscript $B$ denotes that this is the background solution around
which linear fluctuation equations will be solved.

Substituting the ansatz \eqref{Poincdw} into the field equations gives:
\bea
\label{2ndfe}
   \dot{A}^2 - \frac{\k^2}{d(d-1)}({\dot{\td{\phi}}}_B^2 - 2V(\td{\phi}_B)) &=& 0, \\
   \ddot{A} + d\dot{A}^2 + \frac{2\k^2}{d-1} V(\td{\phi}_B) &=& 0, \nono \\
   \ddot{\td{\phi}}_B + d\dot{A} \dot{\td{\phi}}_B - V'(\td{\phi}_B) &=& 0, \nn
\eea
where the dot denotes differentiation with respect to $\tilde{r}$ and
   the prime denotes differentiation with respect to $\td{\phi}$. In
explicitly solving these equations one can use the fact that
these second order equations are solved by any solution of the first
order flow equations \cite{Skenderis:1999mm,DeWolfe:1999cp}:
\bea
\label{1stfe}
   \dot{A} &=& -\frac{\k^2}{d-1} W(\td{\phi}_B), \\
   \dot{\td{\phi}}_B &=& W'(\td{\phi}_B), \nn
\eea
with the potential expressed in terms of a superpotential $W$ as:
\be
\label{suppot}
   V(\td{\phi}_B) = \hp [W'^2 - \frac{d\k^2}{d-1} W^2].
\ee
Conversely, given an explicit solution of \eqref{2ndfe}, which may not 
be asymptotically AdS but $\td{\phi}_B$ should have at most isolated
zeros, one can use
\eqref{1stfe} to define a superpotential $W(\td{\phi}_B)$ \cite{Skenderis:2006jq}.

Now let us consider the backgrounds of interest here, which are
asymptotic to Dp-brane backgrounds.
In these cases, the action \eqref{Dpaction} in the dual frame can be
transformed to the Einstein frame using the transformation
$g_{E} = \exp ( 2 \gamma \phi/(d-1)) g_{dual}$, giving
\be
   S =  - L \int d^{d+1}x \sqrt{G_E} [R_E -\hp (\pa \tilde{\phi})^2 + C e^{-2\g\tilde{\phi}/\nu(d-1)}].
\ee
Here the scalar has been rescaled as
\be
   \tilde{\phi} \equiv \nu \phi, \qquad \nu \equiv \sqrt{2(\frac{d\g^2}{d-1} -\b)},
\ee
so that $\tilde{\phi}$ is canonically normalized.
The metric and dilaton for the decoupled Dp-brane background can then be
written in Einstein frame as
\bea
   ds^2_E &=& d\tilde{r}^2 + (\mu \tilde{r})^{2(\mu+1)/\mu} dx_i dx^i, \nono \\
   \tilde{\phi} &=& -\frac{2\a\n}{\mu}\log(\mu \tilde{r}), \\
   \tilde{r} &=& \frac{\r^{-\mu/2}}{\mu} = \frac{e^{\mu r}}{\mu},
\qquad \mu = -\frac{2\a\g}{d-1} = \frac{(p-3)^2}{p(5-p)}. \nn
\eea
From this solution one can extract the parameters and functions
abstractly defined in \eqref{Poincdw}, \eqref{1stfe} and \eqref{suppot}:
\bea
   \k^2 &=& \hp, \qquad
   A(\tilde{r}) = \frac{\mu+1}{\mu} \log(\mu \tilde{r}), 
\qquad   \td{\phi}_B = \sqrt{\frac{2(\mu+1)(d-1)}{\mu}} \log(\mu\tilde{r}) \nono \\
   V(\td{\phi}_B) &=& -C \exp(-\sqrt{\frac{2\mu}{(\mu+1)(d-1)}} \td{\phi}_B), \nono \\
   W( \td{\phi}_B) &=& -2(d-1)(\mu+1)
   \exp(-\sqrt{\frac{\mu}{2(\mu+1)(d-1)}} \td{\phi}_B). \nn
\eea
Given a more general solution in the dual frame, which
asymptotes to an AdS linear dilaton background, one can similarly
transform it into Einstein frame and extract the corresponding
superpotential etc.

Suppose the fluctuations in the Einstein frame are given by:
\be
g_{E \mu \nu} = g_{E \mu \nu}^{B} + \td{\gamma}_{\mu \nu}; \qquad
\td{\phi} = \td{\phi}^B + \td{\varphi},
\ee
where $\td{S}^i_j \equiv h_B^{ik} \td{\gamma}_{kj}$ is:
\be
\td{S}^i_j =  \td{e}^i_j + \frac{d}{d-1}(\frac{1}{d} \d^i_j -
\frac{\pa^i \pa_j}{\na^2_B}) \td{f} + \frac{\pa^i \pa_j}{\na^2_B} \td{S},
\ee
Then these fluctuations in Einstein frame are related to those in the
dual frame defined in \eqref{dual-frame-fl} via:
\bea
\label{fluctmix}
   \tilde{e}^i_j &=& e^i_j, \qquad
   \tilde{f} = 2 \g \varphi + f, \\
   \td{S} &=&  \frac{2 \g d}{(d-1)} \varphi + S, \qquad
   \nu \td{\varphi} = \varphi, \qquad \td{\gamma}_{rr} = \frac{2 \g
   d}{(d-1)} \varphi. \nn
\eea
Note in particular that the Weyl transformation to the Einstein
   frame takes the fluctuations outside axial gauge: $\td{\gamma}_{rr}
   \neq 0$.

Using \cite{BFS1,Papadimitriou:2004rz}, one can write down
the diagonalised equations of motion for the linear fluctuations in
Einstein frame:
\bea
\label{eomlin}
   &&(\pa^2_{\tilde{r}} + d\dot{A}\pa_{\tilde{r}} - e^{-2A}q^2) \td{e}_j^i = 0, \\
   &&(\pa^2_{\tilde{r}} + [d\dot{A} + 2W \pa_{\td{\phi}}^2 \log W]
\pa_{\tilde{r}} - e^{-2A}q^2) \w = 0, \nono \\
   && \pa_{\tilde{r}} \td{S} = \frac{1}{(d-1)\dot{A}}\left ( e^{-2A}q^2 \td{f} + 2\k^2
(\pa_{\tilde{r}} \td{\phi}_B   \pa_{\tilde{r}} {\varphi} - 
V'(\td{\phi}_B) \td{\varphi}- V(\td{\phi}_B) \td{\gamma}_{rr})\right ), \nn
\eea
where
\be
   \w \equiv \frac{W}{W'} \td{\varphi} + \frac{1}{2\k^2} \td{f},
\ee
and we have Fourier transformed to momentum space, with $q$ being the momentum.

To derive the two point functions we will need to obtain the functional
dependence of the one-point functions on the sources.
The one-point functions are given in terms of the canonical
momenta, with the parts dependent on the fluctuations being given by linear
combinations of radial derivatives of fluctuations.
Hence we write the radial derivatives of the fluctuations $\td{e}^i_j$ and
$\w$ as functionals of the background fields $A$ and $\td{\phi}_B$:
\be
\label{sollinE}
 \pa_{\tilde{r}}  \td{e}^i_j = E(A,\td{\phi}_B) \td{e}^i_j, \qquad
 \pa_{\tilde{r}} {\w} = \Omega(A,\td{\phi}_B) \w.
\ee
The first two equations in (\ref{eomlin}) then become first order
equations for $E$ and $\W$:
\bea
   &&\dot{E} + E^2 + d\dot{A} E - e^{-2A}q^2 = 0, \\
   && \dot{\Omega} + \Omega^2 + [d\dot{A} + 2W \pa_{\td{\phi}}^2 \log W]
\Omega - e^{-2A}q^2 = 0. \nn
\eea
Note that in the case of the Dp-brane backgrounds these equations
 actually coincide since $\pa_\phi^2 \log W = 0$.
Given the solutions for $E$ and $\W$ and omitting terms that contribute to contact terms
one can obtain the required expressions for the radial derivatives of other fluctuations:
\bea
\label{Einstrad}
   \pa_{\tilde{r}} \td{e}^i_j &=& E \td{e}^i_j, \\
   \pa_{\tilde{r}}
\td{\varphi} &=& \W \td{\varphi} + \frac{1}{2\k^2} \frac{W'}{W} \W \td{f}, \nono \\
   \pa_{\tilde{r}}\td{S} &=& -\frac{1}{\k^2} \left[\left(\frac{W'}{W}\right)^2 \W +
\frac{e^{-2A}}{W}q^2\right] \td{f} - 2 \frac{W'}{W} \W \td{\varphi}. \nn
\eea
This completes the diagonalisation of the fluctuation equations in the Einstein
frame. Next one can rewrite these relations in terms of the
  fluctuations and radial derivative in the dual frame as:
\bea
\label{dualrder}
\pa_r \td{e}^i_j &=& e^{\g \phi_B/(d-1)} E e^i_j, \\
   \nu \pa_r \td{\varphi} &=& e^{\g \phi_B/(d-1)}
\left ( \nu^2 \Omega(1+\frac{\g}{\nu \k^2}\frac{W'}{W}) \varphi +
\frac{\nu}{2\k^2}\frac{W'}{W} \Omega f \right), \nn \\
   \pa_r \td{S} &=& e^{\g \phi_B/(d-1)}
\left(  -\frac{1}{\k^2} [(\frac{W'}{W})^2\Omega + \frac{e^{-2A}}{W}q^2] {f}\right. \nono \\
   &&\left.-2 \nu [ (\frac{W'}{W} + \frac{\g}{\nu \k^2} (\frac{W'}{W})^2)
   \Omega + \frac{\g}{\nu \k^2}
   \frac{e^{-2A}}{W} q^2] \varphi \right ). \nn
\eea
Using \eqref{fluctmix} in
\eqref{scalmomfluct}, and applying \eqref{onepoint} one finds that the
   expressions for the one point functions to linear order in the
   fluctuations are:
\bea \label{linear-one}
   \< \cO_{\phi} \> &=& \< \cO_{\phi} \>_B - B ( \nu \pa_r
   \td{\varphi} - \g \pa_r \td{S})_{(2 \sigma)}, \\
 \< T^i_{i} \> &=& \< T^i_{i} \>_{B} - B (d-1) (\pa_r \td{S})_{(2
   \sigma)}, \nn \\
\< T^i_{j,TT} \> &=& \< T^{i}_{j,TT} \>_B + B (\pa_r \td{e}^i_j)_{(2 \sigma)}, \nono
\eea
where $X_{(2 \s)}$ denotes the term of dilatation weight $2 \sigma
   \equiv (d - 2 \a \g)$ in $X$.

To explicitly evaluate these one point functions with linear sources we
now need to determine exact regular solutions for $E$ and $\W$. Up to
this point, we have given completely general expressions, applicable
for all solutions asymptotic to the Dp-brane backgrounds. The actual
background determines the defining differential equations for $E$ and
$\W$. Next we will solve these equations for the specific case of the
decoupled Dp-brane background; as mentioned before, the equations for
$E$ and $\W$ become identical in this case since $\pa_{\td{\phi}}^2
\log W = 0$. The only equation to be solved is thus:
\be
   (\pa^2_{\tilde{r}} + \frac{d(\mu+1)}{\mu \tilde{r}}\pa_{\tilde{r}}
-(\mu \tilde{r}) ^{-2(\mu+1)/\mu)}q^2) \omega = 0.
\ee
The solution which is regular in the interior, $\tilde{r} \rightarrow 0$, is given by
\bea
\label{sollin}
   \omega(\tilde{r}) &=&  (\mu \tilde{r}) ^{-c} K_{\mu c}
   (\frac{q}{(\mu \tilde{r})^{1/\mu}}) \equiv e^{- \s r} K_{\s}(q e^{-r}), \\
   \mu c &=& \half (d - 2 \alpha \gamma) \equiv \sigma, \nono
\eea
where $K$ is the modified Bessel function of the second kind; these are
exactly the same functions found in the previous section. The
   solution for $\Omega$ is then
\be
\Omega = \pa_{\td{r}} \ln (  (\mu \tilde{r}) ^{-c} K_{\mu c}
   (\frac{q}{(\mu \tilde{r})^{1/\mu}}))
\equiv e^{- \mu r} \pa_r \ln ( \chi_{\s}(q,e^{-2r}) ),
\ee
where $\chi_{\sigma}(q,\rho)$ was given in \eqref{chi-func}, and is normalized to approach
one as $\rho \equiv e^{-2r} \rightarrow 0$. The terms appearing in the one point functions
   \eqref{linear-one} follow from taking the projections onto appropriate dilatation
   weight:
\be
(e^{\g \phi_B/(d-1)} \Omega)_{(2 \sigma)} \equiv (e^{\mu r} \Omega)_{(2 \sigma)}
=  - 2 \s \td\chi_{(2 \sigma)}(q).
\ee
where we have used the expansions of $\chi_{\sigma}(q,\rho)$ given in \eqref{chi-func}
and the terms of appropriate dilatation weight, $\td{\chi}_{(2
   \sigma)} (q)$, in these asymptotic
expansions, see \eqref{tilde-chi} and \eqref{tilde-chi2}.

Using \eqref{linear-one} one obtains the renormalised
one point functions to linear order in the sources:
\bea
   \< \cO_{\phi} (q) \> &=& L \td{\chi}_{(2
   \sigma)}(q) \nu (d - 2 \a \g) \left ( - \nu \varphi(q) [1 - \frac{2 \g^2}{(d-1)\nu^2}]^2
\right . \\
&& \qquad \left .
+ f(q) [ \frac{\g}{\nu (d-1)} -
  \frac{2\g^3}{\nu^3 (d-1)^2}]  \right ) \nono \\
\< T^i_i (q) \> &=& 2 L (d -2 \a \g) \td{\chi}_{(2 \sigma)}(q)
\left ( [ - \frac{\g}{\nu} + \frac{2 \g^3}{\nu^3(d-1)}] \nu \varphi(q) +
\frac{\g^2}{\nu^2(d-1)}f(q) \right ), \nn \\
  \< T^i_{j} (q) \>_{TT} &=& L (d - 2 \a \g) \td{\chi}_{(2
   \sigma)}(q) e^i_j(q), \nono
\eea
where we have used $W'/W = - \g/\nu(d-1)$ and $\k^2 = 1/2$.
The first two
expressions can be rewritten as:
\bea
   \< \cO_{\phi} (q) \> &=& L \g \frac{(d -2 \a \g)}{\a (d - 1  - 2 \a
     \g)} \td{\chi}_{(2 \s)} (q)
 \left ( (d-1) \phi_{(0)}(q) + \a f_{(0)}(q) \right ) \\
\< T^i_i (q) \> &=&   -2 L \g \frac{(d -2 \a \g)}{(d - 1  - 2 \a \g)}
\td{\chi}_{(2 \s)} (q)
 \left ( (d-1) \phi_{(0)}(q) + \a f_{(0)}(q) \right ), \nn
\eea
where we have renamed the sources as $\varphi(q) \equiv
\phi_{(0)}(q)$ and $ f(q) \equiv f_{(0)}(q)$ to demonstrate
agreement with the expressions obtained previously in \eqref{1pt_O}
and \eqref{1pt_t}. The two point functions are given as before
by \eqref{2pt-0}.

\section{Applications} \label{eight}

In this section we will present a number of applications of the
holographic methods.

\subsection{Non-extremal D1 branes}

Let us first consider non-extremal D1-branes, and derive the
renormalized vevs and onshell action. The ten-dimensional solution for
non-extremal D1-branes is:
\bea
ds^2 &=& H^{-1/2} (- f dt^2 + dx^2 ) + H^{1/2} (\frac{dr^2}{f} + r^2
d\Omega_7^2); \\
e^{\phi} &=& g_s H^{1/2}; \qquad F_{01r} = g_s^{-1} \pa_r (1 - \frac{Q}{r^6} H^{-1}), \nn
\eea
with
\be
H = 1 + \frac{\mu^6 \sinh^2 \a}{r^6}; \qquad
f = (1 - \frac{\mu^6}{r^6}); \qquad
Q \equiv r_{o}^6 = \mu^6 \sinh \a \cosh \a.
\ee
The extremal limit is reached by taking $\mu \rightarrow 0$ and $\a
\rightarrow \infty$ with $\mu^3 \sinh \a$ fixed. In the near extremal
limit, for which $\mu \ll 1$, the decoupled dual frame metric is
\be
ds^2_{dual} = (g_s N)^{-1/3} \left (  \left ( \frac{r}{r_o} \right )^4 (- f dt^2 + dx^2) + r_{o}^2
  (\frac{dr^2}{r^2 f} +  d\Omega_7^2) \right ).
\ee
Applying the reduction formulae (\ref{reduct}) gives
an asymptotically $AdS_3$ solution of the three-dimensional action:
\bea
ds^2 &=& \frac{d\rho^2}{4 \rho^2 f} + \frac{1}{\rho} (- f dt^2 +
dx^2); \\
e^{-4 \phi/3} &=& \frac{1}{\rho}, \qquad
f = \left (1 - \frac{8 \mu^6}{r_o^9} \rho^{3/2} \right ). \nn
\eea
The inverse Hawking temperature $\beta_{H}$ and the area of the
horizon $A$ are respectively given by
\be
\beta_{H} = \frac{2 \pi r_o^3}{3 \mu^2}; \qquad
A = \frac{8 \pi R_x \mu^4}{r_o^6},
\ee
where the $x$ direction is taken to be periodic with period $2 \pi
R_x$.

Next one can read off the vevs for the stress energy tensor and scalar
operator by bringing the metric into Fefferman-Graham form:
\bea
ds^2 &=& \frac{dz^2}{4 z^2} + \frac{1}{z} \left ( - dt^2 (1 - \frac{16
\mu^6}{3r_o^9} z^{3/2}) + dx^2 (1 + \frac{8
\mu^6}{3r_o^9} z^{3/2}) \right ); \nn \\
e^{- 2  \phi/3} & \equiv & \frac{1}{\sqrt{z}} e^{\kappa} = \frac{1}{\sqrt{z}}
(1 + \frac{4
\mu^6}{3r_o^9} z^{3/2}).
\eea
Then applying (\ref{scalar-d1}) and (\ref{set-d1}) (analytically
continued back to the Lorentzian) the vevs of the
stress energy tensor are:
\be
\langle T_{tt} \rangle = 16 L \frac{\mu^6}{r_o^9}; \qquad
\langle T_{yy} \rangle = 8 L \frac{\mu^6}{r_o^9}; \qquad
    \< \cO \> = - 4 L \frac{\mu^6}{r_o^9},
\ee
with the conformal Ward identity (\ref{d1-ward}) manifestly satisfied. Note that the
mass is given by
\be
M = \int dx \langle T_{tt} \rangle =  L R_x \frac{32 \pi \mu^6}{r_o^9}.
\ee
The renormalized onshell (Euclidean) action $I_E$ is given by
\be
   I_{E} = - L [ \int_{\r \geq \e} d^3 x \sqrt{g} \Phi (R+C) - \int_{\r=\e}
d^2 x \sqrt{h} \Phi ( 2K -4 - R[h])].
\ee
Evaluating this action on the solution gives
\be
I_E = - 2 \pi \beta_{H} R_x L \frac{8 \mu^6}{r_o^9},
\ee
whilst the entropy is
\be
S = 4 \pi L A = L \frac{32 \pi^2 R_x \mu^4}{r_o^6},
\ee
and thus the expected relation
\be
I_E = \beta_{H} M - S
\ee
is satisfied. Note that $M/T_{H}S = 2/3$. This result is in agreement
with the results found in \cite{Klebanov:1996un} for the entropy of non-extremal Dp-branes.
The entropy can be rewritten as
\be
S = \frac{2^4 \pi^{5/2}}{3^3} \frac{N^2}{g_{eff}(T_H)} (V_1 T_H),
\ee
where $V_1 = 2 \pi R_H$ is the spatial volume of the D1 brane and $g_{eff}^2 = g_2^2 N T^{-2}_H$
is the dimensionless effective coupling (with $g_2^2 = g_s/(2 \pi \a')$ the dimensionful
Yang-Mills coupling constant). This is indeed of the form (\ref{SDp}) dictated from
the generalized conformal structure. The overall $N^2$ is due to the fact that the
bulk computation is a tree-level computation.

\subsection{The Witten model of holographic $YM_4$ theory}

As the next application of the formalism let us discuss Witten's
holographic model for four dimensional Yang-Mills theory \cite{Witten:1998zw}. 
An early discussion of holographic computations in this model 
can be found in \cite{Hashimoto:1998if}.
In this model one considers
D4 branes wrapping a circle of size $L_{\t}$ with anti-periodic
boundary conditions for the fermions, which breaks the supersymmetry.
This system at low energies looks like a four-dimensional $SU(N)$
gauge theory, with Yang-Mills coupling $g_4^2 = g_5^2/ L_{\t}$. In the
limit that $\l_{4} = g_4^2 N \gg 1$ there is an effective supergravity
description given by the D4 brane soliton solution,
which (in the string frame) is \cite{Witten:1998zw,Horowitz:1998ha}:
\bea
   ds^2_{st} &=& \left(\frac{r}{r_o}\right)^{3/2} [ \eta_{\a\b} dx^\a dx^\b
     + f(r) d\t^2] +  \left(\frac{r_o}{r}\right)^{3/2}(\frac{dr^2}{f(r)} + r^2 d\Omega_4^2), \nono \\
   e^\phi &=& g_s \left(\frac{r}{r_o}\right)^{3/4}, \qquad F_4 =
   3 g_s^{-1} r_o^3 d\Omega_4, \label{solit} \\
   f(r) &=& 1-\frac{r_{KK}^3}{r^3}, \nono
\eea
where $d \Omega_4$ is the volume form of the $S^4$ and $r_o$ was
defined below \eqref{reduct}. Then $r_{KK}$ is the minimum
value of the radial coordinate and the circle direction $\t$ must have
periodicity $L_{\t} = 4 \pi r_o^{3/2}/( 3r_{KK}^{1/2})$ to
prevent a conical singularity.

By wrapping D8-branes around the $S^4$,
and along the four flat directions, one can model chiral flavors in
the gauge theory \cite{Sakai:2004cn,Sakai:2005yt} and the resulting
Witten-Sakai-Sugimoto model
has attracted considerable attention as a simple holographic model for
a non-supersymmetric four-dimensional gauge theory.
The methods developed in this paper immediately allow one to extract
holographic data from this background, and to quantify
the features of QCD which are well or poorly modeled.

Starting from the ten-dimensional string frame solution, one can move
to the dual frame $ds^2_{dual} = (N e^{\phi})^{-2/3} ds^2$ in which the metric becomes
asymptotically $AdS_6 \times S^4$:
\bea
   ds^2_{dual} &=& (N e^{\phi})^{-2/3} ds^2_{st} = \pi^{2/3} \a' \left
   ( 4  [\frac{d\r^2}{4\r^2 f(\r)}
+ \frac{\eta_{\a\b} dx^\a dx^\b + f(\r) d\t^2}{\r}] + d\Omega_4^2
   \right ), \nn \\
   f(\r) &\equiv& 1- \frac{\r^3}{\r_{KK}^3} = f(r),
\eea
with changed variable $\r = 4 r_o^3/ r$. Comparing with the reduction
given in (\ref{reduct}), one obtains the following six-dimensional
background:
\bea
ds^2 &=& \frac{d\r^2}{4\r^2f(\r)}  + \frac{\eta_{\a\b} dx^\a dx^\b + f(\r)
d\t^2}{\r}; \\
e^{\phi} &=& \frac{1}{\r^{3/4}}, \nn
\eea
which is asymptotically $AdS_6$ with a linear dilaton.

The gauge theory operators dual to the metric and the scalar field are
the five-dimensional stress energy tensor $T_{ij}$ and the gluon operator
${\cal O}$ respectively, which satisfy the dilatation Ward identity
(see \eqref{d4-dilatt} or \eqref{confwi}):
\be
\< T_{i}^{i} \> + \frac{1}{g_5^2} \< {\cal O} \> = 0.
\ee
(There is no anomaly in this case, as both $g_{(0)}$ and
$\kappa_{(0)}$ are constant.) This Ward identity can be
rewritten in terms of operators in the four-dimensional theory
obtained via reduction over the circle: the four-dimensional stress
energy tensor $T^{(4)}_{ab} = L_{\t} T_{ab}$ and the scalar operator
${\cal O}_{\t} = L_{\t} T_{\t\t}$. This gives
\be \label{WIYM}
\< T_{a}^{a} \> + \< {\cal O}_{\t} \> + \frac{1}{g_4^2} \< {\cal O} \>
= 0.
\ee
Consider the dimensional reduction of
the stress energy tensor and gluon operator defined in
\eqref{tij} from five to four dimensions. When the reduction over the
circle preserves supersymmetry, the operator ${\cal O}_{\t}$ coincides with
$-\frac{1}{g_4^2} {\cal O}$ and the four-dimensional stress energy
tensor is traceless. With non-supersymmetric boundary conditions, this 
is not the case anymore, since as we will see shortly the vacuum expectation 
value of the trace of the stress energy tensor is not zero and the vevs
of the two operators are different. With the proper identification of the
relation between $\cO_\t$ and $\cO$, the trace Ward identity would lead to the 
identification of the beta function.

Next one can extract the one point
functions for the stress energy tensor and gluon operators from
the coefficients in the asymptotic
expansion of this solution near the boundary.
To apply the formulae for the holographic vevs, the metric should
first be brought into Fefferman-Graham form by changing the radial variable:
\bea
   \tilde{\r} &=& (1 +  \frac{\r^3}{6\r_{KK}^3}) \r + \cO(\r^5), \\
   ds^2 &=& \frac{d\tilde{\r}^2}{4\tilde{\r}^2} +
\tilde{\r}^{-1} (1+ \frac{\tilde{\r}^3}{6\r_{KK}^3})\eta_{\a\b} dx^\a dx^\b +
\tilde{\r}^{-1} (1- \frac{5 \tilde{\r}^3}{6 \r_{KK}^3}) d\t^2 + \cdots. \nn
\eea
Using \eqref{onepoint}
the one-point function of the scalar operator is thus:
\be
   \<\cO_\phi\> = - 12  L \g \kappa\sub{6} = - \frac{2 L }{3\r^3_{KK}},
\ee
with the vev of the stress energy tensor being:
\be
\< T_{\a\b} \> = \frac{L}{\r^3_{KK}} \eta_{\a\b}; \qquad
\< T_{\t\t} \> =  - 5 \frac{L}{\r^3_{KK}}.
\ee
The gluon condensate can be reexpressed as:
\be
\<\cO_\phi\> = - \frac{2^6 \pi^2 }{3^8} N^2 \frac{\l_4}{L_{\t}^5},
\ee
where recall that $\l_4 = g_4^2 N$ is the four-dimensional 't Hooft
coupling and $L_{\t}$ is the radius of the circle. In terms of the
dimension four operator ${\cal O}$ the condensate is
\be
\< \cO \> =  \frac{2^5 \pi^2 }{3^7} N \frac{\l_4^2}{L_{\t}^4}.
\ee
In comparing results for this holographic model with those of QCD, it would be
natural to match the condensate values, and thus fix $L_{\t}$.

\section{Discussion} \label{nine}

In this paper we have developed precision holography for the non-conformal branes.
We found that all holographic results that were developed earlier in the
context of holography for the conformal branes can be extended to this
more general setup.
All branes under consideration have a near-horizon limit with non-vanishing
dilaton and a metric that (in the string frame)
is conformal to $AdS_{p+2} \times S^{8-p}$.
This implies that there is a frame, the dual frame, where the metric is exactly
$AdS_{p+2} \times S^{8-p}$ (one can cancel the overall conformal
factor by multiplying the metric with the appropriate power of the dilaton).

There are a number of reasons why this frame is distinguished. Firstly,
it is manifest in this frame that there is an effective $(p+1)$-dimensional
gravitational description, obtained by reducing over $S^{8-p}$,
as required by holography. Secondly, the setup becomes the
same as that of holographic RG flows studied earlier. Actually the
bulk solutions do describe an RG flow, albeit a trivial one
driven by the dimension of the coupling constant. Recall that
in the holographic RG flows studied in the past the bulk solution
asymptotically becomes $AdS$, corresponding to the fact that the
dual QFT approaches a fixed point in the UV. The scalar fields
vanish asymptotically, and from the asymptotic fall off one can
infer whether the bulk solution corresponds to a deformation of the
UV Lagrangian by the addition of the operator dual to the corresponding
field or the conformal theory in a non-trivial state characterized
(in part) by the vev of the dual operator. The coefficients in the
asymptotic expansion of the solution determine the coupling constant
multiplying the dual operator in the case of deformations, or the vev
of the dual operator in the case of non-trivial states.

The non-conformal branes are analogous to the case of deformations:
the asymptotic value of the dilaton determines the value of the
coupling constant, which is the (dimensionful) Yang-Mills coupling constant
in the case of D$p$ branes. The main difference is that in the current
context the theory does not flow in the UV to a $(p+1)$-dimensional
fixed point. Rather in the regime where the various approximations
are valid, the theory runs trivially due to the dimensionality of the
coupling constant.

In some cases however we know that a new dimension, the M-theory
dimension, opens up at strong coupling and the theory flows to a
$(p+2)$-dimensional fixed point. This is the case for the IIA
fundamental string and the D4 brane which uplift to the M2
and M5 brane theories, respectively. Here is another instance
that illustrates the preferred status of the dual frame:
the general solution in the dual frame
\bea
ds^2_d &=&  \frac{d\rho^2}{4 \rho^2} + \frac{1}{\rho} g_{ij} dx^i dx^j \\
e^{4 \phi/3} &=& \frac{1}{\rho} e^{2 \k}, \
\eea
lifts to
\be
ds_{d+1}^2 =  \frac{d\rho^2}{4 \rho^2} + \frac{1}{\rho} (g_{ij} dx^i dx^j +
e^{2 \k} dy^2).
\ee
In other words, the dual frame metric in the Fefferman-Graham gauge
in $d$-dimensions
is equal to the $d$-dimensional part of the
metric in $(d+1)$ dimensions in the Fefferman-Graham gauge,
with the dilaton providing the additional dimension.
It was already observed in \cite{Boonstra:1998mp} that the radial
coordinate in the dual frame is identified with the energy of the dual
theory via the UV-IR connection and here we see a more precise formulation
of this statement. The radial direction of the M5 and M2 branes
is also the radial direction in the dual frame of the D4 and F1 branes,
respectively. In more covariant language, the dilatation operator of the
boundary theory is to leading order equal to the radial derivative
of the dual frame metric.

\bigskip

Working in the dual frame, we have systematically developed
holographic renormalization for all non-conformal branes. In
particular, we obtained the general solutions of the field equations
with the appropriate Dirichlet boundary conditions. This allowed us to
identify the volume divergences of the action, and then remove these
divergences with local covariant counterterms. Having defined the
renormalized action, we then proceeded to calculate the holographic
one-point functions which, by further functional differentiation 
w.r.t sources,
yield the higher point functions.
The counterterm actions can be found in \eqref{ct-general} and \eqref{d4-ct1}, whilst the
holographic one point functions are given in \eqref{other-vev} and
\eqref{dppp4}. Note that the
result for the stress energy tensor properly defines the notion of
mass for backgrounds with these asymptotics.

We developed holographic renormalization both in the original
formulation, described in the previous paragraph, and in the radial
Hamiltonian formalism (in section \ref{six}). In the latter, Hamilton-Jacobi theory
relates the variation of the on-shell action w.r.t. boundary conditions,
thus the holographic 1-point functions, to radial canonical momenta.
It follows that one can bypass the on-shell
action and directly compute renormalized correlators using radial
canonical momenta $\pi$, as was developed for asymptotically AdS
spacetimes in \cite{Papadimitriou:2004ap,Papadimitriou:2004rz}. For
explicit calculations, the Hamiltonian method is more efficient
and powerful, as it exploits to the full the underlying symmetry
structure.

Throughout the existence of an underlying generalized
conformal structure plays a crucial role. As we discussed in section
\ref{four}, SYM in $d$ dimensions admits a generalized conformal
structure, in which the action is invariant under Weyl transformations
provided that the coupling constant is also promoted to a background
field $\F_{(0)}$ which transforms appropriately. This background field
can be thought of as a source for a gauge invariant operator $\cO$. 
Then diffeomorphism and Weyl invariance imply Ward identities
for the correlators of the stress energy tensor and the operator $\cO$.
This generalized conformal structure is preserved at
strong coupling, and governs the holographic Ward identities.
In particular, the Dirichlet boundary 
conditions for the dilaton are determined by the field theory source $\F_{(0)}$. 

In the cases of the type IIA fundamental string and D4-branes, all the
holographic results we find are manifestly compatible with the M
theory uplift. In particular, we showed in detail how the asymptotic
solutions, counterterms, one point functions and anomalies descend
from those of M2 and M5 branes. The generalized conformal
structure is also inherited from the higher dimensional conformal
symmetry in these cases. This is exactly analogous to the case of the 
more familiar holographic RG flows, which also have a similar generalized 
conformal structure inherited from the UV fixed point.

Having set up the formalism in full generality, we then proceeded to
discuss a number of examples and applications. In section \ref{seven}
we calculated two point functions of the stress
energy tensor and gluon operator. We computed these two point
functions for the supersymmetric backgrounds, and showed that the
results were consistent with the underlying generalized conformal
structure. In section \ref{2ptgen} we developed a general method for
computing two point functions in any background which asymptotes to the
non-conformal brane background.

In section \ref{eight} we gave several more applications. One was the
explicit evaluation of the mass and action in a non-extremal brane
background. The second was Witten's model for a non-supersymmetric
four-dimensional gauge theory: we computed the dimension four condensates in this model.
One would anticipate that there are many further interesting 
applications of the formalism developed here, to be explored in future work.

\section*{Acknowledgments}

The authors IK and MMT are supported by NWO, via the Vidi grant ``Holography,
duality and time dependence in string theory''. This work was also
supported in part by the EU contract MRTN-CT-2004-512194.

\appendix

\section{Useful formulae} \label{appA}

In this appendix we collect some useful formulae for the asymptotic
expansions.
Given the expansion of the $d$-dimensional metric $g_{ij}$ as
\be
g_{ij} = g_{(0) ij} + \rho g_{(2) ij} + \rho^2 g_{(4) ij} + \cdots
\ee
the inverse metric is given by
\be
g^{-1} = g_{(0)}^{-1} - \rho g_{(0)}^{-1} g_{(2)} g_{(0)}^{-1}
 + \rho^2 (g_{(0)}^{-1} g_{(2)} g_{(0)}^{-1} g_{(2)} g_{(0)}^{-1} -
 g_{(0)}^{-1} g_{(4)} g_{(0)}^{-1}) + \cdots
\ee
Next we compute the expansion of the Christoffel connection,
\be
\G^i_{ij} = \G^i_{(0)ij} + \r \G^i_{(2)ij} + \r^2 \G^i_{(4)ij} + \cdots
\ee
Here $\G^i_{(0)ij}$ is the Christoffel connection of the metric $g_{(0)}$
and
\bea
\G^i_{(2)ij} &=& \frac{1}{2} g_{(0)}^{il}(\nabla_j g_{(2)kl}
+\nabla_k g_{(2)jl}
-\nabla_l g_{(2)jk}) \\
\G^i_{(4)ij} &=& \frac{1}{2}\left(
g_{(0)}^{il}(\nabla_j g_{(4)kl}
+\nabla_k g_{(4)jl}
-\nabla_l g_{(4)jk})
-g_{(2)}^{il}(\nabla_j g_{(2)kl}
+\nabla_k g_{(2)jl}
-\nabla_l g_{(2)jk})\right), \nn
\eea
where $\nabla$ is the covariant derivative in the metric $g_{(0)}$.

From here we then compute the expansion of the associated curvature
\be
R_{ij} = R_{(0) ij} +  \rho R_{(2) ij} + \rho^2 R_{(4) ij} + \cdots
\ee
with $ R_{(0) ij}$ the Ricci tensor of $g_{(0)}$ and
\bea
R_{(2) ij} &=& \frac{1}{2} \left ( \nabla^{k} \nabla_{j} g_{(2) ik} -
\nabla^{i} \nabla_{j} \Tr ( g_{(2)}) + R_{(0) kijl} g_{(2)}^{kl} +
R_{(0) im} g_{(2) j}^{m} \right. \nn \\
&& \left.- \nabla^2 g_{(2) ij} + \nabla_i \nabla^{k}
g_{(2) jk} \right ), \label{appendix1} \\
R_{(4) ij} &=& \frac{1}{2}\left(\frac{1}{2} \nabla^l \Tr g_{(2)} \nabla_l g_{(2) ij}
+  g_{(2)}^{kl} ( R_{kimj} g_{(2) l}^{m} +  R_{kiml}
g_{(2)j}^{m} ) \right. \\
&& +  g_{(2)}^{kl} ( R_{kjmi} g_{(2)l}^{m} +  R_{kjml}
g_{(2)i}^{m} )
+ g_{(2)}^{kl} \nabla_{k} \nabla_{l} g_{(2) ij} + \frac{1}{2} \nabla_{j} g_{(2) lm} \nabla^{l} g_{(2) i}^{m} \nn \\
&& - \frac{1}{2} g_{(2)}^{kl} (\nabla_{i} \nabla_{k} g_{(2) jl} +
\nabla_{j} \nabla_{k} g_{(2) il})
- 2 R_{limj} g_{(4)}^{lm} + R_{im} g_{(4)j}^{m} + R_{jm} g_{(4) i}^m
\nn \\
&& + \frac{1}{4} g_{(2)j}^{l} \nabla_{i} \nabla_{l} \Tr g_{(2)}
+ \frac{1}{4} g_{(2)i}^{l} \nabla_{j} \nabla_{l} \Tr (g_{(2)})
+ \frac{1}{2} \nabla_{i} g_{(2) lm} \nabla^{l} g_{(2) j}^{m}
 \nn \\
&& \left.- \nabla^2 g_{(4) ij}
- \frac{1}{2} \nabla_{i} g_{(2) lm} \nabla_{j} g_{(2)}^{lm}
- \nabla_{m} g_{(2) il} \nabla^{l} g_{(2) j}^{m}
- \nabla_{m} g_{(2) il} \nabla^{m} g_{(2) j}^{l}\right). \nn
\eea

\comment{ In
the case of interest $g_{(2) ij}$ is given by \eqref{firstcoeffs} and therefore
\bea
R_{(2) ij} &=& - \frac{1}{2 (d-1)} \left
( 2 (R_{(0) im} -e^{-\k\sub{0}}\na_i \pa_m e^{\k\sub{0}})
(R_{(0) j}^{m} -e^{-\k\sub{0}}\na^m \pa_j e^{\k\sub{0}}) \right. \\
&& \left. -2 R_{(0) kimj}  (R_{(0)}^{km} - e^{-\k\sub{0}}\na^k
\pa^m e^{\k\sub{0}}) -2 e^{-2\k\sub{0}} \na_i \pa_j e^{\k\sub{0}} \na^2 e^{\k\sub{0}}\right.  \nono \\
&& \left. - \frac{(d-1)}{2d} \nabla_{i} \nabla_{j} (R_{(0)} -2e^{-\k\sub{0}}\na^2 e^{\k\sub{0}})  \right. \nono \\
&& \left .  + (\nabla^2 + \pa^m\k\sub{0} \na_m) ( R_{(0) ij} -
e^{-\k\sub{0}}\na_i \pa_j e^{\k\sub{0}}) \right. \nono \\
&& \left. - 2 \pa_m \k\sub{0} (R^m_{(0)(i}  -
e^{-\k\sub{0}}\na^m \pa_{(i} e^{\k\sub{0}})\pa_{j)} \k_{(0)} -
2 \pa_{i} \k\sub{0} \pa_{j} \k\sub{0} e^{-\k\sub{0}}\na^2 e^{\k\sub{0}} \right. \nono \\
&& \left. - \frac{1}{2d} (\nabla^2 + \pa^m \k\sub{0} \pa_m) (R_{(0)} - 2 e^{-\k\sub{0}}\na^2 e^{\k\sub{0}}) g_{(0)
ij}\right ). \nn
\eea
Ricci and Bianchi identity are useful in deriving this expression:
\be
\nabla_{i} \nabla_{k} R^{i}_{j} = R_{km} R^{m}_{j} - R_{ikmj} R^{im}
+ \half \nabla_{k} \nabla_j R.
\ee}
\section{The energy momentum tensor in the conformal cases} \label{appb}

In this section we streamline the derivation of the vev of the energy-momentum tensor in terms
of the asymptotic coefficients for the conformal cases $D=4$ and $D=6$ given in \cite{deHaro:2000xn}.
The starting point is the expression of the stress energy tensor as sum of two contributions, one
originating from the bulk action and the other from the counterterms, eqns (3.5)-(3.6)-(3.7) of
\cite{deHaro:2000xn}:
\bea
\label{Tregct}
   \<T_{ab}\> &=& 2 L_{D+1}
\lim_{\r \rightarrow 0} (\frac{1}{\r^{D/2-1}} T_{ab}[G]), \\
   T_{ab}[G] &=& T_{ab}^{reg} + T_{ab}^{ct}, \nono \\
   T_{ab}^{reg} &=&  G'_{ab} - G_{ab} \Tr [ G^{-1} G'] - \frac{1-D}{\r} G_{ab}, \nono \\
   T_{ab}^{ct} &=& -\frac{D-1}{\r} G_{ab} + \frac{1}{(D-2)} (R(G)_{ab} - \hp R(G)G_{ab}) \nono \\
       && + \frac{\r}{(D-4)(D-2)^2}[ \Box R(G)_{ab} + 2 R(G)_{a c b d}
     R(G)^{cd} - \frac{D-2}{2(D-1)} D_a D_b R(G) \nono \\
       && - \frac{D}{2(D-1)} R(G) R(G)_{ab} -\hp G_{ab}(R(G)_{cd} R(G)^{cd} - \frac{D}{4(D-1)} R(G)^2\nono \\
       && + \frac{1}{D-1} \Box R(G))] + \frac{1}{2} T_{ab}^{log} \log \r \nono,
\eea
where $L_{D+1} = \frac{1}{16 \pi G_{D+1}}$ with $G_{D+1}$ the Newton
  constant and
$T_{ab}^{log}$ is the stress energy tensor of the action given
   by the conformal
anomaly\footnote{The factor of 1/2 in front of $T_{ab}^{log}$
   corrects a typo in \cite{deHaro:2000xn}.}. Note that for $D=2$ only the first term
in $T_{ab}^{ct}$ applies; for $D=4$ only the first line applies plus
   the logarithmic terms, whilst for $D=6$ all
terms listed are needed and for $D>6$ one would need to include additional terms.

For $D=2$ one immediately obtains the answer
\be
\<T_{ab}\> = 2 L_{D+1}  \left( G_{(2)ab} -G_{(0)ab} \Tr G_{(2)}\right)
\ee
For $D>2$ one can simplify the evaluation of (\ref{Tregct}) by using the
equation of motions (\ref{fg2}) to obtain
\bea
\label{Einstab}
   R_{ab} - \hp R G_{ab} &=& -(D-2) G'_{ab} + (D-2) \Tr (G^{-1} G') G_{ab}\\
      &&  + \r[ 2 G'' - 2G'G^{-1} G'  + \Tr(G^{-1} G') G' \nono \\
      && + (- \Tr(G^{-1} G'') + \Tr(G^{-1} G')^2 - \hp (\Tr G^{-1} G')^2) G]_{ab}. \nono
\eea
Using this identity in \eqref{Tregct} we see that $T_{ab}^{reg}$
      cancels the first line of $T_{ab}^{ct}$ up to the terms
      proportional to $\r$ in \eqref{Einstab}, so $T_{ab}[G]$ is manifestly linear in $\rho$.
It follows that we only need to set $\rho=0$ in the remaining terms to obtain the vev for $D=4$:
\bea
   \<T_{ab}\> &=& 2 L_{D+1} \left (2 G\sub{4}_{ab} - G\sub{2}^2_{ab} + \hp \Tr G\sub{2} G\sub{2}_{ab}\right. \\
   &&\left. + \frac{1}{4} G\sub{0}_{ab}(\Tr (G^{-1} G\sub{2})^2 - (\Tr
      G^{-1} G\sub{2})^2) + 3 H\sub{4}_{ab}\right). \nono
\eea
For $D=6$ one can check straightforwardly that order $\r$ terms in $T_{ab}[G]$ cancel, so there is indeed
a finite limit. To obtain the vev one needs to extract the order $\r^2$ terms. To simplify this
computation we differentiate the field equations (\ref{fg2}) to obtain a formula for the radial derivative
of the Ricci tensor,
\bea
   R'_{ab} &=& R_{(a}^c G'_{b)c} - R_{a c b d} G'^{cd} + D_{(a} D^b
   G'_{b)c} - \hp \Box G'_{ab}+ D_a \pa_b \Tr G' \label{ric'} \\
   &=& \frac{1}{D-2} [-R_{acbd} R^{cd} + \frac{D-2}{4(D-1)} D_a D_b R
     +\hp \Box R_{ab} + \frac{1}{4(D-1)} \Box R G_{ab}  \nono \\
   && + R_a^c R_{bc} - \r[ 4 R\sub{0}_{(a}^c \tilde{C}_{b)c} - 4
       R\sub{0}_{a c b d} \tilde{C}^{cd} - \frac{D-2}{4(D-1)} D_a \pa_b B \nono \\
   && - 2 \Box \tilde{C}_{ab} - \frac{1}{4(D-1)} G\sub{0}_{ab}]] + \cO(\r^2), \nono \\
   \tilde{C}_{ab} &=& (G\sub{4} - \hp G\sub{2}^2 + \frac{1}{4}
   G\sub{2}\Tr G\sub{2})_{ab}, \qquad B = \Tr G\sub{2}^2 - (\Tr G\sub{2})^2. \nono
\eea
Then we note that the terms involving the Riemann tensor and covariant
derivatives enter with the same relative factors as in $T_{ab}^{ct}$,
so we can use (\ref{ric'}) to express $T_{ab}^{ct}$ in terms of
$R'_{ab} = R\sub{2}_{ab} + 2\r R\sub{4}_{ab} + \cdots$, which is
easier to relate to higher expansion coefficients. Indeed, as is
discussed in the next appendix, the coefficient $R\sub{2}_{ab}, R\sub{4}_{ab}$
can be expressed in terms of $G\sub{2}_{ab}$, $G\sub{4}_{ab}$ and $H\sub{6}_{ab}$.

Combining these results and setting $D=6$ we obtain
\be
   \< T_{ab} \> = 2 L_{7} \left( 3 G\sub{6}_{ab} - 3 A\sub{6}_{ab} +
   \frac{1}{8} S_{ab} + \frac{11}{2} H\sub{6}_{ab} \right),
\ee
where $A\sub{6}_{ab}$ and $S_{ab}$ are given by \cite{deHaro:2000xn}
\bea
   S_{ab} &=& \Box C_{ab} + 2 R_{a c b d} C^{cd} + 4(G\sub{2} G\sub{4} - G\sub{4} G\sub{2})_{ab}  \\
    && + \frac{1}{10} (D_a D_b B - G\sub{0}_{ab} \Box B) + \frac{2}{5} G\sub{2}_{ab} B \nono \\
    && + G\sub{0}_{ab}(-\frac{2}{3} \Tr G^3_{(2)} - \frac{4}{15}(\Tr
   G\sub{2})^3 + \frac{3}{5} \Tr G\sub{2} \Tr G^2_{(2)}), \nono \\
   A_{(6)ab} &=& \frac{1}{3}\left( (2G\sub{2}G\sub{4} +
   G\sub{4}G\sub{2})_{ab} - G^3_{(2)ab} + \frac{1}{8} [\Tr G^2_{(2)} -
(\Tr G\sub{2})^2] G\sub{2}_{ab} \right. \nono \\
   && - \Tr G\sub{2}[G\sub{4}_{ab} - \hp G^2_{(2)ab}] - [ \frac{1}{8}
     \Tr G^2_{(2)} \Tr G\sub{2} - \frac{1}{24} (\Tr G\sub{2})^3 \nono \\
   && \left. -\frac{1}{6} \Tr G^3_{(2)} + \hp \Tr (G\sub{2}G\sub{4})] G\sub{0}_{ab} \right), \nono \\
   C_{ab} &=& (G\sub{4} - \hp G\sub{2}^2 + \frac{1}{4} G\sub{2}\Tr
   G\sub{2})_{ab} + \frac{1}{8} G\sub{0}_{ab} B, \qquad B = \Tr G\sub{2}^2 - (\Tr G\sub{2})^2. \nono
\eea
Noting that $L_{7} = N^3/(3 \pi^3)$ and introducing the combination
\be
t_{ab} = G\sub{6}_{ab} - A\sub{6}_{ab} +
   \frac{1}{24} S_{ab}
\ee
the stress energy tensor may be expressed as
\be \label{em-six}
\< T_{ab} \> =  \frac{N^3}{3 \pi^3} (6 t_{ab} + 11  H\sub{6}_{ab}).
\ee
This result includes the term in $H_{\sub{6}_{ab}}$ which was not
given in \cite{deHaro:2000xn}.

\section{Reduction of M5 to D4} \label{appB}

The expansion coefficients for an asymptotically local $AdS_{D+1}$
metric were given in \cite{deHaro:2000xn}. We will be interested in
the case where $D = d+1$, for which the first expansion coefficients are:
\bea
G_{(2)}{}_{ab}  &=&  \frac{1}{d - 1} \left( - R_{(0) ab} + \frac{1}{2d}
R_{(0)} G_{(0)}{}_{ab} \right); \label{g4-m5}  \\
G_{(4) ab} &=& \frac{1}{2(d-3)} \left ( - R_{(2) ab} -2
(G_{(2)}^2)_{ab} + \frac{1}{2} \Tr(G_{(2)}^2) G_{(0)ab} \right ).  \nn
\eea
Using the explicit form of $G_{(2)ab}$ and the $D$-dimensional analogue of \eqref{appendix1} we obtain:
\bea
   R_{(2)ab} &=& -\frac{1}{2(d-1)}\left( 2R_{(0)ac} R^c_{(0)b} -
2 R_{(0) cadb} R_{(0)}^{cd} - \frac{d-1}{2d} D_a D_b R_{(0)}\right.  \\
   && \left.+ D^2 R_{(0)ab} - \frac{1}{2d} D^2 R_{(0)} G_{(0)ab} \right), \nono \\
  G_{(4) ab} &=& - \frac{1}{d-3} \left ( - \frac{1}{8d} D_a D_b R +
\frac{1}{4(d-1)} D_c D^c R_{ab} \right . \nono \\
&& - \frac{1}{8 d (d-1)} D_c D^c R G_{(0) ab} + \frac{1}{2 (d-1)} R^{cd}
R_{acbd} \nn \\
&& - \frac{d-3}{2 (d-1)^2} R_{a}^{c} R_{cb} - \frac{1}{d (d-1)^2} R
R_{ab} \nn \\
&& \left . - \frac{1}{4 (d-1)^2} R^{cd} R_{cd} G_{(0) ab} + 3 \frac{(d+1)}{16 d^2
(d-1)^2} R^2 G_{(0) ab} \right ),  \nn
\eea
where $D_a$ is the covariant derivative in the metric $G_{(0)}$.
Note that $R_{(2)} = 0$, and thus
\be
\Tr G_{(4)} = \frac{1}{4} \Tr (G_{(2)}^2).
\ee
At next order one finds that the trace and the divergence of
$G_{(6)}$ are determined via
\bea
\Tr (G_{(6)}) &=& \frac{2}{3} \Tr (G_{(2)} G_{(4)}) - \frac{1}{6} \Tr
(G_{(2)}^3); \label{m5-tr-cons} \\
D^a G_{(6) ab} &=& D^a A_{(6) ab} + \frac{1}{6} \Tr (G_{(4)} D_b
G_{(2)}); \nn \\
A_{(6) ab} &=& \frac{1}{3}\left(
(2G\sub{2}G\sub{4}+G\sub{4}G\sub{2})_{ab} -
(G^3_{(2)})_{ab} + \frac{1}{8}[\Tr G^2_{(2)} - (\Tr G\sub{2})^2] G\sub{2}_{ab}\right. \nn \\
&& \left. - \Tr G\sub{2}[G\sub{4}_{ab} - \hp (G^2_{(2)})_{ab}] -
[\frac{1}{8}\Tr G^2_{(2)} \Tr G\sub{2} - \frac{1}{24} (\Tr G\sub{2})^3 \right. \nono \\
&& \left. - \frac{1}{6} \Tr G^3_{(2)} + \hp \Tr (G\sub{2}G\sub{4})] G\sub{0}_{ab}\right).
\eea
The logarithmic term in the expansion $H_{(6)}$ is given by
\bea
H_{(6) ab} &=& \frac{1}{6} (R_{(4) ab} +  ( -\Tr(G_{(2)} G_{(4)})
+ \half  \Tr(G_{(2)}^3))  G_{(0)ab} ) \label{log-H} \\
&& - \frac{1}{6} \Tr(G_{(2)}) G_{(4) ab} - \frac{1}{3} (G_{(2)}^3)_{ab} +
\frac{2}{3} (G_{(2)} G_{(4)} + G_{(4)} G_{(2)})_{ab}. \nn
\eea
Note that $H_{(6)}$ is traceless and divergence free.

\bigskip

For the dimensional reduction it is useful to note that the
non-vanishing components of the Riemann tensor can be expressed as
\bea
R(G)_{ijkl} &=& R_{ijkl}; \\
R(G)_{yiyj} &=& - e^{2 \kappa} (\nabla_i \pa_j \kappa + (\pa_i \kappa)
(\pa_j \kappa)), \nn
\eea
and similarly the non-vanishing components of the Ricci tensor are
\bea
R(G)_{ij} &=& R_{ij} - \nabla_{i} \pa_j \kappa - \pa_i \kappa \pa_j
\kappa; \\
R(G)_{yy} &=& e^{2 \kappa} (- \nabla^i \pa_i \kappa - \pa_i \kappa
\pa^i \kappa). \nn
\eea
Let furthermore $S$ be a scalar and $C_{ab}$ a symmetric tensor with $C_{iy}=0$. Then the Laplacian reduces as
\bea
   D^2 S &=& (\na^2 + \pa^i \k \pa_i) S, \\
   D^2 C_{ij} &=& (\na^2 + \pa^l \k \na_l) C_{ij} - 2 \pa_l \k
   \pa_{(i}\k
C_{j)}^l + 2 \pa_i \k \pa_j \k  C^y_y, \nono \\
   D^2 C^y_y &=& (\na^2 + \pa^i \k \pa_i) C^y_y  + 2\pa_i \k \pa_j \k C^{ij} - 2 \pa_i \k \pa^i \k C^y_y. \nono
\eea
Letting $G_{(0) ij} = g_{(0) ij}$ and $G_{(0) yy} =  e^{2 \k_{(0)}}$
one finds that
\bea
R(G)_{(0) ij} &=& R_{(0) ij} - \nabla_{i} \pa_j \k_{(0)} - \pa_i \k_{(0)} \pa_j
\k_{(0)}; \\
R(G)_{(0) yy} &=& e^{2 \k_{(0)}} ( - \nabla^{i} \pa_i \k_{(0)} - \pa_i \k_{(0)} \pa^i
\k_{(0)}),  \nn
\eea
with $R(G)_{(0) yi} =0$. Substituting into (\ref{hssg2}) 
gives\footnote{Round brackets $(ij)$ denote symmetrisation and curly
  brackets $\{ij\}$ traceless symmetrisation of indices.}
\bea
G_{(2)}{}_{ij} &=& \frac{1}{d - 1} \left ( - R_{(0) ij}  +
\frac{1}{2d}  R_{(0)} g_{(0) ij} + (\na_{\{i} \pa_{j\}} \k)_{(0)} +
\pa_{\{i} \k_{(0)} \pa_{j\}} \k_{(0)}) \right ); \\
G_{(2)}{}_{yy} &=&  e^{2 \k_{(0)}} \left (
\frac{1}{2d (d - 1)} R_{(0)} + \frac{1}{d} (\na^2 \k_{(0)} +
(\pa \k_{(0)})^2 ) \right ), \nn
\eea
with $G_{(2)}{}_{yi} = 0$. Now using
\be
G_{yy} = e^{2\k} = e^{(2 \k_{(0)} + 2 \rho \k_{(2)} + \cdots)} = e^{2
  \k_{(0)}} (1 + 2 \rho \k_{(2)} + \cdots)
\ee
one determines $\k_{(2)}$ to be exactly the expression given in
(\ref{firstcoeffs}).

One next shows that $G_{(4) ab}$ in \eqref{g4-m5} reduces as
\be
G_{(4) ij} = g_{(4) ij}; \qquad
G_{(4) yy} = e^{2 \k_{(0)}} (2 \k_{(2)}^2 + 2 \k_{(4)}),
\ee
with $g_{(4) ij}$ and $\k_{(4)}$ given in \eqref{secondcoeffs}. This
follows from the expansion of the six-dimensional curvatures at second order:
\bea
R(G)_{(2) ij} &=& R_{(2) ij} - (\nabla_{i} \pa_j \k)_{(2)} - (\pa_i \k  \pa_j
\k)_{(2)}; \\
R(G)_{(2) yy} &=& - e^{2 \k_{(0)}} (\nabla^{i} \pa_i \k + \pa_i \k \pa^i
\k)_{(2)}  \nn \\
&&  - e^{2 \k_{(0)}} 2 \kappa_{(2)} (\nabla^{i} \pa_i \k + \pa_i \k \pa^i
\k)_{(0)}. \nn
\eea
Reducing (\ref{m5-tr-cons}) gives
\bea
\Tr (G_{(6)}) &=& \Tr (g_{(6)}) + 2 \kappa_{(6)} + \frac{4}{3} \k_{(2)}^3 + 4
\kappa_{(2)} \kappa_{(4)}; \\
&=& \frac{2}{3} \Tr (g_{(2)}g_{(4)}) + \frac{4}{3} \k_{(2)} (\k_{(2)}^2 + 2
\k_{(4)}) - \frac{1}{6} \Tr(g_{(2)}^3), \nn
\eea
and thus gives
\be
\Tr (g_{(6)}) + 2 \kappa_{(6)} = \frac{2}{3} \Tr (g_{(2)}g_{(4)}) -
 \frac{4}{3} \k_{(2)} \k_{(4)}  - \frac{1}{6} \Tr(g_{(2)}^3).
\ee
The reduction of (\ref{log-H}) gives
\be
H_{(6) ij} = h_{(6) ij}; \qquad
H_{(6) yy} = e^{2 \k_{0}} 2 \td{\k}_{(6)},
\ee
with
\bea
h_{(6) ij} &=&  -\frac{1}{12} [ -2R_{(4)ij} + (-\Tr g_{(2)}^3
+ 2 \Tr g_{(2)} g_{(4)} + 8 \k_{2} \k_{4}) g_{(0)ij}
+ 2 (\Tr g_{(2)}) g_{(4)ij}  \nono \\
   && - 8 (g_{(4)} g_{(2)} )_{ij} - 8 (g_{(2)} g_{(4)} )_{ij} + 4 g^3_{(2)ij} + 2 (\na_i \pa_j
\k)_{(4)} + 2 (\pa_i \k \pa_j \k)_{(4)} + 4 \k_{(2)} g_{(4)ij}] \nono \\
\td{\k}_{(6)} &=& - \frac{1}{12} [ (\na^2 \k)_{(4)} +
     (\pa \k)^2_{(4)} +  \Tr g_{(2)} g_{(4)} - \half \Tr g_{(2)}^3
     \\
&& - \k_{(2)} \Tr g_{(2)}^2 + 4 \k_{(4)} \Tr g_{(2)} - 4 \k_{(2)}^3 +
12 \kappa_{(2)} \kappa_{(4)}], \nn
\eea
which agree with the expressions \eqref{logcoeffs}.
In reducing the curvature term $R(G)_{(4) yy}$ one should use the
identities:
\bea
- ( (\na^2 \k) +
     (\pa \k)^2 )_{(0)} &=& - 10 \kappa_{(2)} - \Tr g_{(2)}; \\
- ( (\na^2 \k) +
     (\pa \k)^2 )_{(2)} &=& - 8 \kappa_{(4)} + 6 \kappa_{(2)}^2 + 2
     \kappa_{(2)} \Tr g_{(2)} + \frac{1}{2} \Tr g_{(2)}^2. \nn
\eea

\section{Explicit expressions for momentum coefficients}
\label{D4braneappendix}

In the following we give explicit expressions for the terms in the
expansions of the momenta in eigenfunctions of the dilatation
operator. The expressions given below are applicable for $\beta = 0$
in \eqref{Dpaction} and $d \ge 3$, although in this paper we will use only the case of
$d=5$ (the D4-branes). Here we give $K\sub{2n}_{ij}$ and
$p^{\phi}\sub{2n}$ up to $n=2$; note that $\hat\Phi = e^{\g \phi}$.
These expressions are needed to compute the
anomaly and one point functions for the D4-brane in the Hamiltonian formalism
in section \ref{D4anomaly}\footnote{Round brackets $(ij)$ denote symmetrisation and curly
  brackets $\{ij\}$ traceless symmetrisation of indices.}:
\bea
\g p^{\phi}\sub{2} &=& -\frac{1}{d} [ \frac{1}{2(d-1)} \hat{R} 
+ \hat{\Phi}^{-1} \hna^2 \hat\Phi ], \nono \\
   K\sub{2} &=& \frac{1}{2(d-1)} \hat{R}, \nono \\
   K\sub{2}_{ij} &=& \frac{1}{d-1}[ \hat{R}_{ij} - \frac{1}{2d} \hat{R} h_{ij} -
     \hat\Phi^{-1} \hna_{\{i} \pa_{j\}} \hat\Phi]; \\
\g p^{\phi}\sub{4} &=& \frac{1}{2d(d-1)^2(d-3)}[ -3 \hat{R}_{ij} \hat{R}^{ij}
+ \frac{3(d+1)}{4d} \hat{R}^2 - \frac{3}{d} \hna^2 \hat{R} 
- 3(\hat\Phi^{-1} \hna_{\{i} \pa_{j\}} \hat\Phi)^2 \nono \\
   && -2(d-3)(\hat\Phi^{-1} \hna_i(\hat{R}^{ij} \pa_j \hat\Phi) - \frac{d+2}{2d}
\hat\Phi^{-1} \hna^j(\hat{R} \pa_j \hat\Phi) + \frac{1}{2d} \hat\Phi^{-1} \hna^2 (\hat\Phi \hat{R})) \nono\\
   && - 2d (\hat\Phi^{-1} \hna^i \hna^j \hna_{\{i} \hna_{j\}} \hat\Phi - 2
\hat\Phi^{-1} \na^i (\hat\Phi^{-1} \pa^j \hat\Phi \hna_{\{i} \hna_{j\}} \hat\Phi))], \nono \\
   K\sub{4} &=& -\frac{1}{2(d-3)(d-1)^2} [ - \hat{R}_{ij} \hat{R}^{ij} +
     \frac{d+1}{4d} \hat{R}^2 - \frac{1}{d} \hna^2 \hat{R} 
- (\hat\Phi^{-1} \hna_{\{i} \pa_{j\}} \hat\Phi)^2 \nono \\
   && - 2 \hat\Phi^{-1} \hna^i \hna^j \hna_{\{i} \hna_{j\}} \hat\Phi +4
     \hat\Phi^{-1} \na^i (\hat\Phi^{-1} \pa^j \hat\Phi \hna_{\{i} \hna_{j\}} \hat\Phi)], \nono \\
   K\sub{4}^{ij} &=& \g p^{\phi}\sub{4} h^{ij}
   -\frac{1}{(d-1)^2(d-3)}[ -2R^{ik}{\hat{R}^j}_k + \frac{d+1}{2d} \hat{R} \hat{R}^{ij}
     - 2 \hat\Phi^{-2} \hna^i \pa_k \hat\Phi \hna^{\{j} \pa^{k\}} \hat\Phi \nono \\
   && - \frac{1}{d}(\hna^i \pa^j \hat{R} + \hna^2 \hat{R}^{ij}) + \hat\Phi^{-1} \hna_l X^{ijl}], \nono \\
   X^{ijl} &=& -2 \hna_k (\hat\Phi \hat{R}^{kl}) h^{ij} + 2  \hna^{(i} (
   \hat\Phi \hat{R}^{j)l}) - \hna^l(\hat\Phi \hat{R}^{ij}) \nono \\
   &&+ \frac{d+1}{2d}[\hna^l (\hat\Phi \hat{R}) h^{ij} - h^{l(i} \hna^{j)} (\hat\Phi
     \hat{R})] + 2 \hat\Phi^{-1} \hna^l \pa^{(i} \hat\Phi \pa^{j)} \hat\Phi
   - \hat\Phi^{-1} \hna^{\{i} \pa^{j\}} \hat\Phi \pa^l \hat\Phi \nono \\
   && -\frac{2}{d} \hat\Phi^{-1} h^{l(i} \hna^2 \hat\Phi \pa^{j)} \hat\Phi
+ \frac{1}{d} [ h^{l(i} \hat\Phi \pa^{j)} \hat{R} + \frac{d-1}{2} \hat\Phi
  \pa^l \hat{R} h^{ij} - \hna^l (\hat\Phi \hat{R}^{ij}) \nono \\
   && + 2 \hat\Phi \hna^l \hat{R}^{ij} 
- d \hna^l \hna^2 \hat\Phi h^{ij} + h^{l(i} \hna^{j)} \hna^2 \hat\Phi]. \nono
\eea
Note that the terms $K\sub{2}$ and $K\sub{4}$ correspond to the
(non-logarithmic) counterterms in the action.

\end{document}